\documentclass[10pt,letterpaper]{article}
\usepackage{epsf}
\usepackage{epsfig}
\usepackage{multicol}
\usepackage{times}
\usepackage{amssymb}
\usepackage{amsthm}
\usepackage{url}

\RequirePackage{graphicx}

\usepackage{graphicx}

\def\ba{\begin{array}}
\def\ea{\end{array}}
\newcommand{\beq}{\begin{equation}}
\newcommand{\eeq}{\end{equation}}
\newcommand{\bq}{\begin{eqnarray}}
\newcommand{\eq}{\end{eqnarray}}
\newcommand{\bqn}{\begin{eqnarray*}}
\newcommand{\eqn}{\end{eqnarray*}}
\newcommand{\bdis}{\begin{displaymath}}
\newcommand{\edis}{\end{displaymath}}
\newcommand{\bee}{\begin{enumerate}}
\newcommand{\eee}{\end{enumerate}}
\newtheorem{theorem}{Theorem}
\newtheorem{definition}{Definition}
\newtheorem{lemma}{Lemma}

\newtheorem{proposition}{Proposition}

\title{Towards a Theory of Scale-Free Graphs:\\
       Definition, Properties, and Implications
       (Extended Version)\footnote{The primary version 
	 of this paper is forthcoming from {\em Internet Mathematics},
	 2005.}\\
       \vspace{4mm}
       {\large Technical Report CIT-CDS-04-006,
        Engineering \& Applied Sciences Division\\
    California Institute of Technology,
    Pasadena, CA, USA}}
\author{
Lun Li\footnote{Engineering \& Applied Science Division,
 California Institute of Technology, Pasadena, USA},
David Alderson$^*$, Reiko Tanaka\footnote{RIKEN,
Bio-Mimetic Control Research Center, Nagoya, Japan},
John C. Doyle$^*$, Walter Willinger\footnote{AT\&T
Labs--Research, Florham Park, NJ, USA}
}
\date{Updated: October 2005}

\begin{document}
\maketitle

\begin{abstract}

Although the ``scale-free'' literature is large and
growing, it gives neither a precise definition of
scale-free graphs nor rigorous proofs of many of their
claimed properties. In fact, it is easily shown that the
existing theory has many inherent contradictions and
verifiably false claims. In this paper, we propose a new,
mathematically precise, and structural definition of the
extent to which a graph is scale-free, and prove a series
of results that recover many of the claimed properties while
suggesting the potential for a rich and interesting theory.
With this definition, scale-free (or its opposite,
scale-rich) is closely related to other structural graph
properties such as various notions of self-similarity (or
respectively, self-dissimilarity). Scale-free graphs are
also shown to be the likely outcome of random construction
processes, consistent with the heuristic definitions
implicit in existing random graph approaches.  Our approach
clarifies much of the confusion surrounding the sensational
qualitative claims in the scale-free literature, and offers
rigorous and quantitative alternatives.
\end{abstract}

\begin{multicols}{2}

\section{Introduction}

One of the most popular topics recently within the
interdisciplinary study of complex networks has been the
investigation of so-called ``scale-free'' graphs.
Originally introduced by Barab\'{a}si and
Albert~\cite{BarabasiAlbert99}, scale-free (SF) graphs have
been proposed as generic, yet universal models of network
topologies that exhibit power law distributions in the
connectivity of network nodes. As a result of the apparent
ubiquity of such distributions across many naturally
occurring and man-made systems, SF graphs have been
suggested as representative models of complex systems
ranging from the social sciences (collaboration graphs of
movie actors or scientific co-authors) to molecular biology
(cellular metabolism and genetic regulatory networks) to
the Internet (Web graphs, router-level graphs, and AS-level
graphs).  Because these models exhibit features not easily
captured by traditional Erd\"{o}s-Reny\'{i} random graphs
\cite{ErdosRenyi59}, it has been suggested that the
discovery, analysis, and application of SF graphs may even
represent a ``new science of networks''
\cite{BarabasiBook,dorogovtsev-book2003}.

As pointed out in \cite{BollobasRiordan03,BollobasRiordan04}
and discussed in \cite{fox-keller}, despite the
popularity of the SF network paradigm in the complex systems
literature, the definition of ``scale-free'' in the context
of network graph models has never been made precise, and the
results on SF graphs are largely heuristic and
experimental studies with {\it ``rather little rigorous
mathematical work; what there is sometimes confirms and
sometimes contradicts the heuristic
results''}~\cite{BollobasRiordan03}.
Specific usage of ``scale-free'' to describe graphs can be
traced to the observation in Barab\'{a}si and
Albert~\cite{BarabasiAlbert99} that {\it ``a common
property of many large networks is that the vertex
connectivities follow a scale-free power-law
distribution.''} However, most of the SF literature
\cite{albert_barabasi2002,AlbJeongBar99,AlbJeongBar00,
BarabasiAlbert99,PhysicaA,BarBon-SciAmer03,hier1}
identifies a rich variety of additional (e.g.~topological)
signatures beyond mere power law degree distributions in
corresponding models of large networks. One such feature
has been the role of evolutionary growth or rewiring
processes in the construction of graphs. Preferential
attachment is the mechanism most often associated with
these models, although it is only one of several mechanisms
that can produce graphs with power law degree
distributions.

Another prominent feature of SF graphs in this literature
is the role of highly connected ``hubs.'' Power law degree
distributions alone imply that some nodes in the tail of
the power law must have high degree, but ``hubs'' imply
something more and are often said to ``hold the network
together.'' The presence of a hub-like network core yields
a ``robust yet fragile'' connectivity structure that has
become a hallmark of SF network models. Of particular
interest here is that a study of SF models of the
Internet's router topology is reported to show that {\it
``the removal of just a few key hubs from the Internet
splintered the system into tiny groups of hopelessly
isolated routers''}~\cite{BarBon-SciAmer03}.  Thus,
apparently due to their hub-like core structure, SF
networks are said to be simultaneously robust to the random
loss of nodes (i.e.~``error tolerance'') since these tend
to miss hubs, but fragile to targeted worst-case attacks
(i.e.~``attack vulnerability'') \cite{AlbJeongBar00} on
hubs. This latter property has been termed the ``Achilles'
heel'' of SF networks, and it has featured prominently in
discussions about the robustness of many complex networks.
Albert et al.~\cite{AlbJeongBar00} even claim to {\it
``demonstrate that error tolerance... is displayed {\em
only} by a class of inhomogeneously wired networks, called
scale-free networks''} (emphasis added).  We will use the
qualifier ``SF hubs'' to describe high degree nodes which
are so located as to provide these ``robust yet fragile''
features described in the SF literature, and a goal of this
paper is to clarify more precisely what topological
features of graphs are involved.

There are a number of properties in addition to power law
degree distributions, random generation, and SF hubs that
are associated with SF graphs, but unfortunately, it is
rarely made clear in the SF literature which of these
features define SF graphs and which features are then
consequences of this definition.  This has led to
significant confusion about the defining features or
characteristics of SF graphs and the applicability of these
models to real systems. While the usage of ``scale-free''
in the context of graphs has been imprecise, there is
nevertheless a large literature on SF graphs, particularly
in the highest impact general science journals. For
purposes of clarity in this paper, we will use the term
{\it SF graphs} (or equivalently, {\it SF networks}) to
mean those objects as studied and discussed in this ``SF
literature,'' and accept that this inherits from that
literature an imprecision as to what exactly SF means. One
aim of this paper is to capture as much as possible of the
``spirit'' of SF graphs by proving their most widely
claimed properties using a minimal set of axioms. Another is
to reconcile these theoretical properties with the
properties of real networks, and in particular the
router-level graphs of the Internet.

Recent research into the structure of several important
complex networks previously claimed to be ``scale-free''
has revealed that, even if their graphs could have
approximately power law degree distributions, the networks
in question do not have SF hubs, that the most highly
connected nodes do not necessarily represent an ``Achilles'
heel'', and that their most essential ``robust, yet
fragile'' features actually come from aspects that are only
indirectly related to graph connectivity. In particular,
recent work in the development of a first-principles
approach to modeling the router-level Internet has shown
that the core of that network is constructed from a mesh of
high-bandwidth, low-connectivity routers and that this
design results from tradeoffs in technological, economic,
and performance constraints on the part of Internet Service
Providers (ISPs)~\cite{sigcomm04,PNAS05}. A related line of
research into the structure of biological metabolic
networks has shown that claims of SF structure fail to
capture the most essential biochemical as well as ``robust
yet fragile'' features of cellular metabolism and in many
cases completely misinterpret the relevant biology
\cite{tanaka2005,tanaka-doyle:2004_2}. This mounting
evidence against the heart of the SF story creates a
dilemma in how to reconcile the claims of this broad and
popular framework with the details of specific application
domains (see also the discussion in \cite{fox-keller}). 
In particular, it is now clear that either the
Internet and biology networks are very far from ``scale
free'', or worse, the claimed properties of SF networks are
simply false at a more basic mathematical level,
independent of any purported applications.

The main purpose of this paper is to demonstrate that when
properly defined, ``scale-free networks'' have the potential
for a rigorous, interesting, and rich mathematical theory.
Our presentation assumes an understanding of fundamental
Internet technology as well as comfort with a theorem-proof
style of exposition, but not necessarily any familiarity
with existing SF literature.
While we leave many open questions and conjectures
supported only by numerical experiments, examples, and
heuristics, our approach reconciles the existing
contradictions and recovers many claims regarding the graph
theoretic properties of SF networks.
A main contribution of this paper is the introduction of a
structural metric that allows us to differentiate between
all simple, connected graphs having an identical degree
sequence, particularly when that sequence follows a power law.
Our approach is to leverage related definitions from other
disciplines, where available, and utilize existing methods
and approaches from graph theory and statistics.
While the proposed structural metric is not intended as a
general measure of all graphs, we demonstrate that it yields
considerable insight into the claimed properties of SF
graphs and may even provide a view into {\it the extent to
which a graph is scale-free}.  Such a view has the benefit
of being {\it minimal}, in the sense that it relies on few
starting assumptions, yet yields a rich and general
description of the features of SF networks. While far from
complete, our results are consistent with the main thrust
of the SF literature and demonstrate that a rigorous and
interesting ``scale-free theory'' can be developed, with
very general and robust features resulting from relatively
weak assumptions.  In the process, we resolve some of the
misconceptions that exist in the general SF literature and
point out some of the deficiencies associated with previous
applications of SF models, particularly to technological
and biological systems.

The remainder of this article is organized as follows.
Section \ref{sec:background} provides the basic background
material, including mathematical definitions for scaling
and power law degree sequences, a discussion of related
work on scaling that dates back as far as 1925, and
various additional work on self-similarity in graphs. We
also emphasize here why high variability is a much more
important concept than scaling or power laws per se.
Section \ref{sec:ConvSF} briefly reviews the recent
literature on SF networks, including the failure of
SF methods in Internet applications.
In Section \ref{sec:struct}, we introduce a metric for
graphs having a power-law in their degree sequence, one
that highlights the diversity of such graphs and also
provides insight into existing notions of graph structure
such as self-similarity/self-dissimilarity, motifs, and
degree-preserving rewiring.
Our metric is ``structural''---in the sense that it
depends only on the connectivity of a given graph and not
the process by which the graph is constructed---and can
be applied to any graph of interest.
Then, Section~\ref{sec:prob} connects these structural
features with the probabilistic perspective common in
statistical physics and traditional random graph theory,
with particular connections to graph likelihood, degree
correlation, and assortative/disassortative mixing.
Section \ref{sec:hot} then traces the shortcomings
of the existing SF theory and uses our alternate
approach to outline what sort of potential foundation for a
broader and more rigorous SF theory may be built from
mathematically solid definitions.  We also put the ensuing
SF theory in a broader perspective by comparing it with
recently developed alternative models for the Internet
based on the notion of {\it Highly Optimized Tolerance
(HOT)}~\cite{CD}.  To demonstrate that the
Internet application considered in this paper is
representative of a broader debate about complex systems,
we discuss in Section \ref{sec:bio} another application
area that is very popular within the existing SF
literature, namely biology, and illustrate that there
exists a largely parallel SF vs.~HOT story as well.
We conclude in Section \ref{sec:concl} that many open problems
remain, including theoretical conjectures and the potential
relevance of rigorous SF models to applications other than
technology.

\section{Background}\label{sec:background}

This section provides the necessary background for our
investigation of what it means for a graph to be ``scale-free''.
In particular, we present some basic definitions
and results in random variables, comment on approaches to
the statistical analysis of high variability data, and
review notions of scale-free and self-similarity as they
have appeared in related domains.

While the advanced reader will find much of this section
elementary in nature, our experience is that much of the
confusion on the topic of SF graphs stems from
fundamental differences in the methodological perspectives
between statistical physics and that of mathematics or
engineering.  The intent here is to provide material that
helps to bridge this potential gap in addition to setting
the stage from which our results will follow.

\subsection{Power Law and Scaling Behavior}

\subsubsection{Non-stochastic vs.~Stochastic Definitions}

A finite {\it sequence} $y=(y_1, y_2, \dots, y_n$) of real numbers,
assumed without loss of generality always to be ordered such that
$y_1 \ge y_2 \ge \ldots \ge y_n$, is said to follow a {\it
power law} or {\it scaling relationship} if
\begin{eqnarray} k & = & c {y_k}^{-\alpha},
\label{eq:scaling1}
\end{eqnarray}
where $k$ is (by definition) the {\it rank} of $y_k$,
$c$ is a fixed constant, and $\alpha$ is called the
{\it scaling index}. Since
$\log k = \log (c) - \alpha \log (y_k)$,
the relationship for the rank $k$ versus $y$ appears as a
line of slope $-\alpha$ when plotted on a log-log scale. In
this manuscript, we refer to the relationship
(\ref{eq:scaling1}) as the {\it size-rank} (or {\it
cumulative}) form of scaling. While the definition of
scaling in (\ref{eq:scaling1}) is fundamental to the
exposition of this paper, a more common usage of power laws
and scaling occurs in the context of random variables and
their distributions. That is, assuming an underlying
probability model $P$ for a non-negative random variable
$X$, let $F(x)=P[X \leq x]$ for $x \geq 0$ denote the {\it
(cumulative) distribution function (CDF) of $X$}, and let
$\bar{F}(x)=1-F(x)$ denote the {\it complementary CDF
(CCDF)}. A typical feature of commonly-used distribution
functions is that the (right) tails of their CCDFs decrease
exponentially fast, implying that all moments exist and are
finite. In practice, this property ensures that any
realization $(x_1, x_2, \dots, x_n)$ from an independent
sample $(X_1, X_2, \dots, X_n)$ of size $n$ having the
common distribution function $F$ concentrates tightly
around its (sample) mean, thus exhibiting low variability
as measured, for example, in terms of the (sample) standard
deviation.

In this stochastic context, a random variable $X$
or its corresponding distribution function $F$ is said
to follow a {\it power law} or is {\it scaling} with index
$\alpha>0$ if, as $x\rightarrow\infty$,
\begin{eqnarray}
 P[X>x] = 1-F(x) \approx cx^{-\alpha},
\label{eq:rv-scaling}
\end{eqnarray}
for some constant $0<c<\infty$ and a {\it tail index} $\alpha > 0$.
Here, we write $f(x) \approx g(x)$ as $x\rightarrow\infty$ if
$f(x)/g(x) \rightarrow 1$ as $x\rightarrow\infty$.
For $1 < \alpha <2$, $F$
has infinite variance but finite mean, and for $0<\alpha \leq 1$,
$F$ has not only infinite variance but also infinite mean.
In general, all moments of $F$ of order $\beta\geq\alpha$ are infinite.
Since relationship (\ref{eq:rv-scaling}) implies
$\log(P[X>x]) \approx \log(c) -\alpha \log(x)$,
doubly logarithmic plots of $x$ versus $1-F(x)$ yield
straight lines of slope $-\alpha$, at least for
large $x$.  Well-known examples of power law distributions
include the Pareto distributions of the first and second kind
\cite{johnson}.
In contrast, {\it exponential distributions}
(i.e., $P[X>x]=e^{-\lambda x}$) 
result in approximately straight lines on semi-logarithmic plots.

If the derivative of the cumulative distribution
function $F(x)$ exists, then $f(x) = \frac{d}{dx}F(x)$ is
called the {\it (probability) density function} of $X$ and
implies that the stochastic cumulative form of scaling
or size-rank relationship (\ref{eq:rv-scaling}) has
an equivalent {\it noncumulative} or {\it size-frequency}
counterpart given by
\begin{eqnarray}
f(x) \approx c x^{-(1+\alpha)}
 \label{eq:density}
\end{eqnarray}
which appears similarly as a line of slope
$-(1+\alpha)$ on a log-log scale. However, as discussed in
more detail in Section \ref{sec:freq} below, the use of this
noncumulative form of scaling has been a source of many common
mistakes in the analysis and interpretation of actual data
and should generally be avoided.

Power-law distributions are called scaling distributions
because the sole response to conditioning is a change in scale;
that is, if the random variable $X$ satisfies relationship
(\ref{eq:rv-scaling}) and $x>w$, then the conditional distribution of $X$
given that $X>w$ is given by
\[
P[X > x | X > w] =
\frac{P[X > x]}{P[X > w]} \approx c_1 x^{-\alpha},
\]
where the constant $c_1$ is independent of $x$ and is given by
$c_1 = 1/w^{-\alpha}$. Thus, at least for large values of $x$,
$P[X > x | X > w]$ is identical to the
(unconditional) distribution $P[X > x]$, except
for a change in scale.  In contrast, the
exponential distribution gives
\[
P(X > x | X > w) = e^{-\lambda(x-w)},
\]
that is, the conditional distribution is also identical to
the (unconditional) distribution, except for a change of
location rather than scale.  Thus we prefer the term {\it
scaling} to {\it power law}, but will use them
interchangeably, as is common.

It is important to emphasize again the differences between
these alternative definitions of scaling. Relationship
(\ref{eq:scaling1}) is {\it non-stochastic}, in the sense
that there is no assumption of an underlying probability
space or distribution for the sequence $y$, and in what
follows we will always use the term {\it sequence} to
refer to such a non-stochastic object $y$, and accordingly
we will use {\it non-stochastic} to mean simply the absence
of an underlying probability model.
In contrast, the definitions in (\ref{eq:rv-scaling}) and
(\ref{eq:density}) are  {\it stochastic} and require an
underlying probability model.  Accordingly, when referring
to a random variable $X$ we will explicitly mean an
ensemble of values or realizations sampled from a common
distribution function $F$, as is common usage.  We will
often use the standard and trivial method of viewing a
nonstochastic model as a stochastic one with a singular
distribution.

These distinctions between stochastic and nonstochastic
models will be important in this paper. Our approach allows
for but does not require stochastics.  In contrast, the SF
literature almost exclusively assumes some underlying
stochastic models, so we will focus some attention on
stochastic assumptions. Exclusive focus on stochastic
models is standard in statistical physics, even to the
extent that the possibility of non-stochastic constructions
and explanations is largely ignored. This seems to be the
main motivation for viewing the Internet's router topology
as a member of an ensemble of random networks, rather than
an engineering system driven by economic and technological
constraints plus some randomness, which might otherwise
seem more natural. Indeed, in the SF literature ``random''
is typically used more narrowly than stochastic to mean,
depending on the context, exponentially, Poisson, or
uniformly distributed. Thus phrases like ``scale-free
versus random'' (the ambiguity in ``scale-free''
notwithstanding) are closer in meaning to ``scaling versus
exponential,'' rather than ``non-stochastic versus
stochastic.''

\subsubsection{Scaling and High
Variability}\label{sec:ScalingAndHighVar}

An important feature of sequences that follow the scaling
relationship (\ref{eq:scaling1}) is that they exhibit {\it
high variability}, in the sense that deviations from the
average value or (sample) mean can vary by orders of
magnitude, making the average largely uninformative and not
representative of the bulk of the values.
To quantify the notion of {\it variability}, we use the standard
measure of {\it (sample) coefficient of variation}, which for a given
sequence $y=(y_1, y_2, \dots, y_n)$ is defined as
\begin{eqnarray}
CV(y)={STD}(y)/\bar{y}, \label{eq:cv1a}
\end{eqnarray}
where $\bar{y}=n^{-1}\sum_{k=1}^n y_k$ is the average
size or (sample) mean of $y$ and ${STD}(y)=(\sum_{k=1}^n
(y_k -\bar{y})^2 /(n-1))^{1/2}$ is the (sample) standard
deviation, a commonly-used metric for measuring the
deviations of $y$ from its average $\bar{y}$.
The presence of high variability in a sequence of values often
contrasts greatly with the typical experience of many scientists who work
with empirical data exhibiting {\it low variability}---that is, observations
that tend to concentrate tightly around the (sample) mean and allow for
only small to moderate deviations from this mean value.

A standard ensemble-based measure for quantifying the variability
inherent in a random variable $X$ is the
{\it (ensemble) coefficient of variation CV($X$)} defined as
\begin{eqnarray}
CV(X)=\sqrt{{\rm Var}(X)}/E(X),
\label{eq:cv2}
\end{eqnarray}
where $E(X)$ and $Var(X)$ are the (ensemble) mean and (ensemble)
variance of $X$, respectively.
If $x=(x_1 ,x_2 ,\ldots,x_n )$ is a realization of an independent
and identically distributed (iid) sample of size $n$ taken from the
common distribution $F$ of $X$, it is easy to see that the quantity $CV(x)$
defined in (\ref{eq:cv1a}) is simply an estimate of $CV(X)$.
In particular, if $X$ is scaling with $\alpha < 2$,
then $CV(X)=\infty$, and estimates $CV(x)$ of $CV(X)$
diverge for large sample sizes.
Thus, random variables having a scaling distribution are
extreme in exhibiting high variability.  However, scaling
distributions are only a subset of a larger family of
{\it heavy-tailed distributions} (see \cite{wsc04}
and references therein) that exhibit high variability.
As we will show, it turns out that some of the most celebrated
claims in the SF literature (e.g.~the presence of highly connected
central hubs) have as a necessary condition only the presence of
high variability and not necessarily strict scaling per se.
The consequences of this observation are far-reaching,
especially because it shifts the focus from
scaling relationships, their tail indices, and their
generating mechanisms to an emphasis on heavy-tailed distributions
and identifying the main sources of ``high variability.''


\begin{figure*}[th]
  \begin{center}
  \includegraphics[width=.7\linewidth]{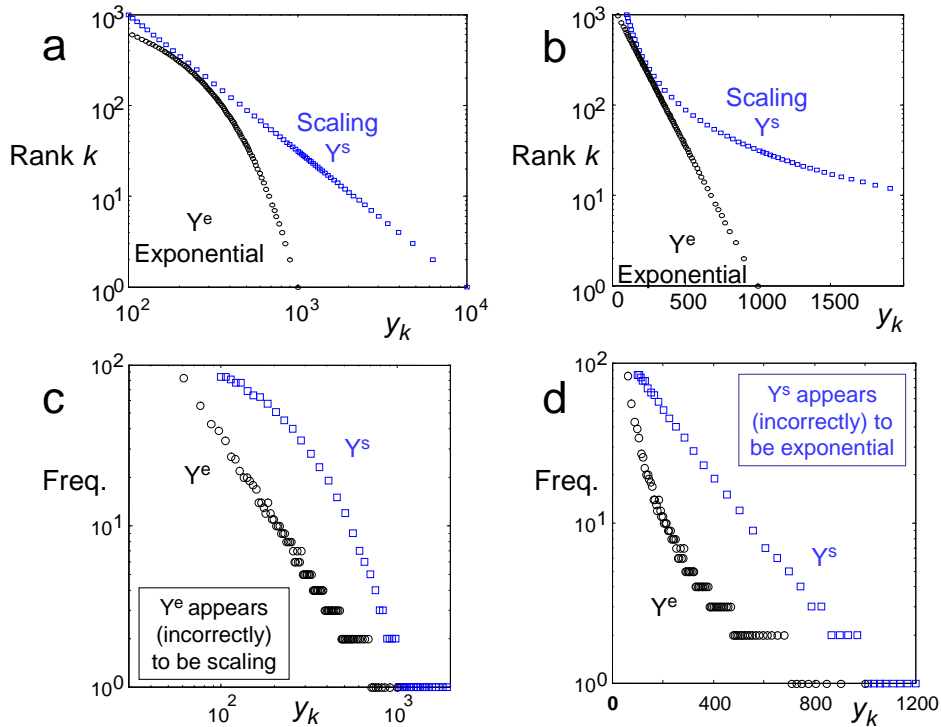}
  \vspace{-2mm}
  \caption{\footnotesize   \label{fig:ExpVsScaling}
   {\sc Plots of exponential $y^e$ (black circles)
   and scaling $y^s$ (blue squares) sequences.}
   \textbf{(a) Doubly logarithmic size-rank plot:} $y^s$ is scaling
(to within integer tolerances) and thus $y^s_k$ versus $k$ is
approximately a straight line.
  \textbf{(b) Semi-logarithmic size-rank plot:}$y^e$ is exponential
   (to within integer tolerances) and thus $y^e_k$ versus $k$
  is approximately a straight line on semi-logarithmic plots
   \textbf{(c) Doubly logarithmic size-frequency plot:} $y^e$ is
   exponential but appears incorrectly to be scaling
  \textbf{(d) Semi-logarithmic size-frequency plot:}$y^s$ is
  scaling but appears incorrectly to be exponential.
  }
  \end{center}
\vspace{-4ex}
\end{figure*}


\subsubsection{Cumulative vs.~Noncumulative log-log Plots}
\label{sec:freq}

While in principle there exists an unambiguous mathematical
equivalence between distribution functions and their
densities, as in (\ref{eq:rv-scaling}) and
(\ref{eq:density}), no such relationship can be assumed to
hold in general when plotting sequences of real or integer
numbers or measured data cumulatively and noncumulatively.
Furthermore, there are good practical reasons to avoid
noncumulative or size-frequency plots altogether (a sentiment
echoed in~\cite{Newman05}), even though they are often used
exclusively in some communities. To illustrate the basic
problem, we first consider two sequences, $y^s$ and $y^e$,
each of length 1000, where $y^s =
(y^s_1,\ldots,y^s_{1000})$ is constructed so that its
values all fall on a straight line when plotted on doubly
logarithmic (i.e., log-log) scale.  Similarly, the values
of the sequence $y^e = (y^e_1,\ldots,y^e_{1000})$ are
generated to fall on a straight line when plotted on
semi-logarithmic (i.e., log-linear) scale. The {\sc
matlab} code for generating these two sequences is available
for electronic download~\cite{matlab-code}.
When ranking the values in each sequence in decreasing order, we
obtain the following unique largest (smallest) values, with their
corresponding frequencies of occurrence given in
parenthesis:
\begin{eqnarray*}
y^s & =& \{10000(1), 6299(1), 4807(1), 3968(1), 3419(1),\ldots\\
&& \ \ \dots, 130(77), 121(77), 113(81), 106(84), 100(84)\},\\
y^e& =& \{1000(1), 903(1), 847(1), 806(1), 775(1),\ldots\\
&& \ \ \dots, 96(39), 87(43), 76(56), 61(83), 33(180)\},
\end{eqnarray*}
and the full sequences are plotted in Figure~\ref{fig:ExpVsScaling}.
In particular, the doubly logarithmic plot in
Figure~\ref{fig:ExpVsScaling}(a) shows the cumulative
or size-rank relationships associated with the sequences
$y^s$ and $y^e$: the largest value of $y^s$ (i.e., 10,000)
is plotted on the x-axis and has rank 1 (y-axis), the
second largest value of $y^s$ is 6,299 and has rank 2,
all the way to the end, where the smallest value of $y^s$
(i.e., 100) is plotted on the x-axis and has rank 1000 (y-axis).
Similarly for $y^e$.  In full agreement with the underlying
generation mechanisms, plotting on doubly logarithmic scale
the rank-ordered sequence of $y^s$ versus rank $k$ results
in a straight line; i.e., $y^s$ is scaling (to within integer
tolerances).  The same plot for the rank-ordered sequence of
$y^e$ has a pronounced concave shape and decreases rapidly for
large ranks---strong evidence for an exponential size-rank
relationship.  Indeed, as shown in Figure~\ref{fig:ExpVsScaling}(b),
plotting on semi-logarithmic scale the rank-ordered sequence
of $y^e$ versus rank $k$ yields a straight line; i.e., $y^e$
is exponential (to within integer tolerances).  The same plot
for $y^s$ shows a pronounced convex shape and decreases very
slowly for large rank values---fully consistent with a scaling
size-rank relationship.  Various metrics for these two sequences are
\[
\begin{tabular}{|c|c|c|}
  \hline
     & $y^e$ & $y^s$ \\ \hline
(sample) mean  & 167 & 267 \\  \hline
(sample) median & 127 & 153 \\ \hline
(sample) STD   & 140 & 504 \\  \hline
(sample) CV    & .84 & 1.89 \\ \hline
\end{tabular}
\]
and all are consistent with exponential and scaling
sequences of this size.


\begin{figure*}[th]
  \begin{center}
  \includegraphics[width=.75\linewidth]{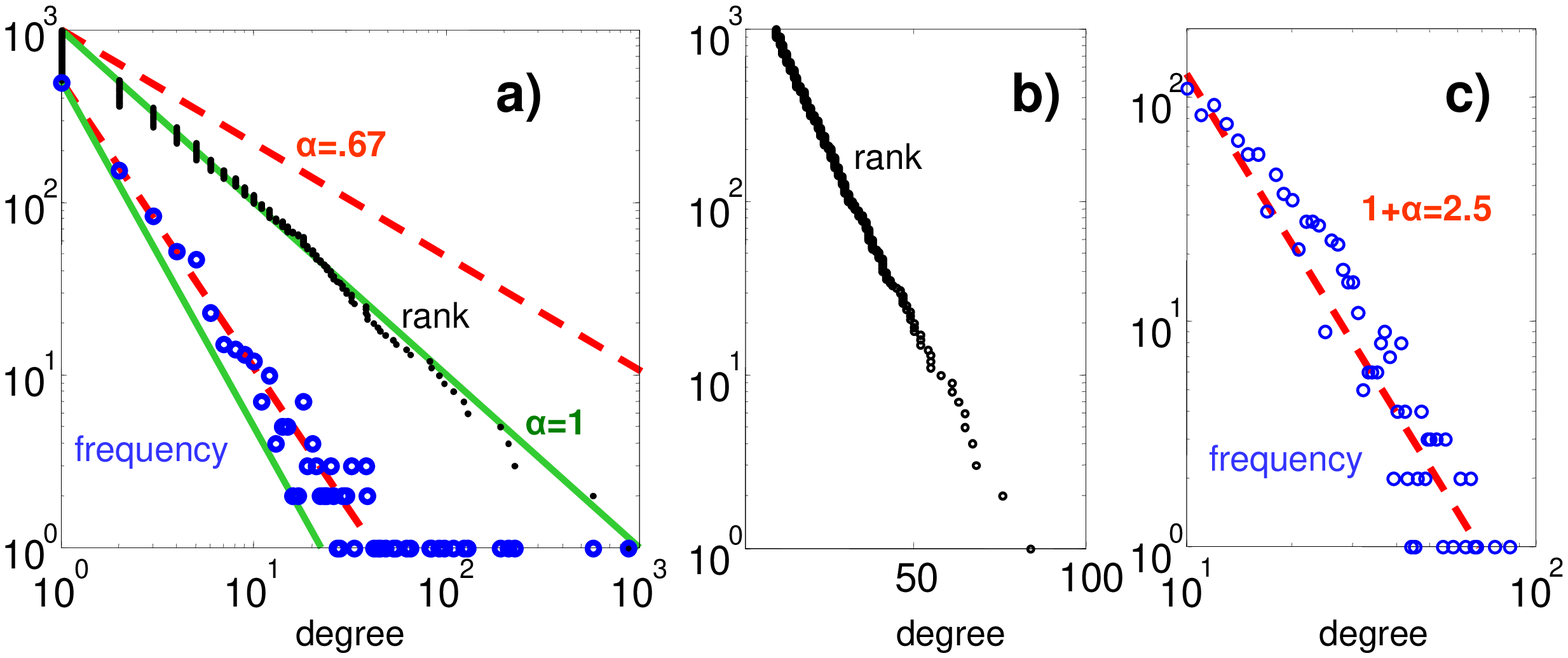}
  \vspace{-2mm}
  \caption{\footnotesize   \label{fig:plot-example}
    {\sc A common error when inferring/estimating scaling behavior.}
     \textbf{(a) 1000 integer data points sampled from the scaling
      distribution $P(X \ge x) = x^{-1}$, for $x \ge 1$.}
      The lower size-frequency plot (blue circles) tends to underestimate
      the scaling index $\alpha$; it supports a slope estimate of about
      -1.67 (red dashed line), implying an $\alpha$-estimate of about
      =0.67 that is obviously inconsistent with the true value of
      $\alpha=1$ (green line). 
      The size-rank plot of the exact same data (upper, black dots) clearly
      supports a scaling behavior and yields an $\alpha$-estimate that is
      fully consistent with the true scaling index $\alpha=1$ (green line).
     \textbf{(b)  1000 data points sampled from an exponential
      distribution plotted on log-linear scale.}  The size-rank plot clearly
      shows that the data are exponential and that scaling is implausible.
     \textbf{(c) The same data as in (b) plotted on log-log scale.}
      Based on the size-frequency plot, it is plausible to infer incorrectly
      that the data are consistent with scaling behavior, with a slope
      estimate of about -2.5, implying an $\alpha$-estimate of about 1.5.
  }
  \end{center}
\vspace{-4ex}
\end{figure*}


To highlight the basic problem caused by the use of
noncumulative or size-frequency relationships, consider
Figure~\ref{fig:ExpVsScaling}(c) and (d) that show on
doubly logarithmic scale and semi-logarithmic scale,
respectively, the non-cumulative or size-frequency plots
associated with the sequences $y^s$ and $y^e$: the largest
value of $y^s$ is plotted on the x-axis and has frequency 1
(y-axis), the second largest value of $y^s$ has also
frequency 1, etc., until the end where the smallest value
of $y^s$ happens to occur 84 times (to within integer
tolerances).  Similarly for $y^e$, where the smallest value
happens to occur 180 times. It is common to conclude
incorrectly from plots such as these, for example, that the
sequence $y^e$ is scaling (i.e., plotting on doubly
logarithmic scale size vs.~frequency results in an
approximate straight line) and the sequence $y^s$ is
exponential (i.e., plotting on semi-logarithmic scale size
vs.~frequency results in an approximate straight
line)---exactly the opposite of what is correctly inferred
about the sequences using the cumulative or size-rank plots
in Figure~\ref{fig:ExpVsScaling}(a) and (b).

In contrast to the size-rank plots of the style in
Figure~\ref{fig:ExpVsScaling}(a)-(b) that depict the raw data
itself and are unambiguous, the use of size-frequency plots
as in Figure~\ref{fig:ExpVsScaling}(c)-(d),
while straightforward to describe low variable data, creates
ambiguities and can easily lead to mistakes when applied to
high variability data.  First, for high precision measurements it is
possible that each data value appears only once in a sample set,
making raw frequency-based data rather uninformative.  To
overcome this problem, a typical approach is to group
individual observations into one of a small number of {\it
bins} and then plot for each bin (x-axis) the relative number of
observations in that bin (y-axis). The problem is that choosing the size and
boundary values for each bin is a process generally left up
to the experimentalist, and this {\it binning process} can
dramatically change the nature of the resulting size-frequency
plots as well as their interpretation (for a concrete example, see
Figure~\ref{fig:mercator} in Section~\ref{sec:measurements}).

These examples have been artificially constructed
specifically to dramatize the effects associated with the
use of cumulative or size-rank vs.~noncumulative or
size-frequency plots for assessing the presence or absence
of scaling in given sequence of observed values. While they
may appear contrived, errors such as those illustrated in
Figure~\ref{fig:ExpVsScaling} are easy to make and are
widespread in the complex systems literature. In fact,
determining whether a realization of a sample of size $n$
generated from one and the same (unknown) underlying
distribution is consistent with a scaling distribution and
then estimating the corresponding tail index $\alpha$ from
the corresponding size-frequency plots of the data is even
more unreliable.  Even under the most idealized
circumstances using synthetically generated pseudo-random
data, size-frequency plots can mislead as shown in the
following easily reproduced numerical experiments. Suppose
that 1000 (or more) integer values are generated by
pseudo-random independent samples from the distribution
$F(x) = 1-x^{-1}$ ($P (X\ge x ) = x^{-1}$) for $x \ge 1$.
For example, this can be done with the {\sc matlab}
fragment \texttt{x=floor(1./rand(1,1000))} where
\texttt{rand(1,1000)} generates a vector of 1000 uniformly
distributed floating point numbers between 0 and 1, and
\texttt{floor} rounds down to the next lowest integer. In
this case, discrete equivalents to equations
(\ref{eq:rv-scaling}) and  (\ref{eq:density}) exist, and
for $x \gg 1$, the density function $f(x)=P[X=x]$ is given
by
\begin{eqnarray*}
 P[X = x] & = & P[X \ge x] - P[X \ge x + 1] \\
  & = & x^{ - 1}  - \left( {x + 1} \right)^{ - 1}  \\
  & \approx & x^{ - 2}.
\end{eqnarray*}
Thus it might appear that the true tail index (i.e.,
$\alpha = 1$) could be inferred from examining either the
size-frequency or size-rank plots, but as illustrated in
Figure~\ref{fig:plot-example} and described in the caption,
this is not the case.

Though there are more rigorous and reliable methods for
estimating $\alpha$ (see for example~\cite{resnick}), the
(cumulative) size-rank plots have significant advantages in
that they show the raw data directly, and possible
ambiguities in the raw data notwithstanding, they are also
highly robust to a range of measurement errors and noise.
Moreover, experienced readers can judge at a glance whether
a scaling model is plausible, and if so, what a reasonable
estimate of the unknown scaling parameter $\alpha$ should
be. For example, that the scatter in the data in
Figure~\ref{fig:plot-example}(a) is consistent with a
sample from $P(X \ge  x) = x^{-1}$ can be roughly
determined by visual inspection, although additional
statistical tests could be used to establish this more
rigorously. At the same time, even when the underlying
random variable $X$ is scaling, size-frequency plots
systematically underestimate $\alpha$, and worse, have a
tendency to suggest that scaling exists where it does not.
This is illustrated dramatically in
Figure~\ref{fig:plot-example}(b)-(c), where exponentially
distributed samples are generated using
\texttt{floor(10*(1-log(rand(1,n))))}.  The size-rank plot in
Figure~\ref{fig:plot-example}(b) is approximately a
straight line on a semilog plot, consistent with an
exponential distribution. The loglog size-frequency plot
Figure~\ref{fig:plot-example}(c) however could be used
incorrectly to claim that the data is consistent with a
scaling distribution, a surprisingly common error in the SF
and broader complex systems literature. Thus even if one a
priori assumes a probabilistic framework, (cumulative)
size-rank plots are essential for reliably inferring and
subsequently studying high variability, and they therefore
are used exclusively in this paper.

\subsubsection{Scaling: More ``normal'' than Normal}

While power laws in event size statistics in many
complex interconnected systems have recently attracted a
great deal of popular attention, some of the aspects of
scaling distributions that are crucial and important for
mathematicians and engineers have been largely ignored in
the larger complex systems literature. This subsection will briefly
review one aspect of scaling that is particularly revealing
in this regard and is a summary of results described in
more detail in \cite{Mandelbrot97,wsc04}.

Gaussian distributions are universally viewed as
``normal'', mainly due to the well-known Central Limit
Theorem (CLT). In particular, the ubiquity of Gaussians is
largely attributed to the fact that they are invariant and
attractors under aggregation of summands, required only to
be independent and identically distributed (iid) and have
finite variance \cite{fellerII}. Another convenient aspect
of Gaussians is that they are completely specified by mean
and variance, and the CLT justifies using these statistics
whenever their estimates robustly converge, even when the
data could not possibly be Gaussian.  For example, much
data can only take positive values (e.g. connectivity) or
have hard upper bounds but can still be treated as
Gaussian. It is understood that this approximation would
need refinement if additional statistics or tail behaviors
are of interest. Exponential distributions have their own
set of invariance properties (e.g.~conditional expectation)
that make them attractive models in some cases. The ease by
which Gaussian data is generated by a variety of mechanisms
means that the ability of any particular model to reproduce
Gaussian data is not counted as evidence that the model
represents or explains other processes that yield
empirically observed Gaussian phenomena. However, a
disconnect often occurs when data have high variability, that
is, when variance or coefficient of variation estimates don't
converge. In particular, the above type of reasoning is
often misapplied to the explanation of data that are
approximately scaling, for reasons that we will discuss
below.

Much of science has focused so exclusively on low
variability data and Gaussian or exponential models that
low variability is not even seen as an assumption. Yet much
real world data has extremely high variability as
quantified, for example, via the coefficient of variation
defined in (\ref{eq:cv2}).  When exploring stochastic
models of high variability data, the most relevant
mathematical result is that the CLT has a generalization
that relaxes the finite variance (e.g.~finite $CV$)
assumption, allows for high variability data arising from
underlying infinite variance distributions, and yields
{\em stable laws} in the limit.  There is a rich and extensive
theory on stable laws (see for example  \cite{StableLaws}), which
we will not attempt to review, but mention only the most
important features.  Recall that a random
variable $U$ is said to have a {\it stable law (with index
$0< \alpha \leq 2$)} if for any $n\geq 2$, there is a
real number $d_n$ such that
\[
U_1 + U_2 +\cdots + U_n \stackrel{d}{=} n^{1/\alpha}U+d_n ,
\]
where $U_1,~U_2,~\ldots,~U_n$ are independent copies of $U$,
and where $\stackrel{d}{=}$ 
denotes equality in distribution.
Following \cite{StableLaws}, the stable laws on the
real line can be represented as a four-parameter family
$S_\alpha (\sigma, \beta, \mu)$, with the {\it index} $\alpha$,
$0<\alpha\leq 2$; the {\it scale parameter} $\sigma>0$; the
{\it skewness parameter} $\beta$, $-1\leq\beta\leq1$; and the
{\it location (shift) parameter} $\mu$, $-\infty<\mu<\infty$.
When $1<\alpha<2$, the shift parameter is the mean,
but for $\alpha\leq1$, the mean is infinite.  There is an abrupt
change in tail behavior of stable laws at the boundary
$\alpha=2$.  While for $\alpha<2$, all stable laws are
scaling in the sense that they satisfy condition
(\ref{eq:rv-scaling}) and thus exhibit infinite variance
or high variability; the case $\alpha=2$ is special and
represents a familiar, not scaling distribution---the
Gaussian (normal) distribution; i.e., $S_2 (\sigma,0,\mu) =
N(\mu,2\sigma^2 )$, corresponding to the finite variance or low
variability case.  While with the exception of Gaussian, Cauchy,
and Levy distributions, the distributions of stable random
variables are not known in closed form, they are known to be
the only fixed points of the renormalization group transformation
and thus arise naturally in the limit of properly normalized sums
of iid scaling random variables. From an unbiased mathematical view,
the most salient features of scaling distributions are this and
additional strong invariance properties (e.g.~to marginalization,
mixtures, maximization), and the ease with which scaling is
generated by a variety of mechanisms
\cite{Mandelbrot97,wsc04}. Combined with the abundant high
variability in real world data, these features suggest that
scaling distributions are in a sense more ``normal'' than
Gaussians and that they are convenient and parsimonious
models for high variability data in as strong a sense as
Gaussians or exponentials are for low variability data.

While the ubiquity of scaling is increasingly recognized
and even highlighted in the physics and the popular
complexity literature
\cite{BakBook,BuchananBook,BarabasiBook,BallBook}, the
deeper mathematical connections and their rich history in
other disciplines have been largely ignored, with
serious consequences.  Models of complexity using
graphs, lattices, cellular automata, and sandpiles
preferred in physics and the standard laboratory-scale
experiments that inspired these models exhibit scaling only
when finely tuned in some way. So even when accepted as
ubiquitous, scaling is still treated as arcane and exotic,
and ``emergence'' and ``self-organization'' are invoked to
explain how this tuning might happen~\cite{ieee-paper}.
For example, that SF network
models supposedly replicate empirically observed scaling
node degree relationships that are not easily captured by
traditional Erd\"{o}s-Reny\'{i} random graphs
\cite{BarabasiAlbert99} is presented as evidence for model
validity. But given the strong invariance properties of
scaling distributions, as well as the multitude of diverse mechanisms
by which scaling can arise in the first place~\cite{Newman05},
it becomes clear that an ability to generate
scaling distributions ``explains'' little, if anything.  Once
high variability appears in real data then scaling
relationships become a natural outcome of the processes
that measure them.


\begin{figure*}[th]
  \begin{center}
  \includegraphics[width=.95\linewidth]{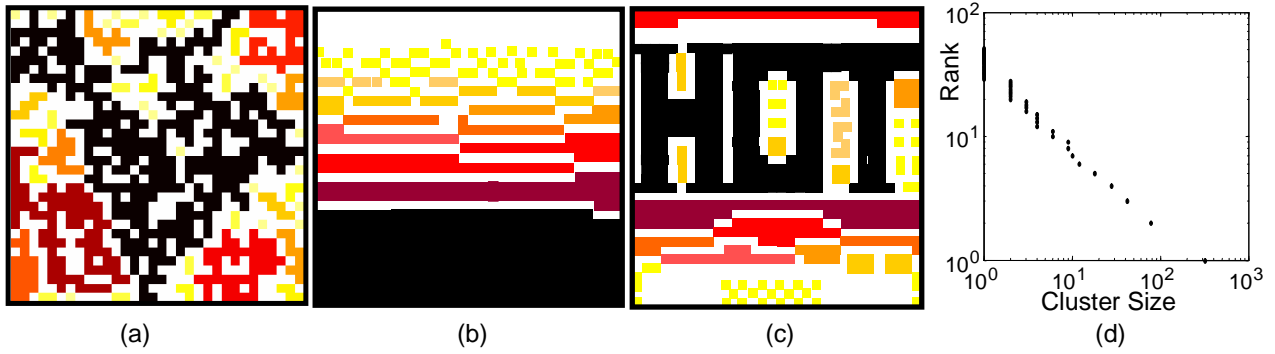}
  \vspace{-2mm}
  \caption{\footnotesize   \label{fig:lattices}
   \textbf{\sc Percolation lattices with scaling cluster sizes.}
   Lattices (a)-(c) have the exact same scaling sequence of
   cluster sizes (d) and the same (critical) density $\approx .59$).
   While random lattice such as in (a) have been be called
   ``scale-free'', the highly structured lattices in (b) or (c)
   typically would not.  This suggests that, even within the framework of
   percolation, scale-free usually means something beyond simple
   scaling of some statistics and refers to geometric or
   topological properties.
  }
  \end{center}
\vspace{-4ex}
\end{figure*}


\subsection{Scaling, Scale-free and Self-Similarity}
\label{sec:sf-vs-scaling}

Within the physics community it is common to refer to functions
of the form (\ref{eq:density}) as {\it scale-free} because they
satisfy the following property
\begin{eqnarray}
f(ax) = g(a) f(x).
\label{eq:scalefree}
\end{eqnarray}
As reviewed by Newman~\cite{Newman05}, the idea is that an
increase by a factor $a$ in the scale or units by which one
measures $x$ results in no change to the overall density
$f(x)$ except for a multiplicative scaling factor.
Furthermore, functions consistent with (\ref{eq:density})
are the {\it only} functions that are scale-free in the
sense of (\ref{eq:scalefree})---free of a characteristic
scale. This notion of ``scale-free'' is clear, and could be
taken as simply another synonym for scaling and power law,
but most actual usages of ``scale-free'' appear to have a
richer notion in mind, and they attribute additional features,
such as some underlying self-similar or fractal geometry or
topology, beyond just properties of certain scalar random
variables.

One of the most widespread and longstanding uses of the
term ``scale-free'' has been in astrophysics to describe
the fractal nature of galaxies. Using a probabilistic
framework, one approach is to model the distribution of
galaxies as a stationary random process and express
clustering in terms of correlations in the distributions
of galaxies (see the review \cite{Fall1979} for an
introduction). In 1977, Groth and Peebles~\cite{Groth1977}
proposed that this distribution of galaxies is well
described by a power-law correlation function, and this has
since been called scale-free in the astrophysics
literature. Scale-free here means that the fluctuation in
the galaxy density have ``non-trivial, scale-free fractal
dimension'' and thus scale-free is associated with fractals
in the spatial layout of the universe.

Perhaps the most influential and revealing notion of
``scale-free'' comes from the study of {\it critical phase
transitions} in physics, where the ubiquity of power laws
is often interpreted as a ``signature'' of a universality
in behavior as well in as underlying generating mechanisms.
An accessible history of the influence of criticality in
the SF literature can found
in~\cite[pp.~73-78]{BarabasiBook}. Here, we will briefly
review criticality in the context of {\it percolation}, as
it illustrates the key issues in a simple and easily
visualized way. Percolation problems are a canonical
framework in the study of statistical mechanics
(see~\cite{StaufferBook} for a comprehensive introduction).
A typical problem consists of a square $n \times n$ lattice
of ``sites'', each of which is either ``occupied'' or
``unoccupied''.  This initial configuration is obtained at
random, typically according to some uniform probability,
termed the {\it density}, and changes to the lattice are
similarly defined in terms of some stochastic process.  The
objective is to understand the relationship among groups of
contiguously connected sites, called {\it clusters}.  One
celebrated result in the study of such systems is the
existence of a {\it phase transition} at a critical density
of occupied sites, above which there exists with high
probability a cluster that spans the entire lattice (termed
a {\it percolating cluster}) and below which no percolating
cluster exists.  The existence of a critical density where
a percolating cluster ``emerges'' is qualitatively similar
to the appearance of a giant connected component in random
graph theory~\cite{Bollobas}.

Figure \ref{fig:lattices}(a) shows an example of a random
square lattice ($n=32$) of unoccupied white sites and a
critical density ($\approx .59$) of occupied dark sites,
shaded to show their connected clusters.  As is consistent
with percolation problems at criticality, the sequence of
cluster sizes is approximately scaling, as seen in Figure
\ref{fig:lattices}(d), and thus there is wide variability
in cluster sizes. The cluster boundaries are fractal, and
in the limit of large $n$, the same fractal geometry occurs
throughout the lattice and on all scales, one sense in
which the lattice is said to be self-similar and
``scale-free''. These scaling, scale-free, and self-similar
features occur in random lattices if and only if (with unit
probability in the limit of large $n$) the density is at
the critical value. Furthermore, at the critical point,
cluster sizes and many other quantities of interest have
power law distributions, and these are all independent of
the details in two important ways. The first and most
celebrated is that they are {\it universal}, in the sense
that they hold identically in a wide variety of otherwise
quite different physical phenomena. The other, which is
even more important here, is that all these power laws,
including the scale-free fractal appearance of the lattice,
is unaffected if the sites are randomly rearranged.  Such
{\it random rewiring} preserves the critical density of
occupied sites, which is all that matters in purely random
lattices.

For many researchers, particularly those unfamiliar with
the strong statistical properties of scaling
distributions, these remarkable properties of critical
phase transitions have become associated with more than
just a mechanism giving power laws.  Rather, power laws
themselves are often viewed as ``suggestive'' or even
``patent signatures'' of criticality and
``self-organization'' in complex systems
generally~\cite{BarabasiBook}.  Furthermore, the concept of
{\it Self-Organized Criticality (SOC)} has been suggested
as a mechanism that automatically tunes the density to the
critical point~\cite{BakBook}.  This has, in turn, given
rise to the idea that power laws alone could be
``signatures'' of specific mechanisms, largely independent
of any domain details, and the notion that such phenomena
are robust to random rewiring of components or elements has
become a compelling force in much of complex systems
research.

Our point with these examples is that typical usage of
``scale-free'' is often associated with some fractal-like
geometry, not just macroscopic statistics that are scaling.
This distinction can be highlighted through the use of the
percolation lattice example, but contrived explicitly to
emphasize this distinction. Consider three percolation
lattices at the critical density (where the distribution of
cluster sizes is known to be scaling) depicted in
Figure~\ref{fig:lattices}(a)-(c). Even though these
lattices have identical cluster size sequences (shown in
Figure~\ref{fig:lattices}(d)), only the random and fractal,
self-similar geometry of the lattice in
Figure~\ref{fig:lattices}(a) would typically be called
``scale-free,'' while the other lattices typically would
not and do not share any of the other ``universal''
properties of critical lattices \cite{CD}. Again, the usual
use of ``scale-free'' seems to imply certain self-similar
or fractal-type features beyond simply following scaling
statistics, and this holds in the existing literature on
graphs as well.

\subsection{Scaling and Self-Similarity in Graphs}

While it is possible to use ``scale-free'' as synonymous
with simple scaling relationships as expressed in
(\ref{eq:scalefree}), the popular usage of this term has
generally ascribed something additional to its meaning, and
the terms ``scaling'' and ``scale-free'' have not been used
interchangeably, except when explicitly used to say that
``scaling'' is ``free of scale.'' When used to describe
many naturally occurring and man-made networks, ``scale
free'' often implies something about the spatial,
geometric, or topological features of the system of
interest (for a recent example of that illustrates this
perspective in the context of the World Wide Web, see \cite{ss-web}).
While there exists no coherent, consistent literature on
this subject, there are some consistencies that we will
attempt to capture at least in spirit.
Here we review briefly some relevant treatments
ranging from the study of river networks to random graphs,
and including the study of network motifs in engineering
and biology.


\begin{figure*}[th]
  \begin{center}
  \includegraphics[width=\linewidth]{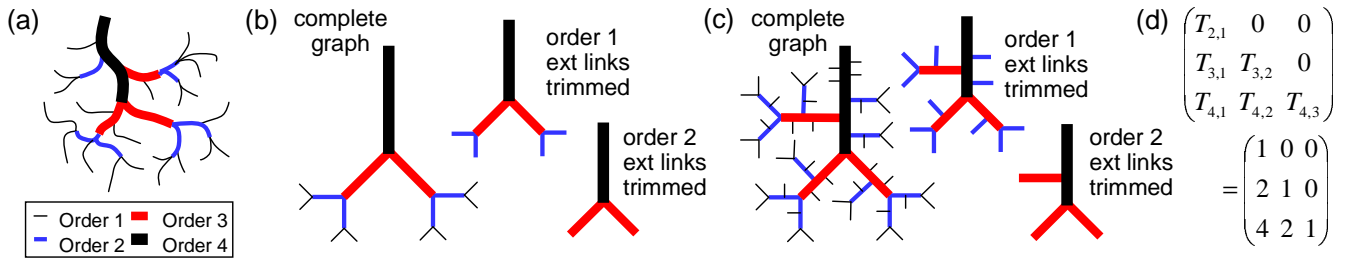}
  \vspace{-4mm}
  \caption{\footnotesize   \label{fig:streams}
   \textbf{\sc Horton-Strahler streams of order 4.}
   (a) Generic stream with segments coded according to
   their order.
   (b) Self-similar tree without side tributaries:
   branching number $b=2$ and $T_k=0$ for all $k$.
   (c) Self-similar tree with side tributaries:
   branching number $b=2$ but $T_k=2^{k-1}$ for $k=1,2,3$.
   (d) Toeplitz matrix of values $T_{\omega, \omega -k} = T_k$,
   representing the side tributaries in (c).
  }
  \end{center}
\vspace{-4ex}
\end{figure*}


\subsubsection{Self-similarity of River Channel Networks}

One application area where self-similar, fractal-like, and
scale-free properties of networks have been considered in great
detail has been the study of geometric regularities arising
in the analysis of tree-branching structures associated
with river or stream channels
\cite{horton1945,strahler1957,hack1957,maranietal1991,
kirchner1993,peckham1996,tarboton1996,dodds_rothman2000}.
Following \cite{peckham1996}, consider a river network
modeled as a tree graph, and recursively assign weights
(the ``Horton-Strahler stream order numbers'') to each link
as follows. First, assign order 1 to all exterior links.
Then, for each interior link, determine the highest order
among its child links, say, $\omega$. If two or more of the
child links have order $\omega$, assign to the parent link
order $\omega+1$; otherwise, assign order $\omega$ to the
parent link. Order $k$ streams or channels are then defined
as contiguous chains of order $k$ links.  A tree whose
highest order stream has order $\Omega$ is called a tree of
order $\Omega$. Using this Horton-Strahler stream ordering
concept, any rooted tree naturally decomposes into a
discrete set of ``scales'', with the exterior links labeled
as order 1 streams and representing the smallest scale or the
finest level of detail, and the order $\Omega$ stream(s)
within the interior representing the
largest scale or the structurally coarsest level of detail.
For example, consider the order 4 streams and their different
``scales'' depicted in Figure~\ref{fig:streams}.

To define topologically self-similar trees, consider the
class of deterministic trees where every stream of order
$\omega$ has $b\geq2$ upstream tributaries of order
$\omega-1$, and $T_{\omega,k}$ side tributaries of order
$k$, with $2\leq\omega\leq\Omega$ and $1\leq k\leq
\omega-1$. A tree is called (topologically) {\it
self-similar} if the corresponding matrix $(T_{\omega,k})$
is a Toeplitz matrix; i.e., constant along diagonals,
$T_{\omega,\omega-k} = T_k$, where $T_k$ is a number that
depends on $k$ but not on $\omega$ and gives the number of
side tributaries of order $\omega-k$.  This definition
(with the further constraint that $T_{k+1}/T_k$ is constant
for all $k$) was originally considered in works by Tokunaga
(see \cite{peckham1996} for references).  Examples of self-similar
trees of order 4 are presented in Figure~\ref{fig:streams}(b-c).

An important concept underlying this ordering scheme can be
described in terms of a recursive ``pruning'' operation that
starts with the removal of the order 1 exterior links.  Such
removal results in a tree that is more coarse and has its own
set of exterior links, now corresponding to the finest level of
remaining detail.  In the next iteration, these order 2 streams
are pruned, and this process continues for a finite number of
iterations until only the order $\Omega$ stream remains.
As illustrated in Figure~\ref{fig:streams}(b-c), successive
pruning is responsible for the self-similar nature of these trees.
The idea is that streams of order $k$ are invariant under the
operation of pruning---they may be relabeled or removed
entirely, but are never severed---and they provide a natural
scale or level of detail for studying the overall structure of
the tree.

As discussed in~\cite{OCN-Rivers-1992a}, early attempts at
explaining the striking ubiquity of Horton-Strahler stream
ordering was based on a stochastic construction in which
{\it ``it has been commonly assumed by hydrologists and geomorphologists
that the topological arrangement and relative sizes of the streams
of a drainage network are just the result of a most probable
configuration in a random environment.''}   However, more recent
attempts at explaining this regularity have emphasized an approach
based on different principles of optimal energy expenditure
to identify the universal mechanisms underlying the evolution
of ``the scale-free spatial organization of a river network''
\cite{OCN-Rivers-1992a,OCN-Rivers-1992b}.
The idea is that, in addition to randomness, necessity in the form
of different energy expenditure principles play a fundamental role in
yielding the multiscaling characteristics in naturally occurring
drainage basins.

It is also interesting to note that while considerable
attention in the literature on river or stream channel
networks is given to empirically observed power law
relationships (commonly referred to as ``Horton's laws of
drainage network composition'') and their physical
explanations, it has been argued in
\cite{kirchner1993,kirchner1994a,kirchner1994b} that these
``laws'' are in fact a very weak test of models or theories
of stream network structures.  The arguments are based on
the observation that because most stream networks (random
or non-random) appear to satisfy Horton's laws
automatically, the latter provide little compelling
evidence about the forces or processes at work in
generating the remarkably regular geometric relationships
observed in actual river networks.  This discussion is akin
to the wide-spread belief in the SF network
literature that since SF graphs exhibit power law
degree distributions, they are capable of capturing a
distinctive ``universal'' feature underlying the evolution
of complex network structures. The arguments provided
in the context of the Internet's physical connectivity
structure~\cite{sigcomm04} are similar in spirit to Kirchner's
criticism of the interpretation of Horton's laws in the
literature on river or stream channel networks.  In
contrast to \cite{kirchner1993} where Horton's laws are
shown to be poor indicators of whether or not stream
channel networks are random, \cite{sigcomm04} makes it
clear that by their very design, engineered networks like
the Internet's router-level topology are essentially non-random,
and that their randomly constructed (but otherwise comparable)
counterparts result in poorly-performing or dysfunctional
networks.

\subsubsection{Scaling Degree Sequence and Degree Distribution}

Statistical features of graph structures that have received
extensive treatment include the size of the largest
connected component, link density, node degree
relationships, the graph diameter, the characteristic path
length, the clustering coefficient, and the betweenness
centrality (for a review of these and other metrics see
\cite{albert_barabasi2002,Newman03,dorogovtsev-book2003}).
However, the single feature that has received the most
attention is the distribution of node degree and whether or
not it follows a power law.

For a graph with $n$ vertices, let $d_i = {\rm
deg}(i)$ denote the degree of node $i$, $1\leq i\leq n$,
and call $D=\{d_1,d_2,\ldots,d_n\}$ the {\it degree
sequence} of the graph, again assumed without loss of
generality always to be ordered $d_1 \ge d_2 \ge \ldots \ge
d_n$. We will say a graph has {\it scaling degree sequence
D} (or {\it $D$ is scaling}) if for all $1 \leq k \leq n_s
\leq n$, $D$ satisfies a {\it power law size-rank
relationship} of the form
 $k \ d_k^{\alpha} = c$,
where $c>0$ and $\alpha > 0$ are constants, and where $n_s$
determines the range of scaling \cite{Mandelbrot97}.  Since
this definition is simply a graph-specific version of
(\ref{eq:scaling1}) that allows for deviations from the
power law relationship for nodes with low connectivity,
we again recognize that doubly logarithmic plots of $d_k$
versus $k$ yield straight lines of slope $-\alpha$, at least
for large $d_k$ values.

This description of scaling degree sequence is general, in
the sense that it applies to any given graph without regard
to how it is generated and without reference to any
underlying probability distributions or ensembles. That is,
a scaling degree sequence is simply an ordered list of
integers representing node connectivity and satisfying
the above scaling relationship. 
In contrast, the SF literature focuses largely on {\it
scaling degree distribution}, and thus a given degree
sequence has the further interpretation as representing a
realization of an iid sample of size $n$ generated from a
common scaling distribution of the type (\ref{eq:rv-scaling}).
This in turn is often induced by some random ensemble of graphs.
This paper will develop primarily a nonstochastic theory
and thus focus on scaling degree sequences, but will
clarify the role of stochastic models and distributions as
well. In all cases, we will aim to be explicit about which
is assumed to hold.

For graphs that are not trees, a first attempt at formally
defining and relating the concepts of ``scaling'' or ``scale-free''
and ``self-similar'' through an appropriately defined notion of
``scale invariance'' is considered by Aiello et al.~and
described in \cite{aielloetal2001}. In short, Aiello et
al.~view the evolution of a graph as a random process of growing
the graph by adding new nodes and links over time.  A model
of a given graph evolution process is then called
``scale-free'' if ``coarse-graining'' in time yields scaled
graphs that have the same power law degree distribution as
the original graph.  Here ``coarse-graining in time''
refers to constructing scaled versions of the original
graph by dividing time into intervals, combining all nodes
born in the same interval into super-nodes, and connecting
the resulting super-nodes via a natural mapping of the
links in the original graph.  For a number of graph growing
models, including the Barab\'{a}si-Albert construction,
Aiello et al.~show that the evolution process is
``scale-free'' in the sense of being invariant with respect
to time scaling (i.e., the frequency of sampling with
respect to the growth rate of the model) and independent of
the parameter of the underlying power law node degree
distribution (see \cite{aielloetal2001} for details).  Note
that the scale invariance criterion considered in
\cite{aielloetal2001} concerns exclusively the degree
distributions of the original graph and its coarse-grained or
scaled counterparts. Specifically, the definition of
``scale-free'' considered by Aiello et al.~is not
``structural'' in the sense that it depends on a
macroscopic statistic that is largely uninformative as far
as topological properties of the graph are concerned.

\subsubsection{Network Motifs}

Another recent attempt at relating the notions of
``scale-free'' and ``self-similar'' for arbitrary graphs
through the more structurally driven concept of
``coarse-graining'' is due to Itzkovitz et
al.~\cite{itzkovitz:alon:2004a}. More specifically, the
main focus in \cite{itzkovitz:alon:2004a} is on
investigating the local structure of basic network building
blocks, termed {\it motifs}, that recur throughout a
network and are claimed to be part of many natural and
man-made systems~\cite{shen-orr:alon:2002,milo:alon:2002}.
The idea is that by identifying motifs that appear in a
given network at much higher frequencies than in comparable
random networks, it is possible to move beyond studying
macroscopic statistical features of networks (e.g.~power
law degree sequences) and try to understand some of the
networks' more microscopic and structural features. The
proposed approach is based on simplifying complex network
structures by creating appropriately coarse-grained
networks in which each node represents an entire pattern
(i.e., network motif) in the original network. Recursing on
the coarse-graining procedure yields networks at different
levels of resolution, and a network is called
``scale-free'' if the coarse-grained counterparts are
``self-similar'' in the sense that the same coarse-graining
procedure with the same set of network motifs applies at
each level of resolution. When applying their approach to
an engineered network (electric circuit) and a biological
network (protein-signaling network), Itzkovitz et al.~found
that while each of these networks exhibits well-defined
(but different) motifs, their coarse-grained counterparts
systematically display very different motifs at each level.

A lesson learned from the work in
\cite{itzkovitz:alon:2004a} is that networks that have
scaling degree sequences need not have coarse-grained
counterparts that are self-similar. This further motivates
appropriately narrowing the definition of ``scale-free'' so
that it does imply some kind of self-similarity. In fact,
the examples considered in \cite{itzkovitz:alon:2004a}
indicate that engineered or biological networks may be the
opposite of ``scale-free'' or ``self-similar''---their
structure at each level of resolution is different, and the
networks are ``scale-rich'' or ``self-dissimilar.''  As
pointed out in \cite{itzkovitz:alon:2004a}, this
observation contrasts with prevailing views based on
statistical mechanics near phase-transition points which
emphasize how self-similarity, scale invariance, and power
laws coincide in complex systems.  It also suggests that
network models that emphasize the latter views may be
missing important structural features
\cite{itzkovitz:alon:2004a,itzkovitz:alon:2004b}. A more
formal definition of {\it self-dissimilarity} was recently
given by Wolpert and Macready
\cite{wolpert:macready:2000,wolpert:macready:2004} who
proposed it as a characteristic measure of complex systems.
Motivated by a data-driven approach, Wolpert and Macready
observed that many complex systems tend to exhibit
different structural patterns over different space and time
scales. Using examples from biological and economic/social
systems, their approach is to consider and quantify how
such complex systems process information at different
scales.  Measuring a system's self-dissimilarity across
different scales yields a complexity ``signature'' of the
system at hand.  Wolpert and Macready suggest that by
clustering such signatures, one obtains a purely
data-driven, yet natural, taxonomy for broad classes of
complex systems.

\subsubsection{Graph Similarity and Data Mining}

Finally, the notion of graph similarity is fundamental to
the study of attributed graphs (i.e., objects that have an
internal structure that is typically modeled with the help
of a graph or tree and that is augmented with attribute
information).  Such graphs arise as natural models for
structured data observed in different database applications
(e.g., molecular biology, image or document retrieval). The
task of extracting relevant or new knowledge from such
databases (``data mining'') typically requires some notion
of {\it graph similarity} and there exists a vast
literature dealing with different graph similarity measures
or metrics and their properties \cite{graph-ss1,graph-ss2}.
However, these measures tend to exploit graph features
(e.g., a given one-to-one mapping between the vertices of
different graphs, or a requirement that all graphs have to
be of the same order) that are specific to the application
domain. For example, a common similarity measure for graphs
used in the context of pattern recognition is the edit
distance \cite {edit-dist}. In the field of image
retrieval, the similarity of attributed graphs is often
measured via the vertex matching distance
\cite{vm-distance}. The fact that the computation of many
of these similarity measures is known to be NP-complete has
motivated the development of new and more practical
measures that can be used for more efficient similarity
searches in large-scale databases (e.g., see
\cite{em-distance}).

\section{The Existing SF Story}\label{sec:ConvSF}

In this section, we first review the existing SF literature
describing some of the most popular models and their most
appealing features. This is then followed by a brief a critique
of the existing theory of SF networks in general and in the
context of Internet topology in particular.

\subsection{Basic Properties and Claims}\label{sec:SFclaims}

The main properties of SF graphs that appear in the
existing literature can be summarized as

\begin{enumerate}
\addtolength{\itemsep}{-0.5ex} \item SF networks have
scaling (power law) degree distribution.

\item SF networks can be generated by certain random
processes, the foremost among which is preferential
attachment.

\item SF networks have highly connected ``hubs'' which
``hold the network together'' and give the ``robust yet
fragile'' feature of error tolerance but attack
vulnerability.

\item SF networks are generic in the sense of being
preserved under random degree preserving rewiring.

\item SF networks are self-similar.

\item SF networks are universal in the sense of not
depending on domain-specific details.

\end{enumerate}

\noindent This variety of features suggest the potential
for a rich and extensive theory. Unfortunately, it is
unclear from the literature which properties are necessary
and/or sufficient to imply the others, and if any
implications are strict, or simply ``likely'' for an
ensemble. Many authors apparently define scale-free in
terms of just one property, typically scaling degree
distribution or random generation, and appear to claim that
some or all of the other properties are then consequences.
A central aim of this paper is to clarify exactly what
options there are in defining SF graphs and
deriving their additional properties. Ultimately, we
propose below in Section \ref{sec:sf-def} a set of minimal
axioms that allow for the preservation of the most common
claims. However, first we briefly review the existing
treatment of the above properties, related historical
results, and shortcomings of the current theory,
particularly as it has been frequently applied to the
Internet.

The ambiguity regarding the definition of ``scale-free''
originates with the original papers
\cite{BarabasiAlbert99,AlbJeongBar00}, but have continued since.
Here SF graphs appear to be defined both as graphs with
scaling or power law degree distributions and as being
generated by a stochastic
construction mechanism based on {\it incremental growth}
(i.e.~nodes are added one at a time) and {\it preferential
attachment} (i.e.~nodes are more likely to attach to nodes
that already have many connections). Indeed, the apparent
equivalence of scaling degree distribution and preferential
attachment, and the ability of thus-defined (if ambiguously
so) SF network models to generate node degree statistics
that are consistent with the ubiquity of empirically
observed power laws is the most commonly cited evidence
that SF network mechanisms and structures are in some sense
universal \cite{AlbJeongBar99, AlbJeongBar00, BarabasiBook,
BarabasiAlbert99, hier1}.

Models of preferential attachment giving rise to power law
statistics actually have a long history and are at least 80
years old. As presented by Mandelbrot \cite{Mandelbrot97},
one early example of research in this area was the work of
Yule \cite{Yule25}, who in 1925 developed power law models
to explain the observed distribution of species within
plant genera. Mandelbrot \cite{Mandelbrot97} also documents
the work of Luria and Delbr\"{u}ck, who in 1943 developed a
model and supporting mathematics for the explicit
generation of scaling relationships in the number of
mutants in old bacterial populations \cite{LurDel43}. A
more general and popular model of preferential attachment
was developed by Simon \cite{Simon55} in 1955 to explain
the observed presence of power laws within a variety of
fields, including economics (income distributions, city
populations), linguistics (word frequencies), and biology
(distribution of mutants in bacterial cultures).
Substantial controversy and attention surrounded these
models in the 1950s and 1960s \cite{Mandelbrot97}.  A
recent review of this history can also be found in
\cite{Mitzenmacher03}.  By the 1990s though these models
had been largely displaced in the popular science
literature by models based on critical phenomena from
statistical physics \cite{BakBook}, only to resurface
recently in the scientific literature in this context of
``scale-free networks'' \cite{BarabasiAlbert99}. Since
then, numerous refinements and modifications to the
original Barab\'{a}si-Albert construction have been
proposed and have resulted in SF network models that can
reproduce power law degree distributions with any $\alpha
\in [1,2]$, a feature that agrees empirically with many
observed networks \cite{albert_barabasi2002}. Moreover, the
largely empirical and heuristic studies of these types of
``scale-free'' networks have recently been enhanced by a
rigorous mathematical treatment that can be found in
\cite{BollobasRiordan03} and involves a precise version of
the Barab\'{a}si-Albert construction.

The introduction of SF network models, combined with the
equally popular (though less ambiguous) ``small world"
network models~\cite{WattsStrogatz}, reinvigorated the use
of abstract random graph models and their properties
(particularly node degree distributions) to study a
diversity of complex network systems. For example,
Dorogovtsev and Mendes \cite[p.76]{dorogovtsev-book2003}
provide a ``standard programme of empirical research of a
complex network'', which for the case of undirected graphs
consist of finding 1) the degree distribution; 2) the
clustering coefficient; 3) the average shortest-path
length. The presumption is that these features
adequately characterize complex networks. Through the
collective efforts of many researchers, this approach has
cataloged an impressive list of real application networks,
including communication networks (the WWW and the
Internet), social networks (author collaborations, movie
actors), biological networks (neural networks, metabolic
networks, protein networks, ecological and food webs),
telephone call graphs, mail networks, power grids and
electronic circuits, networks of software components, and
energy landscape networks (again, comprehensive reviews of
these many results are widely available
\cite{albert_barabasi2002,BarabasiBook,Newman03,dorogovtsev-book2003,
pastor-satorras-book2004}). While very different in detail,
these systems share a common feature in that their degree
distributions are all claimed to follow a power law,
possibly with different tail indices.

Regardless of the definitional ambiguities, the use of
simple stochastic constructions that yield scaling degree
distributions and other appealing graph properties
represent for many researchers what is arguably an ideal
application of statistical physics to explaining and
understanding complexity. Since SF models have
their roots in statistical physics, a key assumption is
always that any particular network is simply a realization
from a larger ensemble of graphs, with an explicit or
implicit underlying stochastic model. Accordingly, this
approach to understanding complex networks has focused on
those networks that are most likely to occur under an
assumed random graph model and has aimed at identifying or
discovering macroscopic features that capture the
``essence'' of the structure underlying those networks.
Thus preferential attachment offers a general and hence
attractive ``microscopic'' mechanism by which a growth
process yields an ensemble of graphs with the
``macroscopic'' property of power law node degree
distributions \cite{PhysicaA}. Second, the resulting
SF topologies are ``generic.'' Not only is any
specific SF graph the generic or likely element
from such an ensemble, but also {\it ``... an important
property of scale-free networks is that [degree preserving]
random rewiring does not change the scale-free nature of
the network''} (see Methods Supplement to
\cite{JeongTombor00}).
Finally, this ensemble-based approach has an appealing kind
of ``universality'' in that it involves no model-specific
domain knowledge or specialized ``design'' requirements and
requires only minimal tuning of the underlying model parameters.

Perhaps most importantly, SF graphs are
claimed to exhibit a host of startling ``emergent''
consequences of universal relevance, including intriguing
self-similar and fractal properties (see below for
details), small-world characteristics \cite{AmaralPNAS00},
and ``hub-like'' cores.  Perhaps the central claim for
SF graphs is that they have hubs, what we term SF
hubs, which ``hold the network together.''  As noted, the
structure of such networks is highly vulnerable (i.e., can
be fragmented) to attacks that target these hubs
\cite{AlbJeongBar00}.  At the same time, they are resilient
to attacks that knock out nodes at random, since a randomly
chosen node is unlikely to be a hub and thus its removal
has minimal effect on network connectivity. In the context
of the Internet, where SF graphs have been proposed as models
of the router-level Internet~\cite{barabasi-internet},
this has been touted ``the Achilles' heel
of the Internet'' \cite{AlbJeongBar00}, a vulnerability
that has presumably been overlooked by networking
engineers.  Furthermore, the hub-like structure of
SF graphs is such that the epidemic threshold is
zero for contagion phenomena
\cite{PasVes-Epidemic01,NatureVirus01,
VirusProneNet01,pastor-satorras-book2004}, thus suggesting
that the natural way to stop epidemics, either for computer
viruses/worms or biological epidemics such as AIDS, is to
protect these hubs \cite{ALB-Epidemic02,LincolnWORM03}.
Proponents of this modeling framework have further
suggested that the emergent properties of SF graphs
contributes to truly universal behavior in complex networks
\cite{BianconiBarabasi01} and that preferential attachment
as well is a universal mechanism at work in the evolution
of these networks \cite{JeongNeda03,dorogovtsev-book2003}.


\begin{figure*}[th]
  \begin{center}
  \includegraphics[width=\linewidth]{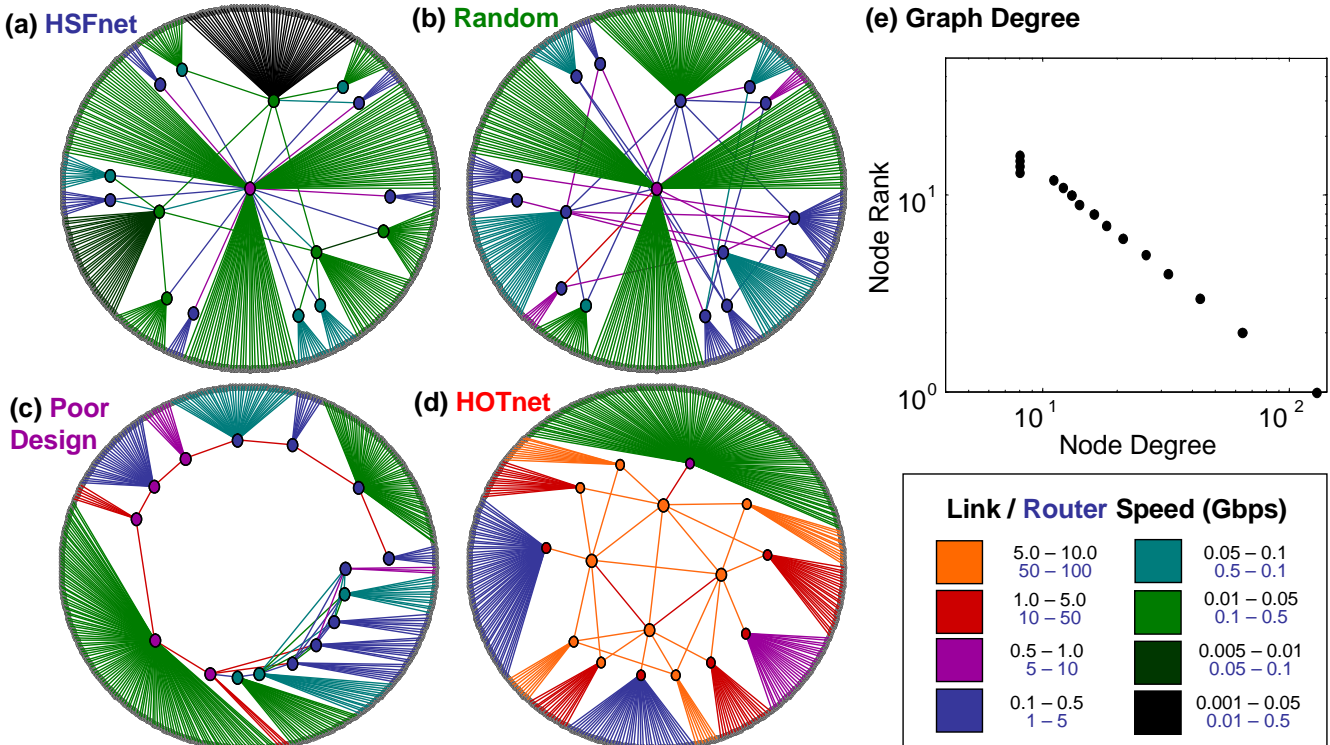}
  \vspace{-4mm}
  \caption{\footnotesize   \label{fig:toynet}
   {\sc Network graphs having exactly the same number of nodes and links,
   as well as the same (power law) degree sequence}.
   As toy models of the router-level Internet, all graphs are subject to
   same router technology constraints and the same traffic demand model
   for routers at the network periphery.
   \textbf{(a) Hierarchical scale-free (HSF) network:} Following
    roughly a recently proposed construction that combines scale-free
    structure and inherent modularity in the sense of exhibiting an
    hierarchical architecture \cite{hier2}, we start with
    a small 3-pronged cluster and build a 3-tier network a la
    Ravasz-Barab\'{a}si,
    adding routers at the periphery roughly in a preferential manner.
   \textbf{(b) Random network:} This network is obtained from the
    HSF network in (a) by performing a number of pairwise random
    degree-preserving rewiring steps.
   \textbf{(c) Poor design:} In this heuristic
    construction, we arrange the interior routers in a line, pick
    a node towards the middle to be the high-degree, low bandwidth
    bottleneck, and establish connections between high-degree and
    low-degree nodes.
   \textbf{(d) HOT network:} The construction mimics the build-out of a
    network by a hypothetical ISP.  It produces a 3-tier network hierarchy
    in which the high-bandwidth, low-connectivity routers live in the
    network core while routers with low-bandwidth and high-connectivity
    reside at the periphery of the network.
   \textbf{(e) Node degree sequence for each network.}  Only $d_i > 1$ shown.
  }
  \end{center}
\vspace{-4ex}
\end{figure*}


\subsection{A Critique of Existing Theory}

The SF story has successfully captured the interest
and imagination of researchers across disciplines, and with
good reason, as the proposed properties are rich and
varied.  Yet the existing ambiguity in its mathematical
formulation and many of its most essential properties has
created confusion about what it means for a network to be
``scale-free'' \cite{fox-keller}.
One possible and apparently popular interpretation is that
scale-free means simply graphs with scaling degree {\it sequences},
and that this alone implies all other features
listed above.  We will show that this is incorrect, and in
fact none of the features follows from scaling alone. Even
relaxing this to random graphs with scaling degree {\em
distributions} is by itself inadequate to imply any further
properties.  A central aim of this paper is to clarify the
reasons why these interpretations are incorrect, and propose
minimal changes to fix them. The opposite extreme
interpretation is that scale-free graphs are defined as
having all of the above-listed properties. We will show
that this is possible in the sense that the set of such
graphs is not empty, but as a definition this leads to two
further problems. Mathematically, one would prefer fewer
axioms, and we will rectify this with a minimal definition.
We will introduce a structural metric that provides a view of
the extent to which a graph is scale-free and from which all
the above properties follow, often with necessary and
sufficient conditions. The other problem is that the
canonical examples of apparent SF networks, the
Internet and biological metabolism, are then very far from
scale-free in that they have {\em none} of the above
properties except perhaps for scaling degree distributions. This
is simply an unavoidable conflict between these properties
and the specifics of the applications, and cannot be fixed.

As a result, a rigorous theory of SF graphs must
either define scale-free more narrowly than scaling degree
sequences or distributions in order to have nontrivial
emergent properties, and thus lose central claims of
applicability, or instead define scale-free as merely
scaling, but lose all the universal emergent features that
have been claimed to hold for SF networks. We will
pursue the former approach because we believe it is most
representative of the spirit of previous studies and also
because it is most inclusive of results in the existing
literature.  At the most basic level, simply to be a
nontrivial and novel concept, scale-free clearly must mean
more than a graph with scaling degree sequence or distribution.
It must capture some
aspect of the graph itself, and not merely a sequence of
integers, stochastic or not, in which case the SF
literature and this paper would offer nothing new. Other
authors may ultimate choose different definitions, but in
any case, the results in this paper clarify for the first
time precisely what the graph theoretic alternatives are
regarding the implications of any of the possible
alternative definitions.  Thus the definition of the word
``scale-free'' is much less important than the mathematical
relationship between their various claimed properties, and
the connections with real world networks.

\subsection{The Internet as a Case Study}

To illustrate some key points about the existing claims
regarding SF networks as adopted in the popular
literature and their relationship with scaling degree
distributions, we consider an application to the Internet where
graphs are meant to model Internet connectivity at the
router-level. For a meaningful explanation of empirically
observed network statistics, we must account for network
design issues concerned with technology constraints,
economic factors, and network performance \cite{sigcomm04,PNAS05}.
Additionally, we should annotate the nodes and links in
connectivity-only graphs with domain-specific information
such as router capacity and link bandwidth in such a way
that the resulting annotated graphs represent technically
realizable and functional networks.

\subsubsection{The SF Internet}

Consider the simple toy model of a ``hierarchical'' SF
network {\it HSFnet} shown in Figure~\ref{fig:toynet}(a),
which has a ``modular'' graph constructed according to a
particular type of preferential attachment \cite{hier2} and
to which are then preferentially added degree-one end
systems, yielding the power law degree sequence shown in
Figure~\ref{fig:toynet}(e). This type of construction has
been suggested as a SF model of both the Internet and
biology, both of which are highly hierarchical and modular
\cite{hier1}.  The resulting graph has all the features
listed above as characteristic of SF networks and
is easily visualized and thus convenient for our
comparisons. Note that the highest-degree nodes in the tail
of the degree sequence in Figure~\ref{fig:toynet}(e)
correspond to the SF hub nodes in the SF network {\it
HSFnet}, Figure~\ref{fig:toynet}(a). This confirms the
intuition behind the popular SF view that power law degree
sequences imply the existence of SF hubs that are crucial for
global connectivity. If such features were true for the
real Internet, this finding would certainly be startling
and profound, as it directly contradicts the Internet's
legendary and most clearly understood robustness property,
i.e., it's high resilience to router failures
\cite{ddclark}.

Figure~\ref{fig:toynet} also depicts three other networks
with the exact same degree sequence as {\it HSFnet}.  The
variety of these graphs suggests that the set of all
connected simple graphs (i.e., no self-loops or parallel
links) having exactly the same degree sequence shown in
Figure~\ref{fig:toynet}(e) is so diverse that its elements
appear to have nothing in common as graphs beyond what
trivially follows from having a fixed (scaling) degree
sequence. They certainly do not appear to share any of the
features summarized above as conventionally claimed for
SF graphs. Even more striking are the differences
in their structures and annotated bandwidths (i.e.,
color-coding of links and nodes in
Figure~\ref{fig:toynet}). For example, while the graphs in
Figure~\ref{fig:toynet}(a) and (b) exhibit the type of hub
nodes typically associated with SF networks, the graph in
Figure~\ref{fig:toynet}(d) has its highest-degree nodes
located at the networks' peripheries.  We will show this
provides concrete counterexamples to the idea that power law
degree sequences imply the existence of SF hubs. This then
creates the obvious dilemma as to the concise meaning of a
``scale-free graph'' as outlined above.

\subsubsection{A Toy Model of the Real Internet}\label{sec:toynets}

In terms of using SF networks as models for the Internet's
router-level topology, recent Internet research has
demonstrated that the real Internet is nothing like
Figure~\ref{fig:toynet}(a), size issues notwithstanding,
but is at least qualitatively more like the network shown
in Figure~\ref{fig:toynet}(d).  We label this network {\it
HOTnet} (for {\it Heuristically Optimal Topology}), and
note that its overall power law in degree sequence comes
from high-degree routers at the network periphery
that aggregate the traffic of
end users having low bandwidth demands, while supporting
aggregate traffic flows with a mesh of low-degree core
routers \cite{sigcomm04}.  In fact, as we will discuss
in greater detail in Section 6, there is little evidence
that the Internet as a whole has scaling degree or even
high variability, and much evidence to the contrary, for
many of the existing claims of scaling are based on
a combination of relying on highly ambiguous data and
making a number of statistical errors, some of them similar to those
illustrated in Figures~\ref{fig:ExpVsScaling} and
\ref{fig:plot-example}. What is true is that a network like
{\it HOTnet} is consistent with existing technology, and
could in principle be the router level graph for some small
but plausible network.  Thus a network with a scaling
degree sequence in its router graph is plausible even if
the actual Internet is not scaling. It would however look
qualitatively like {\it HOTnet} and nothing like {\it HSFnet}.

To see in what sense {\it HOTnet} is heuristically optimal,
note that from a network design perspective, an important
question is how well a particular topology is able to carry
a given demand for traffic, while fully complying with
actual technology constraints and economic factors.  Here,
we adopt as standard metric for {\it network performance}
the maximum throughput of the network under a ``gravity
model'' of end user traffic demands \cite{gravity}.  The
latter assumes that every end node $i$ has a total
bandwidth demand $x_i$, that two-way traffic is exchanged
between all pairs $(i,j)$ of end nodes $i$ and $j$, the
flow $X_{ij}$ of traffic between $i$ and $j$ is given by
$X_{ij}=\rho x_i x_j$, where $\rho$ is some global
constant, and is otherwise uncorrelated from all other
flows. Our performance measure for a given network $g$ is
then its maximum throughput with gravity flows, computed as
\begin{equation}
\mbox{\em Perf}(g) = {\max_{\rho} \sum_{ij} X_{ij}},
~~~s.t.~R X \leq B , \label{eq:perf}
\end{equation}
where $R$ is the routing matrix obtained using standard
shortest path routing. $R = [R_{kl}]$, with $R_{kl}=1$ if
flow $l$ passes through router $k$, and $R_{kl}=0$
otherwise. $X$ is the vector of all flows $X_{ij}$, indexed
to match the routing matrix $R$, and $B$ is a vector
consisting of all router bandwidth capacities.

An appropriate treatment of router bandwidth capacities
represented in $B$ is important for computing network
performance and merits additional explanation.  Due to
fundamental limits in technology, routers must adhere to
flow conservation constraints in the total amount of
traffic that they process per unit of time.  Thus, routers
can support a large number of low bandwidth connections or
a smaller number of high bandwidth connections.  In many
cases, additional routing overhead actually causes the
total router throughput to decrease as the number of
connections gets large, and we follow the presentation in
\cite{sigcomm04} in choosing the term $B$ to correspond
with an abstracted version of a widely deployed Cisco
product (for details about this abstracted constraint and
the factors affecting real router design, we refer the
reader to \cite{Alderson04,sigcomm04}).

The application of this network performance metric to the
four graphs in Figure~\ref{fig:toynet} shows that although
they have the same degree sequence, they are very different
from the perspective of network engineering, and that these
differences are significant and critical.  For example, the
SF network {\it HSFnet} in Figure~\ref{fig:toynet}(a)
achieves a performance of $\mbox{\em Perf}(HSFnet) = 6.17
\times 10^{8}$ bps, while the
HOT network {\it HOTnet} in Figure~\ref{fig:toynet}(d)
achieves a performance of $\mbox{\em Perf}(HOTnet) = 2.93
\times 10^{11}$ bps, which is greater by more than two
orders of magnitude. The reason for this vast difference is
that the HOT construction explicitly incorporates the
tradeoffs between realistic router capacities and economic
considerations in its design process while the SF
counterpart does not.

The actual construction of {\it HOTnet} is fairly
straightforward, and while it has high performance, it is
not formally optimal.  We imposed the constraints that it
must have exactly the same degree sequence as {\it HSFnet},
and that it must satisfy the router degree/bandwidth
constraints. For a graph of this size the design then
easily follows by inspection, and mimics in a highly
abstracted way the design of real networks. First, the
degree one nodes are designated as end-user hosts and
placed at the periphery of the network, though geography
per se is not explicitly considered in the design. These
are then maximally aggregated by attaching them to the
highest degree nodes at the next level in from the
periphery, leaving one or two links on these ``access
router'' nodes to attach to the core. The lowest degree of
these access routers are given two links to the core, which
reflects that low degree access routers are capable of
handling higher bandwidth hosts, and such high value
customers would likely have multiple connections to the
core. At this point there are just 4 low degree nodes left,
and these become the highest bandwidth core routers, and
are connected in a mesh, resulting in the graph in
Figure~\ref{fig:toynet}(d). While some rearrangements are
possible, all high performance networks using a gravity
model and the simple router constraints we have imposed
would necessarily look essentially like {\it HOTnet}. They
would all have the highest degree nodes connected to degree
one nodes at the periphery, and they would all have a
low-degree mesh-like core.

Another feature that has been highlighted in the SF literature
is the attack vulnerability of high degree hubs.  Here again,
the four graphs in Figure~\ref{fig:toynet} are illustrative of
the potential differences between graphs having the same degree
sequence.  Using the performance metric defined in (\ref{eq:perf}),
we compute the performance of each graph without disruption
(i.e., the complete graph), after the loss of high degree nodes,
and after the loss of the most important (i.e., worst case) nodes.
In each case, when removing a node we also remove any corresponding
degree-one end-hosts that also become disconnected, and we compute
performance over shortest path routes between remaining nodes
but in a manner that allows for rerouting.  We find that for
{\it HSFnet}, removal of the highest degree nodes does in fact
disconnect the network as a whole, and this is equivalent to the
worst case attack for this network.  In contrast, removal of the
highest degree nodes results in only minor disruption to
{\it HOTnet}, but a worst case attack (here, this is the removal of
the low-degree core routers) does disconnect the network.
The results are summarized below.

\begin{center}
\setlength{\tabcolsep}{1mm}
\begin{tabular}{|c|c|c|c|}
\hline
{\footnotesize Network} & {\footnotesize Complete} &
  {\footnotesize High Degree} & {\footnotesize Worst Case}\\
{\footnotesize Performance}  & {\footnotesize Graph} &
  {\footnotesize Nodes Removed} & {\footnotesize Nodes Removed}\\
\hline
{\footnotesize HSFnet} & {\footnotesize $5.9197e+09$} &
  {\footnotesize Disconnected} & {\footnotesize = `High Degree' case} \\
\hline
{\footnotesize HOTnet} & {\footnotesize $2.9680e+11$} &
  {\footnotesize $2.7429e+11$} & {\footnotesize Disconnected} \\
\hline
\end{tabular}
\end{center}

This example thus illustrates two important points.  The
first is that {\it HSFnet} does indeed have all the graph
theoretic properties listed above that are attributed to SF
networks, including attack vulnerability, while {\it
HOTnet} has none of these features except for scaling
degree. Thus the set of graphs that have the standard
scale-free attributes is neither empty nor trivially
equivalent to graphs having scaling degree. The second
point is that the standard SF models are in all important
ways exactly the opposite of the real Internet, and fail to
capture even the most basic features of the Internet's
router-level connectivity. While the intuition behind these
claims is clear from inspection of Figure~\ref{fig:toynet}
and the performance comparisons, full clarification of
these points requires the results in the rest of this paper
and additional details on the Internet
\cite{Alderson04,sigcomm04,PNAS05}. These observations naturally
cast doubts on the relevance of conventional SF models in
other application areas where domain knowledge and specific
functional requirements play a similarly crucial role as in
the Internet context. The other most cited SF example is
metabolic networks in biology, where many recent SF studies
have focused on abstract graphs in which nodes represent
metabolites, and two nodes are ``connected'' if they are
involved in the same reaction. In these studies, observed
power laws for the degree sequences associated with such
graphs have been used to claim that metabolic networks are
scale-free \cite{BarOltBio04}. Though the details are far
more complicated here than in the Internet story above,
recent work in \cite{tanaka2005} that is summarized in Section
\ref{sec:bio} has shown there is a
largely parallel story in that the SF claims are completely
inconsistent with the actual biology, despite their
superficial appeal and apparent popularity.

\section{A Structural Approach}\label{sec:struct}

In this section, we show that considerable insight into the
features of SF graphs and models is available from a metric that
measures the extent to which high-degree nodes connect to
other high-degree nodes.  As we will show, such a metric is
both necessary and useful for explaining the extreme differences
between networks that have identical degree sequence, especially
if it is scaling.  By focusing on a graph's structural properties
and not on not how it was generated, this approach does
not depend on an underlying random graph model but is
applicable to any graph of interest.

\subsection{The $s$-Metric}

Let $g$ be an undirected, simple, connected graph having
$n=|\mathcal{V}|$ nodes and $l=|\mathcal{E}|$ links, where
$\mathcal{V}$ and $\mathcal{E}$ are the sets of nodes and
links, respectively. As before, define $d_i$ to be the
degree of node $i \in \mathcal{V}$, $D = \{d_1, d_2, \dots,
d_n\}$ to be the degree sequence for $g$ (again assumed to
be ordered), and let $G(D)$ denote the set of all connected
simple graphs having the same degree sequence $D$. Note
that most graphs with scaling degree will be neither simple
nor connected, so this is an important and nontrivial
restriction.  Even with these constraints, it is clear
based on the previous examples that the elements of $G(D)$
can be very different from one another, so that in order to
constitute a non-trivial concept, ``scale-free'' should
mean more than merely that $D$ is scaling and should depend
on additional {\em topological} or {\it structural}
properties of the elements in $G(D)$.

\begin{definition} \label{def:s-metric}
For any graph $g$ having fixed degree sequence $D$,
we define the metric 
\begin{equation} \label{eq:s-metric}
s(g)= \sum_{(i,j)\in \mathcal{E}} d_i d_j .
\end{equation}
\end{definition}

Note that $s(g)$ depends only on the graph $g$ and not
explicitly on the process by which it is constructed.
Implicitly, the metric $s(g)$ measures the extent to which the
graph $g$ has a ``hub-like'' core and is maximized when
high-degree nodes are connected to other high-degree nodes.
This observation follows from the
{\it Rearrangement Inequality} \cite{rearrange}, which
states that if $a_1\geq a_2 \geq \cdots \geq a_n$ and
$b_1\geq b_2 \geq \cdots \geq b_n$, then for any
permutation $(a'_1, a'_2, \cdots, a'_n)$ of $(a_1, a_2,
\cdots, a_n)$, we have
\begin{eqnarray*}
a_1b_1+a_2b_2+\cdots+a_nb_n&\geq&
a'_1b_1+a'_2b_2+\cdots+a'_nb_n\\
&\geq&a_nb_1+a_{n-1}b_2+\cdots+a_1b_n.
\end{eqnarray*}
Since high $s(g)$-values are achieved only by connecting
high-degree nodes to each other, and low $s(g)$-values are
obtained by connecting high-degree nodes only to low-degree
nodes, the $s$-metric moves beyond simple statements concerning
the presence of ``hub'' nodes (as is true for any degree sequence
$D$ that has high variability) and attempts to quantify what
role such hubs play in the overall structure of the graph.
In particular, as we will show below, graphs with relatively high
$s(g)$ values have a ``hub-like core'' in the sense that these
hubs play a central role in the overall connectivity of the
network. We will also demonstrate that the metric
$s(g)$ provides a view that is not only
mathematically convenient and rigorous, but also
practically useful as far as what it means for a graph to
be ``scale-free''.


\begin{figure*}[th]
  \begin{center}
  \includegraphics[width=.65\linewidth]{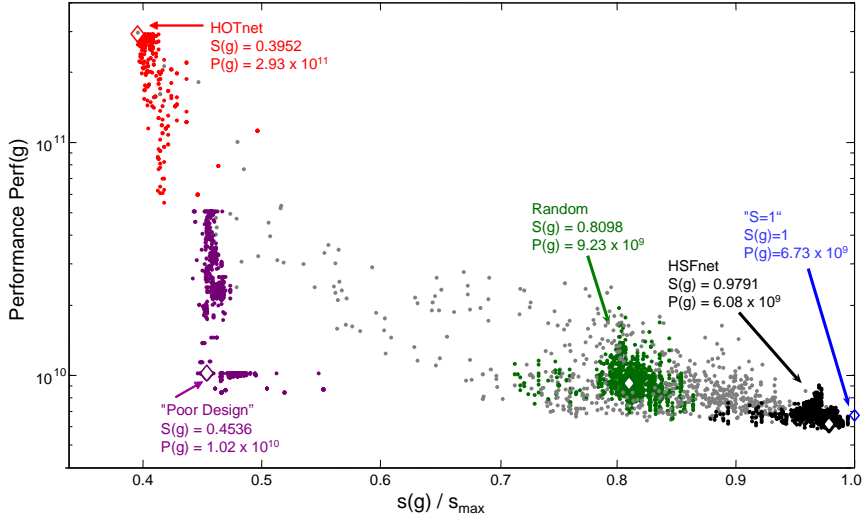}
  \vspace{-4mm}
  \caption{\footnotesize   \label{fig:perf-llh}
   {\sc Exploration of the space of connected network graphs
   having exactly the same (power law) degree sequence}.
   Values for the four networks are shown together with the values for
   other networks obtained by pairwise degree-preserving rewiring.
   Networks that are ``one-rewiring'' away from their starting point
   are shown in a corresponding color, while other networks obtained
   from more than one rewiring are shown in gray.
   Ultimately, only a careful
   design process explicitly incorporating technological constraints,
   traffic demands, or link costs yields high-performance networks.
   In contrast, equivalent networks resulting from even carefully
   crafted random constructions result in poor-performing networks.
  }
  \end{center}
  \vspace{-4ex}
\end{figure*}


\subsubsection{Graph Diversity and the $\mbox{\em Perf}(g)$ vs.~$s(g)$ Plane}

Although our interest in this paper will be in graphs for
which the degree sequence $D$ is scaling, we can compute
$s(g)$ with respect to any ``background'' set $G$ of graphs,
and we need not restrict the set to
scaling or even to connected or simple graphs.  Moreover,
for any background set, there exists a graph whose
connectivity maximizes the $s$-metric defined in
(\ref{eq:s-metric}), and we refer to this as an
``$s_{\max}$ graph''.  The $s_{\max}$ graphs for different
background sets are of interest since they are essentially
unique and also have the most ``hub-like'' core structure.
Graphs with low $s$-values are also highly relevant, but
unlike $s_{\max}$ graphs they are extremely diverse with
essentially no features in common with each other or with
other graphs in the background set except the degree
sequence $D$.

Graphs with high variability and/or scaling in their degree
sequence are of particular interest, however, and not simply
because of their association with SF models. Intuitively, scaling
degrees appear to create great ``diversity'' in $G(D)$.
Certainly the graphs in Figure~\ref{fig:toynet} are
extremely diverse, despite having identical scaling degree
$D$, but to what extent does this depend on $D$ being
scaling? As a partial answer, note that at the extremes of
variability are $m$-regular graphs with $CV(D) = 0$, which have
$D=\{m,m,m,\ldots,m \}$ for some $m$, and perfect star-like
graphs with $D=\{n-1,1,1,1,\ldots,1\}$, which have maximal
$CV(D) \approx \sqrt{n}/2$. In both of these extremes all
graphs in $G(D)$ are isomorphic and thus have only one
value of $s(g)$ 
for all $g \in G(D)$ so from this measure the space $G(D)$
of graphs lacks any diversity. In contrast, when $D$ is scaling
with $\alpha < 2$, $CV(D)\rightarrow \infty$ and it is easy to
construct $g$ such that $s(g)/s_{\max} \rightarrow 0$ as $n
\rightarrow \infty$, suggesting a possibly enormous
diversity in $G(D)$.

Before proceeding with a discussion of some of the features
of the $s$-metric as well as for graphs having high $s(g)$
values, we revisit the four toy networks in
Figure~\ref{fig:toynet} and consider the combined
implications of the performance-oriented metric $\mbox{\em
Perf}(g)$ introduced in (\ref{eq:perf}) and the
connectivity-specific metric $s(g)$ defined above.
Figure~\ref{fig:perf-llh} is a projection of $g \in G(D)$
onto a plane of $\mbox{\em Perf}(g)$ versus $s(g)$ and will
be useful throughout in visualizing the extreme diversity
in the set $G(D)$ for $D$ in Figure~\ref{fig:toynet}.  Of
relevance to the Internet application is that graphs with
high $s(g)$-values tend to have low performance, although a
low $s(g)$-value is no guarantee of good performance, as
seen by the network in Figure~\ref{fig:toynet}(c) which has
both small $s(g)$ and small $\mbox{\em Perf}(g)$. The
additional points in the $\mbox{\em Perf}(g)$ vs.~$s(g)$
plane involve degree preserving rewiring and will be
discussed in more detail below.

These observations undermine the claims in the SF
literature that are based on scaling degree alone implying
any additional graph properties.  On the other hand, they
also suggest that the sheer diversity of $G(D)$ for scaling
$D$ makes it an interesting object of study.  We won't
further compare $G(D)$ for scaling versus non-scaling $D$
or attempt to define ``diversity'' precisely here, though
these are clearly interesting topics. We will focus on
exploring the nature of the diversity of $G(D)$ for scaling
$D$ such as in Figure~\ref{fig:toynet}.

In what follows, we will provide evidence that graphs with
high $s(g)$ enjoy certain self-similarity properties, and
we also consider the effects of random degree-preserving
rewiring on $s(g)$.  In so doing, we argue that the $s$-metric,
as well as many of the other definitions and properties
that we will present, are of interest for any graph or any set
of graphs.  However, we will continue to focus our attention
primarily on simple connected graphs having scaling degree sequences.
The main reason is that many applications naturally have simple
connected graphs. For example, while the Internet protocols
in principle allow router connectivity to be nonsimple, it
is relatively rare and has little impact on network
properties.  Nevertheless, using other sets in many cases is
preferable and will arise naturally in the sequel.
Furthermore, while our interest will be on simple, connected
graphs with scaling degree sequence, we will often specialize
our presentation to trees, in order to simplify the development
and maximize contact with the existing SF literature.  To this
end, we will exploit the construction of the $s_{\max}$ graph
to sketch some of these relationships in more detail.

\subsubsection{The $s_{\max}$ Graph and Preferential Attachment}
\label{sec:smax}

Given a particular degree sequence $D$, it is possible to
construct the $s_{\max}$ graph of $G(D)$ using a
deterministic procedure, and both the generation process
and its resulting structure are informative about the
$s(g)$ metric.  Here, we describe this construction at a
high level of abstraction (with all details deferred to
Appendix~\ref{sec:smax-appendix}) in order to provide appropriate
context for the discussion of key features that is to follow.

The basic idea for constructing the $s_{\max}$ graph is to
order all potential links $(i,j)$ for all $i,j \in \mathcal{V}$
according to their {\it weight} $d_i d_j$ and then add them one
at a time in a manner that results in a simple, connected graph
having degree sequence $D$.  While simple enough in concept, this
type of ``greedy'' heuristic procedure may have difficulty
achieving the intended sequence $D$ due to the global
constraints imposed by connectivity requirements.
While the specific conditions under which this procedure is
guaranteed to yield the $s_{\max}$ graph are deferred to
Appendix~\ref{sec:smax-appendix}, we note that this type of
construction works well in practice for the networks under
consideration in this paper, particularly those in
Figure~\ref{fig:toynet}.

In cases where the intended degree sequence $D$ satisfies
$\sum_i d_i = 2(n-1)$, then all simple connected graphs
having degree sequence $D$ correspond to trees (i.e., acyclic
graphs), and this simple construction procedure is
guaranteed to result in an $s_{\max}$ graph.  Acyclic
$s_{\max}$ graphs have several nice properties that we will
exploit throughout this presentation.  It is
worth noting that since adding links to a tree is equivalent
to adding nodes one at a time, construction of acyclic $s_{\max}$
graphs can be viewed essentially as a type of deterministic
preferential attachment.
Perhaps more importantly, by its construction
the $s_{\max}$ tree has a natural ordering within its overall
structure, which we now summarize.

Recall that a tree can be organized into hierarchies
by designating a single vertex as the ``root'' of the
tree from which all branches emanate.
This is equivalent to assigning a direction to each arc
such that all arcs flow away from the root.  As a result,
each vertex of the graph
becomes naturally associated with a particular ``level''
of the hierarchy, adjacent vertices are separated by a
single level, and the position of a vertex within the
hierarchy is in relation to the root.
For example, assuming the root of the tree is at level 0
(the ``highest'' level), then its neighbors are at level 1
(``below'' level 0), their other neighbors in turn are at
level 2 (``below'' level 1), and so on.

Mathematically, the choice of the root vertex is an
arbitrary one, however for the $s_{\max}$ tree, the
vertex with largest degree sits as the natural root and is
the most ``central'' (a notion we will formalize below).
With this selection, two vertices $u,v \in \mathcal{V}$ that
are directly connected to each other in the acyclic $s_{\max}$
graph have the following relative position within the
hierarchy.
If $d_u \geq d_v$, then vertex $u$ is one level ``above''
vertex $v$ (alternatively, we say that vertex $u$ is ``upstream''
of vertex $v$ or that vertex $v$ is ``downstream'' from vertex $u$).
Thus, moving up the hierarchy of the tree (i.e., upstream)
means that vertex degrees are (eventually) becoming larger,
and moving down the hierarchy (i.e., downstream) means that
vertex degrees are (eventually) becoming smaller.

In order to illustrate this natural ordering within the
$s_{\max}$ tree, we introduce the following notation.
For any vertex $v \in \mathcal{V}$, let $\mathcal{N}(v)$
denote the set of neighboring vertices for $v$, where for
simple connected graphs $|\mathcal{N}(v)| = d_v$.
For an acyclic graph $g$, define $\tilde{g}^{(v)}$ to be
the {\it subgraph (subtree) of vertex $v$}; that is,
$\tilde{g}^{(v)}$ is the subtree containing vertex $v$ along
with all downstream nodes.
Since the notion of upstream/downstream is relative
to the overall root of the graph, for convenience
we will additionally use the notation $\tilde{g}^{(v,u)}$ to
represent the {\it subgraph of the vertex $v$ that is itself
connected to upstream neighbor vertex $u$}.
The (ordered) degree sequence of the subtree $\tilde{g}^{(v)}$
(equivalently for $\tilde{g}^{(v,u)}$) is then
$D(\tilde{g}^{(v)}) = \{ d^{(v)}_1, d^{(v)}_2, \dots \}$,
where $d^{(v)}_1 = d_v$ and the rest of the sequence
represents the degrees of all downstream nodes.
$D(\tilde{g}^{(v)})$ is clearly a subsequence of $D(g)$.
Finally, let $\mathcal{E}(\tilde{g}^{(v)})$ denote the set
of edges in the subtree $\tilde{g}^{(v)}$.

For this subtree, we define its $s$-value as
\begin{eqnarray}
s(\tilde{g}^{(v,u)}) & = & d_v d_u +
\sum_{(j,k) \in \mathcal{E}(\tilde{g}^{(v)})} d_j d_k. \label{eq:s-tree}
\end{eqnarray}
This definition provides a natural decomposition for the
$s$-metric, in that for any vertex $v \in \mathcal{V}$, we can write
\[ s(g) = \sum_{k \in \mathcal{N}(v)} s(\tilde{g}^{(k,v)}). \]
Furthermore, the $s$-value for any subtree can be defined as a
recursive relationship on its downstream subtrees, specifically
\begin{eqnarray*} s(\tilde{g}^{(v,u)})
& = & d_v d_u + \sum_{k \in \mathcal{N}(v) \backslash u}
s(\tilde{g}^{(k,v)}).
\end{eqnarray*}

\begin{proposition} \label{prop:smax-tree-ordering}
Let $g$ be the $s_{\max}$ acyclic graph corresponding to degree
sequence $D$. Then for two vertices $u,v \in \mathcal{V}$ with
$d_u > d_v$ it necessarily follows that
  \renewcommand{\labelenumi}{(\alph{enumi})}
  \begin{enumerate}
  \item
    vertex $v$ cannot be upstream from vertex $u$;
  \item
    the number of vertices in $\tilde{g}^{(v)}$ cannot be greater
    than the number of vertices in $\tilde{g}^{(u)}$ (i.e.,
    $|D(\tilde{g}^{(u)})| \geq |D(\tilde{g}^{(v)})|$);
  \item the degree sequence of $\tilde{g}^{(u)}$ dominates that of
    $\tilde{g}^{(v)}$ (i.e., $d^{(u)}_1 \geq d^{(v)}_1,
    d^{(u)}_2 \geq d^{(v)}_2, \dots$); and
  \item $s(\tilde{g}^{(u)}) > s(\tilde{g}^{(v)})$.
  \end{enumerate}
\end{proposition}

\noindent Although we do not prove each of these statements formally,
each of parts (a)-(d) is true by simple contradiction.  Essentially,
if any of these statements is false, there is a rewiring operation
that can be performed on the graph $g$ that increases its $s$-value,
thereby violating the assumption that $g$ is the $s_{\max}$ graph.
See Appendix~\ref{sec:smax-appendix} for additional information.

\begin{proposition} \label{prop:smax-subtree}
Let $g$ be the $s_{\max}$ acyclic graph corresponding to degree
sequence $D$.  Then it necessarily follows that for each
$v \in \mathcal{V}$ and any $k \neq v \in \mathcal{V}$,
the subgraph $\tilde{g}^{(v)}$ maximizes
$s(g^{(v,k)})$ for the degree sequence $D(\tilde{g}^{(v)})$.
\end{proposition}

\noindent The proof of Proposition~\ref{prop:smax-subtree}
follows from an inductive argument that starts with the
leaves of the tree and works its way upstream.
Essentially, in order for a tree to be the $s_{\max}$
acyclic graph, then each of its branches must be the
$s_{\max}$ subtree on the corresponding degree subsequence,
and this must hold at all levels of the hierarchy.


\begin{figure*}[t]
  \begin{center}
  \includegraphics[width=.45\linewidth]{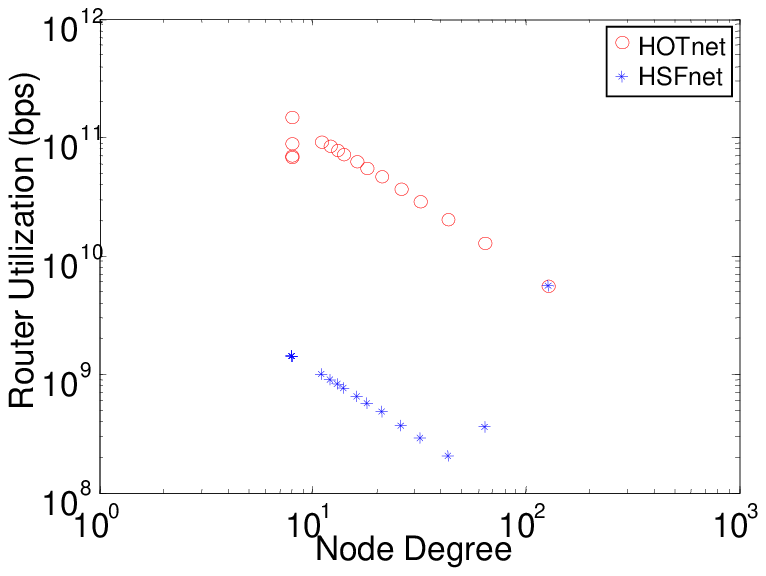}
  \includegraphics[width=.45\linewidth]{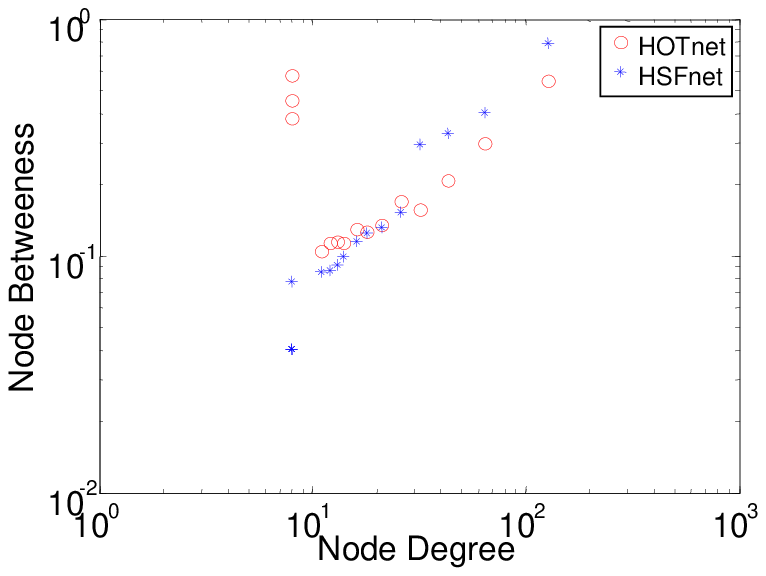}
  \vspace{-4mm}
  \caption{\label{fig:toynet-util}
  Left: The centrality of nodes as defined by total traffic throughput.
  The most ``central'' nodes in {\it HOTnet} are the low-degree core
  routers while the most ``central'' node in {\it HSFnet} is the
  highest-degree ``hub''. The {\it HOTnet} throughputs are
  close to the router bandwidth constraints.
  Right: The betweenness centrality versus node degree for
  non-degree-one nodes from both the {\it HSFnet} and {\it HOTnet}
  graphs in Figure~\ref{fig:toynet}.  In {\it HSFnet}, node
  centrality increases with node degree, and the highest degree nodes
  are the most ``central''.  In contrast, many of the most
  ``central'' nodes in  {\it HOTnet} have low degree, and the highest
  degree nodes are significantly less ``central'' than in {\it
  HSFnet}.
  }
  \end{center}
  \vspace{-4ex}
\end{figure*}


\subsection{The $s$-Metric and Node Centrality}

While considerable attention has been devoted to network node
degree sequences in order to measure the structure of complex
networks, it is clear that such sequences alone are insufficient
to characterize the aggregate structure of a graph.
Figure~\ref{fig:toynet} has shown that high degree nodes can
exist at the periphery of the network or at its core, with
serious consequences for issues such as network performance
and robustness in the presence of node loss.
At the same time, it is clear from the $s_{\max}$ construction
procedure that graphs with the largest $s(g)$ values will have
their highest degree nodes located in the network core.
Thus, an important question relates to the {\it centrality}
of individual high-degree nodes within the larger network and
how this relates, if at all, to the $s$-metric for graph
structure.  Again, the answer to this question helps to
quantify the role that individual ``hub'' nodes play
in the overall structure of a network.

There are several possible means for measuring node centrality,
and in the context of the Internet, one such measure is the
total throughput (or {\it utilization}) of a node when the
network supports its maximum flow as defined in (\ref{eq:perf}).
The idea is that under a gravity model in which traffic demand
occurs between all node pairs, nodes that are highly utilized are
central to the overall ability of the network to carry traffic.
Figure~\ref{fig:toynet-util} shows the utilization of individual
nodes within {\it HSFnet} and {\it HOTnet}, when each network
supports its respective maximum flow, along with the corresponding
degree for each node.  The picture for {\it HOTnet} illustrates that
the most ``central'' nodes are in fact low-degree nodes, which
correspond to the core routers in Figure~\ref{fig:toynet}(c).
In contrast, the node with highest utilization in {\it HSFnet}
is the highest degree node, corresponding to the ``central hub''
in Figure~\ref{fig:toynet}(a).

Another, more graph theoretic, measure of node centrality
is its so-called {\it betweenness} (also known as
{\it betweenness centrality}), which is most often
calculated as the fraction of shortest paths between node
pairs that pass through the node of interest
\cite{dorogovtsev-book2003}.  Define $\sigma_{st}$ to be
the number of shortest paths between two nodes $s$ and $t$.
Then, the betweenness centrality of any vertex $v$ can be
computed as
\[ C_b(v) = \frac{\sum_{s < t \in \mathcal{V}} \sigma_{st}(v)}
                 {\sum_{s < t \in \mathcal{V}}\sigma_{st}}, \]
where $\sigma_{st}(v)$ is the number of paths
between $s$ and $t$ that pass through node $v$.
In this manner, betweenness centrality provides a measure of the
traffic {\it load} that a node must handle.  An alternate
interpretation is that it measures the influence that an individual
node has in the spread of information within the network.

Newman \cite{newman-bc-2003} introduces a more general
measure of betweenness centrality that includes the flow
along all paths (not just the shortest ones), and based on
an approach using random walks demonstrates how this
quantity can be computed by matrix methods. Applying this
alternate metric from \cite{newman-bc-2003} to the simple
annotated graphs in Figure~\ref{fig:toynet}, we observe
in Figure~\ref{fig:toynet-util} that the high-degree nodes
in {\it HSFnet} are the most central, and in fact this
measure of  betweeness centrality increases with node degree.
In contrast, most of the nodes in {\it HOTnet} that are
central are not high degree nodes, but the low-degree
core routers.

Understanding the betweenness centrality of individual nodes is
considerably simpler in the context of trees. Recall
that in an acyclic graph there is exactly one path
between any two vertices, making the calculation of
$C_b(v)$ rather straightforward.  Specifically, observe
that $\sum_{s < t \in \mathcal{V}} \sigma_{st}
= n(n-1)/2$ and that for each $s \neq v \neq t \in \mathcal{V}$,
$\sigma_{st}(v) \in \{0,1\}$.
This recognition facilitates the following more general
statement regarding the centrality of high-degree nodes in
the $s_{\max}$ acyclic graph.
\begin{proposition} \label{prop:centrality-tree}
Let $g$ be the $s_{\max}$ acyclic graph for degree sequence $D$,
and consider two nodes $u,v \in \mathcal{V}$ satisfying $d_u > d_v$.
Then, it necessarily follows that $C_b(u) > C_b(v)$.
\end{proposition}

\noindent The proof of Proposition~\ref{prop:centrality-tree}
can be found in Appendix~\ref{sec:smax-appendix}, along with
the proof of the $s_{\max}$ construction.
Thus, the highest degree nodes in the $s_{\max}$ acyclic graph
are the most central.
More generally for graphs that are not trees, we believe
that there is a direct relationship between high-degree ``hub''
nodes in large-$s(g)$ graphs and a ``central'' role in overall
network connectivity, but this has not been formally proven.

\subsection{The $s$-Metric and Self-Similarity}

When viewing graphs as multiscale objects, natural
transformations that yield simplified graphs are pruning of
nodes at the graph periphery
and/or collapsing of nodes, although these are only
the simplest of many possible ``coarse-graining''
operations that can be performed on graphs. These
transformations are of particular interest because they are
often inherent in measurement processes that are aimed at
detecting the connectivity structure of actual networks. We
will use these transformations to motivate that there is a
plausible relationship between high-$s(g)$ graphs and
self-similarity, as defined by these simple operations. We
then consider the transformation of random pairwise
degree-preserving (link) rewiring that suggests a more
formal definition of the notion of a self-similar graph.

\subsubsection{Graph Trimming by Link Removal}

Here, we consider the properties of $s_{\max}$ graphs
under the operation of graph trimming, in which
links are removed from the graph one at a time.
Recall that by construction, the links in the $s_{\max}$
graph are selected from a list of potential links (denoted
as $(i,j)$ for $i,j \in \mathcal{V}$) that are ordered
according to their weights $d_i d_j$.   Denote the
(ordered) list of links in the $s_{\max}$ graph as
$\mathcal{E} = \{ (i_1, j_1), (i_2, j_2), \dots, (i_l, j_l)
\}$, and consider a procedure that removes links in reverse
order, starting with $(i_l, j_l)$. Define $\tilde{g}_k$ to
be the remaining graph after the removal of all but the
first $k-1$ links, (i.e., after removing $(i_l, j_l),
(i_{l-1}, j_{l-1}), \dots, (i_{k+1}, l_{k+1}), (i_{k},
l_{k})$). The remaining graph will have a partial degree
sequence $\tilde{D}_k = \{ d_1^{'}, d_2^{'}, \dots, d_k^{'}
\}$, where $d_m^{'} \leq d_m, m = 1, 2, \dots k$, but the
original ordering is preserved, i.e., $d_1^{'} \geq d_2^{'}
\geq \dots \geq d_k^{'}$.  This last statement holds
because when removing links starting with the smallest $d_i
d_j$, nodes will ``lose'' links in reverse order according
to their node degree.

Observe for trees that removing a link is equivalent to
removing a node (or subtree), so we could have equivalently
defined this process in terms of ``node pruning''.
As a result, for acyclic $s_{\max}$ graphs, it is easy to
see the following.

\begin{proposition} \label{prop:pruning}
Let $g$ be an acyclic $s_{\max}$ graph
satisfying ordered degree sequence
$D=\{d_1, d_2, \ldots, d_n\}$.  For $1\leq k \leq n$,
denote by $\tilde{g}_{k}$ the acyclic graph
obtained by removing (``trimming'') in
order nodes $n, n-1, \ldots, k+1$ from $g$.
Then,
$\tilde{g}_{k}$ is the $s_{\max}$ graph for degree
sequence $\tilde{D}_k = \{ d_1^{'}, d_2^{'}, \ldots, d_k^{'} \}$.
\end{proposition}

\noindent The proof of Proposition~\ref{prop:pruning}
follows directly from our
proof of the construction of the $s_{\max}$ graph for trees
(see Appendix~\ref{sec:smax-appendix}).
More generally, for graphs exhibiting large
$s(g)$-values, properly defined graph operations of
link-trimming appear to yield simplified graphs
with high s-values, thus suggesting a broader
notion of self-similarity or invariance under
such operations.  However, additional work remains
to formalize this notion.

\subsubsection{Coarse Graining By Collapsing Nodes}

A kind of {\it coarse graining} of a graph can be obtained
for producing simpler graphs by collapsing existing nodes
into aggregate or super nodes and removing any duplicate
links emanating from the new nodes. Consider the case of a
tree $g$ having degree sequence $D=\{d_1, d_2, \dots,
d_n\}$ satisfying $d_1 \geq d_2 \geq \dots \geq d_n$ and
connected in a manner such that $s(g)=s_{\max}$. Then, as
long as node aggregation proceeds in order with the degree
sequence (i.e.~aggregate nodes $1$ and $2$ into $1'$, then
aggregate nodes $1'$ and $3$ into $1''$, and so on), all
intermediate graphs $\tilde{g}$ will also have
$s(\tilde{g})=s_{\max}$.  To see this, observe that
for trees, when aggregating nodes $1$ and $2$, we have
an abbreviated degree sequence $D' = \{ d_1^{'}, d_3, \
\dots, d_n\}$, where $d_1^{'} = d_1 + d_2 -2$.  Provided
that $d_2 \geq 2$ then we are guaranteed to have
$d_1^{'} \geq d_3$, and the overall ordering of $D'$ is
preserved.  Similarly when aggregating nodes $1^{'}$ and
$3$ we have abbreviated degree sequence $D^{''} =
\{ d_1^{''}, d_4, \dots, d_n \}$, where
$d_1^{''} = d_1 + d_2 + d_3 - 4$.  So as long as
$d_3 \geq 2$ then $d_1^{''} \geq d_4$ and ordering of $D^{''}$
is preserved.  And in general, as long as each new node is
aggregated in order and satisfies $d_i \geq 2$, then we are
guaranteed to maintain an ordered degree sequence.  As a
result, we have proved the following proposition.

\begin{proposition} For acyclic $g \in G(D)$ with
$s(g)=s_{\max}$, coarse graining according to the above
procedure yields smaller graphs $g' \in G(D')$ that are
also the $s_{\max}$ graphs of this truncated degree distribution.
\end{proposition}

For cyclic graphs, this type of node aggregation operation
maintains $s_{\max}$ properties only if the resulting
degree sequence remains ordered, i.e.~$d_{1'} \geq d_3 \geq
d_4$ after the first coarse graining operation and $d_{1''}
\geq d_4 \geq d_5$ after the second coarse graining
operation, etc.  It is relatively easy to generate cases
where arbitrary node aggregation violates this condition
and the resulting graph is no longer self-similar in the
sense of having a large $s(g)$-value.  However, when this
condition is satisfied, the resulting simpler graphs seem
to satisfy a broader self-similar property.
Specifically, for high-s(g) graphs $g\in G(D)$, properly
defined graph operations of coarse-graining appear to
yield simplified graphs in $G(D)$ with high s-values
(i.e., such graphs are self-similar or invariant under
proper coarse-graining), but this has not been proved.

These are of course not the only coarse graining, pruning,
or merging processes that might be of interest, and for
which $s_{\max}$ graphs are preserved, but they are perhaps the
simplest to state and prove.

\subsection{Self-Similar and Self-Dissimilar} \label{sec:ssandsds}

While graph transformations such as link trimming or node
collapse reflect some aspects of what it means for a graph
to be self-similar, the graph transformation of random
pairwise degree-preserving link rewiring offers additional
notions of self-similarity which potentially are even
richer and also connected with the claim in the SF
literature that SF graphs are preserved under such
rewirings.


\begin{figure*}[t]
  \begin{minipage}[t]{.42\textwidth}
    \includegraphics[width=.95\linewidth]{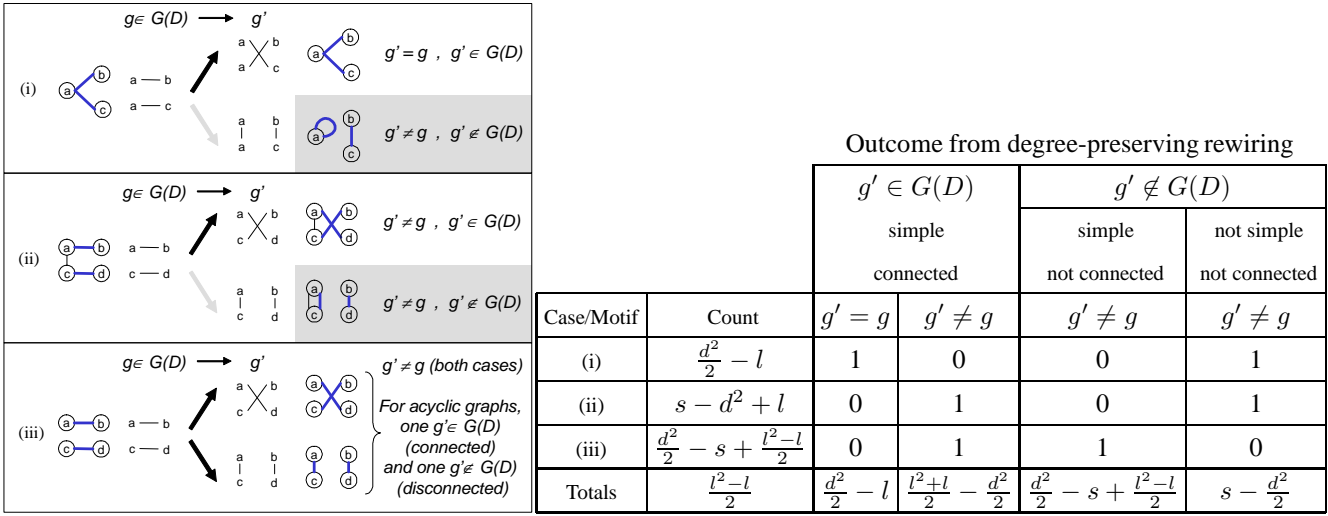}
  \end{minipage}
  \hspace{-5mm}
  \begin{minipage}[t]{.5\textwidth}
    \renewcommand{\arraystretch}{1.5}
    \setlength{\tabcolsep}{1mm}
    \begin{tabular}{|c|c|c|c|c|c|}
    \multicolumn{2}{c}{} &
      \multicolumn{4}{c}{Outcome from degree-preserving rewiring}\\
    \cline{3-6}
    \multicolumn{2}{c}{} &
      \multicolumn{2}{|c|}{$g' \in G(D)$} & \multicolumn{2}{|c|}{$g'
      \not\in G(D)$} \\
    \cline{5-6}
    \multicolumn{2}{c}{} &
      \multicolumn{2}{|c|}{ \footnotesize simple} & { \footnotesize
      simple} & { \footnotesize not simple}\\
    \multicolumn{2}{c}{} &
      \multicolumn{2}{|c|}{ \footnotesize connected} & { \footnotesize
      not connected} & { \footnotesize not connected}\\
    \hline
    {\footnotesize Case/Motif} & { \footnotesize Count} & $g' = g$ &
      $g' \neq g$ & $g' \neq g$ & $g' \neq g$ \\
    \hline
    {\footnotesize (i)} & $\frac{d^2}{2} - l$ & 1 & 0 & 0 & 1\\
    \hline
    {\footnotesize (ii)} & $s - d^2 + l$ & 0 & 1 & 0 & 1\\
    \hline
    {\footnotesize (iii)} & $\frac{d^2}{2} - s + \frac{l^2 - l}{2}$ & 0
      & 1 & 1 & 0\\
    \hline
    {\footnotesize Totals} & $\frac{l^2-l}{2}$ & $\frac{d^2}{2}-l$ &
      $\frac{l^2+l}{2} - \frac{d^2}{2}$ & $\frac{d^2}{2}-s+\frac{l^2-l}{2}$
      & $s - \frac{d^2}{2}$\\
    \hline
    \end{tabular}
  \end{minipage}
  \caption{\label{fig:rewire}
    {\it (a) Three possible subgraph-based motifs in
    degree-preserving rewiring in acyclic graphs.}
    Blue links represents links to be rewired.  Rewiring
    operations that result in non-simple graphs (shaded) are
    assumed to revert to the original configuration. Thus defined,
    rewiring of motif (i) does not result in any new graphs,
    rewiring of motif (ii) results in one possible new graph, and
    rewiring of motif (iii) results in two possible new graphs.
    {\it (b) The numbers of the three motifs
    and successively the number for each
    possible rewiring outcome.}  We distinguish between
    equal, not equal but connected and simple, not connected
    but simple, and not simple graphs
    that are similar to each graph with the given motif
    selected for rewiring.
    The total number of cases (column sum) is $(l^2 -l)/2$,
    while the total number (row sum) of outcomes is twice that at
    $l^2 -l$.
    Here, we use the abbreviated notation $d^2 = \sum_k {d_k}^2$
    and $s=s(g)$, with $l$ equal to the number of links in the graph.
  }
\vspace{-4ex}
\end{figure*}


\subsubsection{Subgraph-Based Motifs} \label{sec:rewire}

For any graph $g \in G(D)$, consider the set of local
degree-preserving rewirings of distinct pairs of links.
There are ${l \choose 2}=l(l-1)/2$ pairs of different
links on which degree preserving rewiring can occur.
Each pair of links defines its own network subgraph, and in
the case where $g$ is an acyclic graph (i.e.~a tree), these
form three distinct types of subgraphs, as shown in
Figure~\ref{fig:rewire}(a). Using the notation $d^2 =
\sum{d_k}^2$, $s=s(g)$ we can enumerate the number of these
subgraphs as follows:
\begin{enumerate}
\item The two links share a common node.
  There are $\sum_{i=1}^n {d_i \choose 2} = \frac{1}{2} d^2 - l$
  possible ways that this can occur.
\item The links have two nodes that are connected by a
third link.
  There are $\sum_{(i,j) \in \mathcal{E}} (d_i-1)(d_j-1) =
  s - d^2 + l$ possible ways that this can occur.
\item The links have end points that do not
share any direct connections.
  There are ${l \choose 2} - \sum_{i=1}^n {d_i \choose 2} -
  \sum_{(i,j) \in \mathcal{E}} (d_i-1)(d_j-1) = \frac{1}{2} d^2 - s +
  \frac{1}{2}(l^2 - 2)$ possible ways that this can occur.
\end{enumerate}
\noindent Collectively, these three basic subgraphs account
for all possible ${l \choose 2} = l(l-1)/2$ pairs of
different links.  The subgraphs in cases (i) and (ii) are
themselves trees, while the subgraph in case (iii) is not.
We will refer to these three cases for subgraphs as
``motifs'', in the spirit of \cite{milo:alon:2002}, noting
that our notion of subgraph-based motifs is motivated by
the operation of random rewiring to be discussed below.

The simplest and most striking feature of the relationship
between motifs and $s(g)$ for acyclic graphs is
that we can derive formulas for the number of subgraph-based
(local) motifs (and the outcomes of rewiring) entirely in
terms of $d^2$, $s=s(g)$, and $l$.
Thus, for example, we can see that graphs having
higher $d^2$ (equivalently higher $CV$) values have fewer
of the second motif.  If we fix $D$, and thus $l$ and $d^2$,
for all graphs of interest, then the only remaining
dependence is on $s$, and graphs with higher
$s(g)$-values contain fewer disconnected (case iii) motifs.
This can be interpreted as a motif-level connection between
$s(g)$ and self-similarity, in that graphs with higher $s(g)$
contain more motifs that are themselves trees, and thus more
similar to the graph as a whole.  Graphs having lower $s(g)$
have more motifs of type (iii) that are disconnected and
thus dissimilar to the graph as a whole. Thus
high-$s(g)$ graphs have this ``motif self-similarity,''
low-$s(g)$ graphs have ``motif self-dissimilarity'' and we
can precisely define a measure of this kind of
self-similarity and self-dissimilarity as follows.

\begin{definition} \label{def:selfsim}
For a graph $g \in G(D)$, another measure of the {\it extent to
which $g$ is self-similar} is the metric $ss(g)$ defined as
the number of motifs (cases i-ii) that are themselves
connected graphs. Accordingly, the measure of self-dissimilarity
$sd(g)$ is then the number of motifs (case iii) that are
disconnected.
\end{definition}

For trees, $ss(g)=s-d^2/2$ and $sd(g)=-s+(l^2-l+d^2)/2$, so
this local motif self-similarity (self-dissimilarity) is
essentially equivalent to high-$s(g)$ (resp.~low-$s(g)$). As
noted previously, network motifs have already been used as
a way to study self-similarity and coarse graining
\cite{itzkovitz:alon:2004a,itzkovitz:alon:2004b}.
There, one defines a recursive procedure by which node
connectivity patterns become represented as a single node
(i.e.~a different kind of motif), and it was shown that many
important technological and biological networks were
self-dissimilar, in the sense coarse-grained counterparts
display very different motifs at each level of abstraction.
Our notion of motif self-similarity is much simpler, but
consistent, in that the Internet has extremely low $s(g)$
and thus minimally self-similar at the motif level.  The
next question is whether high $s(g)$ is connected with
``self-similar'' in the sense of being preserved under rewiring.

\subsubsection{Degree-preserving Rewiring}

We can also connect $s(g)$ in several ways with the effect
that local rewiring has on the global structure of graphs
in the set $G(D)$.  Recall the above process by which two
network links are selected at random for degree-preserving
rewiring, and note that when applied to a graph $g \in G(D)$,
there are four possible distinguishable outcomes:
\begin{enumerate}

\item $g' = g$ with $g' \in G(D)$: the new graph $g'$ is
{\it equal} to the original graph $g$ (and therefore also a
simple, connected graph in $G(D)$);

\item $g' \neq g$ with $g' \in G(D)$: the new graph $g'$ is
not equal to $g$, but is still a simple, connected graph in
the set $G(D)$ (note that this can include $g'$ which are
isomorphic to $g$);

\item $g' = g$ with $g' \not\in G(D)$: the new graph $g'$
is still simple, but is not connected;

\item $g' = g$ with $g' \not\in G(D)$: the new graph $g'$
is no longer simple (i.e.~it either contains self-loops
or parallel links).

\end{enumerate}
There are two possible outcomes from the
rewiring of any particular pair of links, as shown in
Figure~\ref{fig:rewire}(a) and this yields a total of
$2 {l\choose 2} = l(l-1)$ possible outcomes of the rewiring
process.  In our discussion here, we ignore isomorphisms
and assume that all non-equal graphs are different.

We are ultimately interested in retaining within our new
definitions the notion that high $s(g)$ graphs are somehow
preserved under rewiring provided this is sufficiently
random and degrees are preserved.
Scaling is of course trivially preserved by any
degree-preserving rewiring, but high $s(g)$ value is not.
Again, Figure~\ref{fig:toynet} provides a clear example,
since successive rewirings can take any of these graphs to
any other. More interesting for high $s(g)$ graphs is the
effect of {\it random} rewiring. Consider again the
$\mbox{\em Perf}(g)$ vs.~$s(g)$ plane from
Figure~\ref{fig:perf-llh}. In addition to the four networks
from Figure~\ref{fig:toynet}, we show the $\mbox{\em
Perf}(g)$ and $s(g)$ values for other graphs in $G(D)$
obtained by degree-preserving rewiring from the initial
four networks. This is done by selecting uniformly and
randomly from the $l(l-1)$ different rewirings of the
$l(l-1)/2$ different pairs of links, and restricting
rewiring outcomes to elements of $G(D)$ by resetting all
disconnected or nonsimple neighbors to equal. Points that
match the color of one of the four networks are only one
rewiring operation away, while points represented in gray
are more than one rewiring operation away.

The connections of the results in
Figure~\ref{fig:rewire}(b) to motif counts is more
transparent however than to the consequences of successive
rewiring. Nevertheless, we can use the results in
Figure~\ref{fig:rewire}(b) to describe related ways in
which low $s(g)$ graphs are ``destroyed'' by random
rewiring. For any graph $g$, we can enumerate among all
possible pairs of links on which degree preserving rewiring
can take place and count all those that result in equal or
non-equal graphs. In Figure \ref{fig:rewire}, we consider
the four cases for degree-preserving rewiring of acyclic
graphs, and we count the number of ways each can occur. For
motifs (i) and (ii), it is possible to check locally for
outcomes that produce non-simple graphs and these cases
correspond to the shaded outcomes in
Figure~\ref{fig:rewire}(a). If we a priori exclude all such
nonsimple rewirings, then there remain a total of
$l(l-1)-s+d^2/2$ simple similar neighbors of a tree.  We
can define a measure of local rewiring self-dissimilarity
for trees as follows.

\begin{definition}
For a tree $g \in G(D)$, we measure the extent to which $g$
is self-dissimilar under local rewiring by the metric
$rsd(g)$ defined as the number of simple similar neighbors
that are disconnected graphs.
\end{definition}

For trees, $rsd(g)=sd(g)=-s+(l^2-l+d^2)/2$, so this local
rewiring self-dissimilarity is identical to motif
self-dissimilarity and directly related to low $s(g)$ values.
This is because only motif (iii) results in simple but not
connected similar neighbors.

\subsection{A Coherent Non-Stochastic Picture}

Here, we pause to reconsider the features/claims for SF graphs
in the existing literature (Section~\ref{sec:SFclaims}) in light
of our structural approach to graphs with scaling degree sequence $D$.
In doing so, we make a simple observation: high-$s(g)$ graphs exhibit
most of the features highlighted in the SF literature, but low-$s(g)$
graphs do not, and this provides insight into the diversity of
graphs in the space $G(D)$.
Perhaps more importantly, given a graph with scaling degree $D$
the $s(g)$ metric provides a ``litmus test'' as to whether or
not the existing SF literature might be relevant to the network
under study.

By definition, all graphs in $G(D)$ exhibit power laws in their
node degrees provided that $D$ is scaling.  However,
preferential attachment mechanisms typically yield only high-$s(g)$
graphs---indeed the $s_{\max}$ construction uses what is
essentially the ``most preferential'' type of attachment mechanism.
Furthermore, while all graphs having scaling degree sequence $D$
have high-degree nodes or ``hubs'', only for high-$s(g)$ graphs
do such hubs tend to be critical for overall connectivity.
While it is certainly possible to construct a graph with low $s(g)$
and having a central hub, this need not be the case, and
our work to date suggests that most low-$s(g)$ graphs do not have
the type of central hubs that create an ``Achilles' heel''.
Additionally, we have illustrated that high-$s(g)$ graphs
exhibit striking self-similarity properties, including that
they are largely preserved under appropriately defined
graph transformations of trimming, coarse graining and
random pairwise degree-preserving rewiring. In the case of
random rewiring, we offered numerical evidence and
heuristic arguments in support of the conjecture that in
general high-$s(g)$ graphs are the likely outcome of
performing such rewiring operations, whereas low-$s(g)$
graphs are unlikely to occur as a result of this process.

Collectively, these results suggest that a definition of
``scale-free graphs'' that restricts graphs to having
{\it both} scaling degree $D$ {\it and} high-$s(g)$ results in a
coherent story.  It recovers all of the structural results
in the SF literature and provides a possible explanation why
some graphs that exhibit power laws in their node degrees
do not seem to satisfy other properties highlighted in the
SF literature.
This non-stochastic picture represents what is arguably a
reasonable place to stop with a theory for ``scale-free'' graphs.
However, from a graph theoretic perspective, there is considerable
more work that could be done.
For example, it may also be possible to expand the discussion
of Section \ref{sec:ssandsds} to account more comprehensively
for the way in which local motifs are transformed into
one another and to relate our attempts more directly to
the approach considered in \cite{milo:alon:2002}.
Elaborating on the precise relationships and providing a
possible interpretation of motifs as capturing a kind of local
as well as global self-similarity property of an underlying graph
remain open interesting problems.
Additionally, we have also seen that the use of
degree-preserving rewiring among connected graphs provides
one view into the space $G(D)$.  However, the geometry of
this space is still complicated, and additional work is
required to understand its remaining features. For example,
our work to date suggests that for scaling $D$ it is
impossible to construct a graph that has both high
$\mbox{\em Perf}(g)$ and high $s(g)$, but this has not been
proven.  In addition, it will be useful to understand the
way that degree-preserving rewiring causes one to ``move''
within the space $G(D)$ (see for example,
\cite{markovchain-plrg,farkas2004}).

It is important to emphasize that the purpose of the $s(g)$
metric is to provide insight into the structure of ``scale
free'' graphs and {\it not as a general metric for
distinguishing among all possible graphs}.  Indeed, since
the metric fails to distinguish among graphs having low
$s(g)$, it provides little insight other than to say that
there is tremendous diversity among such graphs. However,
if a graph has high $s(g)$, then we believe that there
exist strong properties that can be used to understand the
structure (and possibly, the behavior) of such systems.  In
summary, if one wants to understand ``scale-free graphs'',
then we argue that $s(g)$ is an important metric and highly
informative. However, for graphs with low $s(g)$ then
this metric conveys limited information.

Despite the many appealing features of a theory that
considers only non-stochastic properties, most of the SF
literature has considered a framework that is inherently
stochastic. Thus, we proceed next with a stochastic version
of the story, one that connects more directly with the
existing literature and common perspective on SF graphs.

\section{A Probabilistic Approach}\label{sec:prob}

While the introduction and exploration of the $s$-metric
fits naturally within standard studies of graph theoretic
properties, it differs from the SF literature in that our
structural approach does not depend on a probability model
underlying the set of graphs of interest.
The purpose of this section is to compare our approach with
the more conventional probabilistic and ensemble-based
views. For many application domains, including the
Internet, there seems to be little motivation to assume
networks are samples from an ensemble, and our treatment
here will be brief while trying to cover this broad
subject. Here again, we show that the $s(g)$ metric is
potentially interesting and useful, as it has a direct
relationship to notions of graph likelihood, graph degree
correlation, and graph assortativity.
This section also highlights the striking differences in
the way that randomness is treated in physics-inspired
approaches versus those shaped by mathematics and
engineering.

The starting point for most probabilistic approaches to the study
of graphs is through the definition of an appropriate {\it statistical
ensemble} (see for example \cite[Section 4.1]{dorogovtsev-book2003}).

\begin{definition} A {\it statistical ensemble of graphs} is defined by
\renewcommand{\labelenumi}{(\roman{enumi})}
\begin{enumerate}
\item a set $G$ of graphs $g$, and
\item a rule that associates a real number (``probability'')
$0\leq P(g)\leq 1$ with each graph $g\in G$ such that
$\sum_{g\in G} P(g) = 1$.
\end{enumerate} \label{def:graph-ensemble}
\end{definition}
\noindent To describe an ensemble of graphs, one can either
assign a specific weight to each graph or define some
process (i.e., a stochastic generator) which results in a
weight.  For example, in one basic model of random graphs,
the set $G$ consists of all graphs with vertex set
$V=\{1,2,\ldots,n\}$ having $l$ edges, and each element in
$G$ is assigned the same probability $1/{n \choose l}$. In
an alternative random graph model, the set $G$ consists of
all graphs with vertex set $V=\{1,2,\ldots,n\}$ in which
the edges are chosen independently and with probability
$0<p<1$. In this case, the probability $P(g)$ depends on
the number of edges in $g$ and is given by $P(g)=p^l
(1-p)^{n-l}$, where $l$ denotes the number of edges in
$g\in G$.

The use of stochastic construction procedures to assign
statistical weights has so dominated the study of graphs
that the assumption of an underlying probability model
often becomes implicit.  For example, consider the four
graph construction procedures listed in
\cite[p.22]{dorogovtsev-book2003} that are claimed to form
{\it ``the basis of network science,''} and include (1) classical
random graphs due to Erd\"{o}s and Reny\'{i}
  \cite{ErdosRenyi59};
(2) equilibrium random graphs with a given degree distribution such as
  the {\it Generalized Random Graph (GRG)} method \cite{ChungLu03};
(3) ``small-world networks'' due to Watts and Strogatz
  \cite{WattsStrogatz}; and
(4) networks growing under the mechanism of preferential linking due
  to Barab\'{a}si and Albert \cite{BarabasiAlbert99} and made
  precise in \cite{BollobasRiordan03}.
All of these construction mechanisms are inherently {\it
stochastic} and provide a natural means for assigning, at
least in principle, probabilities to each element
in the corresponding space of realizable graphs.
While deterministic (i.e., non-stochastic) construction
procedures have been considered \cite{BarRavVic01}, their
study has been restricted to the treatment of deterministic
preferential attachment mechanisms that result in
pseudofractal graph structures.
Graphs resulting from other types of deterministic
constructions are generally ignored in the context of
statistical physics-inspired approaches since within the
space of all feasible graphs, their likelihood of occurring
is typically viewed as vanishingly small.

\subsection{A Likelihood Interpretation of $s(g)$}
\label{sec:likelihood}

Using the construction procedure associated with the {\em
general model of random graphs with a given expected degree
sequence} considered in \cite{ChungLu03} (also called the
{\it Generalized Random Graph (GRG) model} for short) we
show that the $s(g)$ metric allows for a more familiar
ensemble-related interpretation as {\it (relative)
likelihood} with which the graph $g$ is constructed
according to the GRG method. To this end, the GRG model is
concerned with generating graphs with given {\it expected}
degree sequence $D = \{d_1,\ldots d_n\}$ for vertices
$1,\ldots,n$.  The link between vertices $i$ and $j$ is
chosen independently with probability $p_{ij}$, with
$p_{ij}$ proportional to the product $d_i d_j$ (i.e.
$p_{ij}=\rho d_i d_j$, where $\rho$ is a sufficiently small
constant), and this defines a probability measure $P$ on
the space of all simple graphs and thus induces a
probability measure on $G(D)$ by conditioning on
having degree $D$. The construction is fairly general and
can recover the classic Erd\"{o}s-R\'{e}nyi random graphs
\cite{ErdosRenyi59} by taking the expected degree sequence
to be $\{pn,pn,\ldots,pn\}$ for constant $p$. As a result
of choosing each link $(i,j)\in \mathcal{E}$ with a
probability that is proportional to $d_i d_j$ in the GRG
model, different graphs are typically assigned different
probabilities under $P$.  This generation method is closely
related to the {\it Power Law Random Graph (PLRG)} method
\cite{PLRG}, which also attempts to replicate a given
(power law) degree sequence. The PLRG method involves
forming a set $L$ of nodes containing as many distinct
copies of a given vertex as the degree of that vertex,
choosing a random matching of the elements of $L$, and
applying a mapping of a given matching into an appropriate
(multi)graph. It is believed that the PLRG and GRG models
are {\it ``basically asymptotically equivalent, subject to
bounding error estimates''}~\cite{PLRG}. Defining the {\it
likelihood} of a graph $g\in G(D)$ as the
logarithm of its probability under the measure $P$, we can
show that the log likelihood (LLH) of a graph $g \in
G(D)$, can be computed as
\beq LLH(g) \approx \kappa +\rho \ s(g),
\eeq
where $\kappa$ is a constant.

Note that the probability of any graph $g $ under $P$
is given by \cite{ParkNewman}
\begin{eqnarray*}
P(g)&=&
\displaystyle{\prod_{(i,j)\in \mathcal{E}} p_{ij}
\prod_{(i,j)\notin \mathcal{E}} (1-p_{ij})},
\end{eqnarray*}
and using the fact that under the GRG model, we have
$p_{ij}=\rho d_i d_j$, where $D = (d_1,\ldots d_n)$ is the given
degree sequence, we get
\begin{eqnarray*}
P(g)&=&
\rho^l \displaystyle{\prod_{i\in \mathcal{V}}
d_i^{d_i}
\prod_{(i,j)\notin \mathcal{E}} (1-\rho d_i d_j)} \\
&=& \rho^l \displaystyle{\prod_{i\in \mathcal{V}}
d_i^{d_i} \frac{
\prod_{i,j \in \mathcal{V}} (1-\rho d_i d_j)}
{\prod_{(i,j)\in \mathcal{E}} (1-\rho d_i d_j)}}.
\end{eqnarray*}
Taking the log, we obtain
\begin{eqnarray*}
\log P(g) &=& l \log \rho
+ \displaystyle{\sum_{i\in \mathcal{V}} d_i \log d_i
+ \sum_{i,j \in \mathcal{V}} \log (1-\rho d_i d_j)}\\
& & - {\sum_{(i,j)\in \mathcal{E}} \log (1-\rho d_i d_j)}.
\end{eqnarray*}
Defining
\[
\kappa = l \log\rho +\displaystyle{\sum_{i\in \mathcal{V}}
d_i \log d_i + \sum_{i,j \in \mathcal{V}} \log (1-\rho d_i d_j)},
\]
we observe that $\kappa$ is constant for fixed degree sequence $D$.
Also recall that $\log(1 + a) \approx a$ for $|a|<<1$.
Thus, if $\rho$ is sufficiently small so that $p_{ij} =
\rho d_i d_j << 1$, we get
\[ LLH(g)=\log P(g)  \approx
\kappa + \sum_{(i,j)\in \mathcal{E}}\rho d_i d_j. \] This
shows that the graph likelihood $LLH(g)$ can be made
proportional to $s(g)$ and thus we can interpret
$s(g)/s_{\max}$ as {\it relative likelihood} of $g \in
G(D)$, for the $s_{\max}$-graph has the highest likelihood
of all graphs in $G(D)$. Choosing $\rho=1 / \sum_{i\in
\mathcal{V}} d_i = 1/2l$ in the GRG formulation results in
the expectation
\[
E(d_i) = \sum_{j=1}^{n} p_{ij} = \sum_{j=1}^{n} \rho d_i d_j =
\rho d_i \sum_{j=1}^{n} d_j = d_i.
\]
However, this $\rho$ may not have $p_{ij} = \rho d_i d_j <<
1$ and can even make $p_{ij}> 1$, particularly in cases
when the degree sequence is scaling. Thus $\rho$ must often
be chosen much smaller than $\rho=1 / \sum_{i\in
\mathcal{V}} d_i = 1/2l$ to ensure that $p_{ij} << 1$ for
all nodes $i,j$. In this case, the ``typical'' graph resulting
from this construction with have degree sequence much less
than $D$, however this sequence will be proportional to the
desired degree sequence, $E(d_i) \propto d_i$.

While this GRG construction yields a probability
distribution on  $G(D)$ by conditioning on having degree
sequence $D$, this is not an efficient, practical method to
generate members of $G(D)$, particularly when $D$ is
scaling and it is necessary to choose $\rho << 1/2l$.
The appeal of the GRG
method is that it is easy to analyze and yields
probabilities on $G(D)$ with clear interpretations. All
elements of $G(D)$ will have nonzero probability with log
likelihood proportional to $s(g)$.  But even the $s_{max}$
graph may be extremely unlikely, and thus a naive Monte
Carlo scheme using this construction would rarely yield any
elements in $G(D)$.  There are many conjectures in the SF
literature that suggest that a wide variety of methods,
including random degree-preserving rewiring, produce
``essentially the same'' ensembles. Thus it may be possible
to generate probabilities on $G(D)$ that can both be
analyzed theoretically and also provide a practical scheme
to generate samples from the resulting ensemble.  While we
believe this is plausible, it's rigorous resolution is well
beyond the scope of this paper.


\begin{figure*}[th]
  \begin{center}
  \includegraphics[width=.35\linewidth]{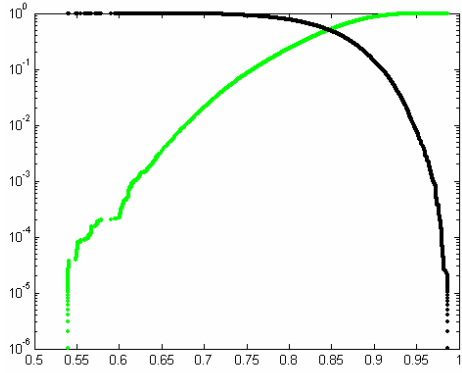}
  \vspace{-2mm}
  \caption{\footnotesize   \label{fig:cumf}
   {\sc Results from Monte Carlo generation of preferential
   attachment graphs having 1000 nodes.}  For each trial,
   we compute the value $s(g)$ and then renormalize to $S(g)$
   against the $s_{\max}$ graph having the same degree sequence.  Both
   the CDF and CCDF are shown.  In comparison, the {\it HOTnet} graph
   has $S(HOTnet)=0.3952$ and $S(HSFnet)=0.9791$.
  }
  \end{center}
\vspace{-4ex}
\end{figure*}


\subsection{Highly Likely Constructions}

The interpretation of $s(g)$ as (relative) graph likelihood
provides an explicit connection between this structural
metric and the extensive literature on random graph models.
Since the GRG method is a general means of generating
random graphs, we can in principle generate random
instances of ``scale-free'' graphs with a prescribed power
law degree sequence, by using GRG as described above and
then conditioning on that degree sequence. (And more
efficient, practical schemes may also be possible.) In the
resulting probability distribution on the space of graphs
$G(D)$, high-$s(g)$ graphs with hub-like core structure are
literally ``highly likely'' to arise at random, while
low-$s(g)$ graphs with their high-degree nodes residing at
the graphs' peripheries are ``highly unlikely'' to result
from such stochastic construction procedures.

While graphs resulting from stochastic preferential
attachment construction may have a different underlying
probability model than GRG-generated graphs, both result
in simple graphs having approximate scaling relationships
in their degree distributions.  One can understand the manner
in which high-$s(g)$ graphs are ``highly
likely'' through the use of a simple Monte Carlo simulation
experiment.  Recall that the toy graphs in
Figure~\ref{fig:toynet} each contained 1000 nodes and that
the graph in Figure~\ref{fig:toynet}(b) was ``random'' in
the sense that it was obtained by successive arbitrary
rewirings of {\it HSFnet} in Figure~\ref{fig:toynet}(a). An
alternate approach to generating random graphs having a
power law in their distribution of node degree is to use
the type of preferential attachment mechanism first
outlined in~\cite{BarabasiAlbert99} and consider the
structural features that are most ``likely'' among a large
number of trials.  Here, we generate 100,000 graphs each
having 1000 nodes and measure the $s$-value of each.  It is
important to note that successive graphs resulting from
preferential attachment will have different node degree
sequences (one that is undoubtedly different from the
degree sequence in Figure~\ref{fig:toynet}(e)), so a raw
comparison of $s(g)$ is not appropriate.  Instead, we
introduce the normalized value $S(g) = s(g)/s_{\max}$ and
use it to compare the structure of these graphs.  Note that
this means also generating the $s_{\max}$ graph associated
with the particular degree sequence for the graph resulting
from each trial.  Fortunately, the construction procedure
in Appendix~\ref{sec:smax-appendix} makes this
straightforward, and so in this manner we obtain the
normalized $S$-values for 100,000 graphs resulting from the
same preferential attachment procedure. Plotting the CDF
and CCDF of the $S$-values for these graphs in
Figure~\ref{fig:cumf}, we observe a striking picture: all
of the graphs resulting from preferential attachment had
values of $S$ greater than 0.5, most of the graphs had
values $0.6 < S(g) < 0.9$, and a significant number had
values $S(g) > 0.9$.  In contrast, the graphs in
Figure~\ref{fig:toynet} had values: $S(HSFnet) = 0.9791$,
$S(Random) = 0.8098$, $S(HOTnet) = 0.3952$, and
$S(PoorDesign) = 0.4536$. Again, from the perspective of
stochastic construction processes, low-$S$ values typical
of HOT constructions are ``very unlikely'' while high-$S$
values are much more ``likely'' to occur at random.

With this additional insight into the $s$-values associated
with different graphs, the relationship in the $\mbox{\em
Perf}(g)$ vs.~$s(g)$ plot of Figure \ref{fig:perf-llh} is
clearer. Specifically, high-performance networks resulting
from a careful design process {\it are vanishingly rare
from a conventional probabilistic graph point of view}. In
contrast, the likely outcome of random graph constructions
(even carefully handcrafted ones) are networks that have
extremely poor performance or lack the desired
functionality (e.g., providing connectivity) altogether.

\subsection{Degree Correlations}

Given an appropriate statistical ensemble of graphs, the
expectation of a random variable or random vector $X$ is
defined as
\begin{equation}
\langle X \rangle = \sum_{g \in G} X(g) P(g).
\end{equation}
For example, for $1 \leq i \leq n$, let $D_i$ be the random variable
denoting the degree of node $i$ for a graph $g\in G$ and let
$D = \{D_1, D_2, \dots, D_n\}$ be the random vector representing
the node degrees of $g$. Then the {\it degree distribution}
is given by
\[
P (k) \equiv P(\{g \in G: D_i (g)=k;i=1,2,\ldots,n\})
\]
and can be written in terms of an expectation of a random variable, namely
\[
P (k) = \frac{1}{n} \left\langle \sum_{i=1}^n \delta [ D_i - k ]
\right\rangle
\]
where
\bdis
\delta[D_i(g)-k]=\left\{\begin{array}{ll} 1
& \mbox{if node } i \mbox{ of graph } g \mbox{ has degree } k \\
0 & \mbox{ otherwise.} \end{array} \right. \edis

One previously studied topic has been the
correlations between the degrees of connected nodes. To
show that this notion has a direct relationship to the
$s(g)$ metric, we follow \cite[Section
4.6]{dorogovtsev-book2003} and define the degree
correlation between two adjacent vertices having respective
degree $k$ and $k'$ as follows.

\begin{definition} The {\it degree correlation} between two
neighbors having degrees $k$ and $k'$ is defined by
\beq
P (k,k')=\frac{1}{n^2} \left \langle \sum_{i,j=1}^{n}
\delta[d_i-k]a_{ij}\delta[d_j-k'] \right \rangle
\eeq
where the $a_{ij}$ are elements of the network node adjacency matrix
such that
\bdis
a_{ij}=\left\{\begin{array}{ll} 1 & \mbox{if nodes } i,j \mbox{ are
  connected} \\ 0 & \mbox{ otherwise} \end{array} \right.
\edis
and where the random variables $\delta[D_i-k]$ are as above.
\end{definition}
\noindent As an expectation of indicator-type random variables,
$P (k,k')$ can be interpreted as the probability that a
randomly chosen link connects nodes of degrees $k$ and
$k'$, therefore $P (k,k')$ is also called the
``degree-degree distribution'' for links.
Observe that for a given graph $g$ having degree sequence $D$,
\bqn
s(g)
&=&\sum_{(i,j) \in \mathcal{E}} d_i d_j \\
&=&\sum_{(i,j) \in \mathcal{E}} \sum_{k \in D}
 k \delta[d_i-k]  \sum_{k' \in D}  \delta[d_j-k'] k'  \\
&=& \sum_{(i,j) \in \mathcal{E}} \sum_{k \in D} \sum_{k' \in D}
k \delta[d_i-k] \delta[d_j-k'] k' \\
&=&\frac{1}{2} \sum_{k,k' \in D}  k k' \sum_{i,j=1}^{n}
\delta[d_i-k]a_{ij}\delta[d_j-k']    \\
\eqn

\noindent Thus, there is an inherent relationship between the structural
metric $s(g)$ and the degree-degree distribution, which we formalize
as follows.

\begin{proposition} \label{prop:s-ensemble}
\beq
\langle s \rangle =\frac{n^2}{2}\sum_{k,k'} k k' P (k,k').
\eeq
\end{proposition}

\noindent \textbf{Proof of Proposition~\ref{prop:s-ensemble}:}
For fixed degree sequence $D$,
\bqn
\langle s \rangle
&=&
\left\langle
\frac{1}{2} \sum_{k,k' \in D}  k k' \sum_{i,j=1}^{n}
\delta[d_i-k]a_{ij}\delta[d_j-k'] \right\rangle \\
&=&
\frac{1}{2} \sum_{k,k' \in D}  k k' \left\langle
\sum_{i,j=1}^{n}
\delta[d_i-k]a_{ij}\delta[d_j-k'] \right\rangle \\
&=&
\frac{n^2}{2} \sum_{k,k' \in D}  k k' P (k,k').
\eqn

This result shows that for an ensemble of graphs having
degree sequence $D$, the expectation of $s$ can be written
purely in terms of the degree correlation.
While other types of correlations
have been considered (e.g., the correlations associated
with clustering or loops in connectivity), degree
correlations of the above type are the most obviously
connected with the $s$-metric.

\subsection{Assortativity/Disassortativity of Networks}
\label{sec:assort}

Another ensemble-based notion of graph degree correlation
that has been studied is the measure $r(g)$ of {\it
assortativity} in networks as introduced by
Newman~\cite{Newman02}, who describes {\it assortative
mixing} ($r>0$) as {\it ``a preference for high-degree vertices
to attach to other high-degree vertices''} and {\it
disassortative mixing} ($r<0$) as the converse, where
{\it ``high-degree vertices attach to low-degree ones.''} Since
this is essentially what we have shown $s(g)$ measures, the
connection between $s(g)$ and assortativity $r(g)$ should
be and ultimately is very direct. As with all concepts in
the SF literature, assortativity is developed in the
context of an ensemble of graphs, but Newman provides a
sample estimate of assortativity of any given graph $g$.
Using our notation, Newman's formula~\cite[Eq.~4]{Newman02}
can be written as
\begin{equation}
r(g) = \frac{\left[  \sum_{(i,j)\in \mathcal{E}}d_i d_j
\right] - \left[ \sum_{i\in \mathcal{V}} \frac{1}{2} d_i^2
\right]^2/l}
 {\left[  \sum_{ i\in \mathcal{V}} \frac{1}{2}d_i^3 \right]
- \left[  \sum_{ i\in \mathcal{V}} \frac{1}{2} d_i^2
\right]^2/l}  , \label{assort_str}
\end{equation}
where $l$ is the number of links in the graph. Note that
the first term of the numerator of $r(g)$ is precisely
$s(g)$, and the other terms depend only on $D$ and not on
the specific graph $g \in G(D)$.  Thus $r(g)$ is linearly
related to $s(g)$.  However, when we compute $r(g)$ for the
graphs in Figure~\ref{fig:toynet} the values all are in the
interval $[-0.4815,-0.4283]$. Thus all are roughly equally
disassortative and $r(g)$ seems not to distinguish between
what we have viewed as extremely different graphs. The
assortativity interpretation appears to directly contradict
both what appears obvious from inspection of the graphs,
and the analysis based on $s(g)$. Recall that for $S(g) =
s(g)/s_{\max}$ the graphs in Figure~\ref{fig:toynet} had
$S(HSFnet) = 0.979$ and $S(HOTnet) = 0.395$, with
high-degree nodes in {\it HSFnet} attached to other
high-degree nodes and in {\it HOTnet} attached to
low-degree nodes.

The essential reason for this apparent conflict is that $-1
\le r(g) \le 1$ and $0<S(g) \le 1$ are normalized against a
different ``background set'' of graphs. For $S(g) =
s(g)/s_{\max}$ here, we have computed $s_{\max}$
constrained to simple, connected graphs, whereas $r(g)$
involves no such constraints.  The $r=0$ graph with the same
degree sequence as {\it HSFnet} and {\it HOTnet} would be
non-simple---having, for example, the highest degree ($d_1$) node
highly connected to itself (with multiple self-loops) and with
multiple parallel connections to the other high-degree nodes
(e.g.~multiple links to the $d_2$ node). The corresponding
$r=1$ graph would be both non-simple and disconnected---having
the highest degree ($d_1$) node essentially connected {\em
only} to itself. So {\it HSFnet} could be thought of as
assortative when compared with graphs in $G(D)$, but
dissassortative when compared with all graphs. To emphasize
this distinction, the description of {\it assortative
mixing} ($r>0$) could be augmented to ``high-degree
vertices attach to other high-degree vertices, including
self-loops.'' Since high variability, simple, connected
graphs will all typically have $r(g)<0$, this measure is
less useful than simply comparing raw $s(g)$ for this class
of graphs.  Thus conceptually, $r(g)$ and $s(g)$ have the
same aim, but with different and largely incomparable
normalizations, both of which are interesting.

We will now briefly sketch the technical details behind the
normalization of $r(g)$. The first term of the denominator
$\sum_{i \in \mathcal{V}} d_i^3/2l$ is equal to $s_{\max}$
for ``unconstrained'' graphs (i.e., those not restricted to
be simple or even connected; see
Appendix~\ref{sec:smax-appendix} for details), and the
normalization term in the denominator can be understood
accordingly as this $s_{\max}$.  The term $\left( \sum_{
i \in \mathcal{V}} d_i^2/2 \right)^2/l$ can be interpreted
as the ``center'' or zero assortativity case, again for
unconstrained graphs.
Thus, the perfectly assortative graph can be viewed as the
$s_{\max}$ graph (within a particular background set $G$),
and the assortativity of graphs is measured relative to the
$s_{\max}$ graph, with appropriate centering.

Newman's development of assortativity~\cite{Newman02}
is motivated by a definition that works both for an ensemble
of graphs and as a sample-based metric for individual graphs.
Accordingly, his definition depends on $Q(k,k')$, the
joint distribution of the {\em remaining degrees} of the
two vertices at either end of a randomly selected link
belonging to a graph in an ensemble.  That is, consider a
physical process by which a graph is selected from a
statistical ensemble and then a link is arbitrarily chosen
from that graph. The question of assortativity can then be
understood in terms of some (properly normalized)
statistical average between the degrees of the nodes at
either end of the link.  We defer the explicit connection
between the ensemble-based and sample-based notions of
assortativity and our structural metric $s(g)$ to
Appendix~\ref{sec:assort-appendix}.

\section{SF Graphs and the Internet Revisited}\label{sec:hot}

Given the definitions of $s(g)$, the various
self-similarity and high likelihood features of high-$s(g)$
graphs, as well as the extreme diversity of the set of
graphs $G(D)$ with scaling degree $D$, we look to
incorporate this understanding into a theory of SF graphs
that recovers both the spirit and existing results, while
making rigorous the notion of what it means for a graph to
be ``scale-free''. To do so, we first trace the exact
nature of previous misconceptions concerning the SF
Internet, introduce an updated definition of a scale-free
graph,  clarify what statements in the SF literature can be
recovered, and briefly outline the prospects for applying
properly defined SF models in view of alternative
theoretical frameworks such as HOT (Highly
Optimized/Organized Tolerance/Tradeoffs).
In this context, it is also important to understand the
popular appeal that the SF approach has had. One reason is
certainly its simplicity, and we will aim to preserve that
as much as possible as we aim to replace largely heuristic
and experimental results with ones more mathematical in nature.
The other is that it
relies heavily on methods from statistical physics, so much
so that replacing them with techniques that are shaped by
mathematics and engineering will require a fundamental
change in the way complex systems such as the Internet are
viewed and studied.

The logic of the existing SF theory and its central claims
regarding the Internet consists of the following steps:
\begin{enumerate}
\item The claim that measurements of the Internet's
router-level topology can be reasonably modeled with a
graph $g$ that has scaling degree sequence $D$.
\item The assertion, or definition, that a graph $g$ with
scaling degree sequence $D$ is a scale-free graph.
\item The claim that scale-free graphs have a host of
``emergent'' features, most notably the presence of several
highly connected nodes (i.e.~``hubs'') that are critical to
overall network connectivity and performance.
\item The conclusion that the Internet is therefore
scale-free, and its ``hubs,'' through which most traffic
must pass, are responsible for the ``robust yet fragile''
feature of failure tolerance and attack vulnerability.
\end{enumerate}
In the following, we revisit the steps of this logic
and illustrate that the conclusion in Step 4 is based on a
series of misconceptions and errors, ranging in scope from
taking highly ambiguous Internet measurements at face value to
applying an inherently inconsistent SF theory to an engineered
system like the Internet.

\subsection{Scaling Degree Sequences and the Internet}
\label{sec:measurements}

The Internet remains one of
the most popular and highly cited application areas where
power laws in network connectivity have ``emerged
spontaneously'', and the notion that this increasingly
important information infrastructure exhibits a signature
of self-organizing complex systems has generated
considerable motivation and enthusiasm for SF networks.
However, as we will show here, this basic observation is
highly questionable, and at worst is the simple result of
errors emanating from the misinterpretation of available
measurements and/or their naive and inappropriate
statistical analysis of the type critiqued in
Section~\ref{sec:ScalingAndHighVar}

To appreciate the problems inherent in the available data,
it is important to realize that Internet-related
connectivity measurements are notorious for their
ambiguities, inaccuracies, and incompleteness. This is due
in part to the multi-layered nature of the Internet
protocol stack (where each level defines its own
connectivity), and it also results from the efforts of
Internet Service Providers (ISPs) who intentionally obscure
their network structure in order to preserve what they
believe is a source of competitive advantage.
Consider as an example the router-level connectivity of the
Internet, which is intended to reflect (physical) one-hop
distances between routers/switches.  Although information about
this type of connectivity is typically inferred from
{\it traceroute} experiments which record successive
IP-hops along paths between selected network host computers
(see for example the Mercator \cite{GovTang00}, Skitter
\cite{Skitter}, and Rocketfuel \cite{Rocketfuel02} projects),
there remain a number of challenges when trying to reverse-engineer
a network's physical infrastructure from traceroute-based
measurements.
The first challenge is that IP connectivity is an abstraction
(at ``Layer 3'') that sits on top of physical connectivity
(at ``Layer 2''), so traceroute is unable to record directly
the network's physical structure, and its measurements are
highly ambiguous about the dependence between these two layers.
Such ambiguity in Internet connectivity persists even at higher
layers of the protocol stack, where connectivity becomes increasingly
virtual, but for different reasons (see for example
Section~\ref{sec:virtual-internet} below for a discussion of the
Internet's AS and Web graphs).

To illustrate how the somewhat subtle interactions among the different
layers of the Internet protocol stack can give the (false)
appearance of high connectivity at the IP-level, recall how
at the physical layer the use of Ethernet technology
near the network periphery or Asynchronous Transfer Mode (ATM) technology
in the network core can give the appearance of high IP-connectivity
since the physical topologies associated with these technologies
may not be seen by IP-based traceroute.  In such cases, machines
that are connected to the same Ethernet or ATM network may have the
illusion of direct connectivity from the perspective of IP, even
though they are separated by an entire network (potentially spanning
dozens of machines or hundreds of miles) at the physical level.  In an
entirely different fashion, the use of ``Layer 2.5 technologies''
such as Multiprotocol Label Switching (MPLS) tend to mask a
network's physical infrastructure and can give the illusion
of one-hop connectivity at Layer 3.  Note that in both cases,
it is the explicit and intended design of these technologies to
hide the physical network connectivity from IP.
Another practical problem when interpreting
traceroute data is to decide which IP addresses/interface cards
(and corresponding DNS names) refer to the same router, a process
known as  {\it alias resolution}~\cite{Spring-Alias04}.  While
one of the contributing factors to the high fidelity of the
current state-of-the-art Rocketfuel maps is the use of an improved
heuristic for performing alias resolution~\cite{Rocketfuel02},
further ambiguities remain, as pointed out for example
in~\cite{renata:imc2003}.
Yet another difficulty when dealing with traceroute-derived
measurements has been considered
in~\cite{LakhinaByers03,AchlioptasClauset03}
and concerns a potential bias whereby
IP-level connectivity is inferred more easily and accurately
the closer the routers are to the traceroute source(s).  Such bias
possibly results in incorrectly interpreting power law-type degree
distributions when the true underlying connectivity structure is
a regular graph (e.g., Erd\"{o}s-Reny\'{i}~\cite{ErdosRenyi59}).


\begin{figure*}[t]
  \begin{center}
  \includegraphics[width=.25\linewidth]{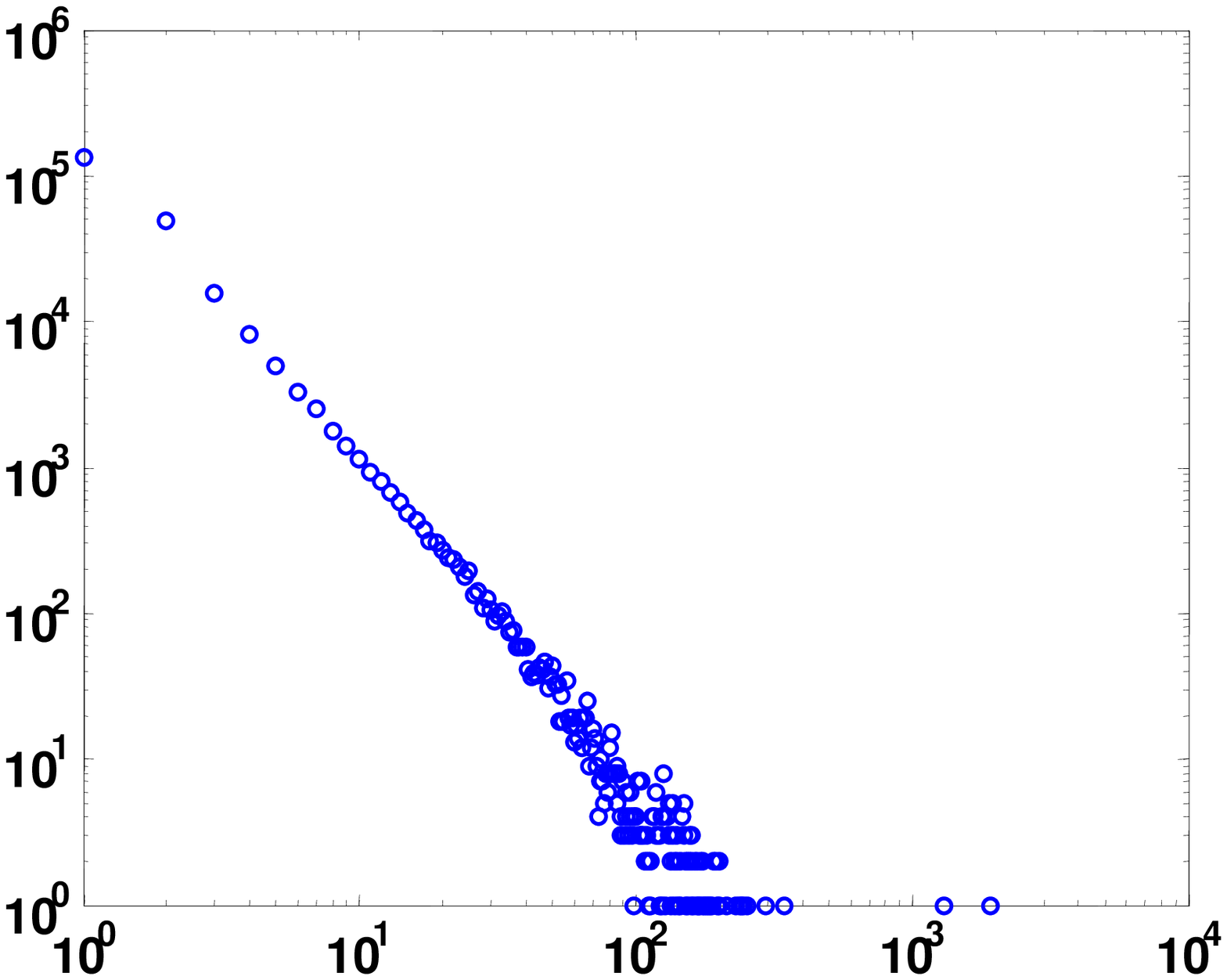}
  \hspace{-2mm}
  \includegraphics[width=.25\linewidth]{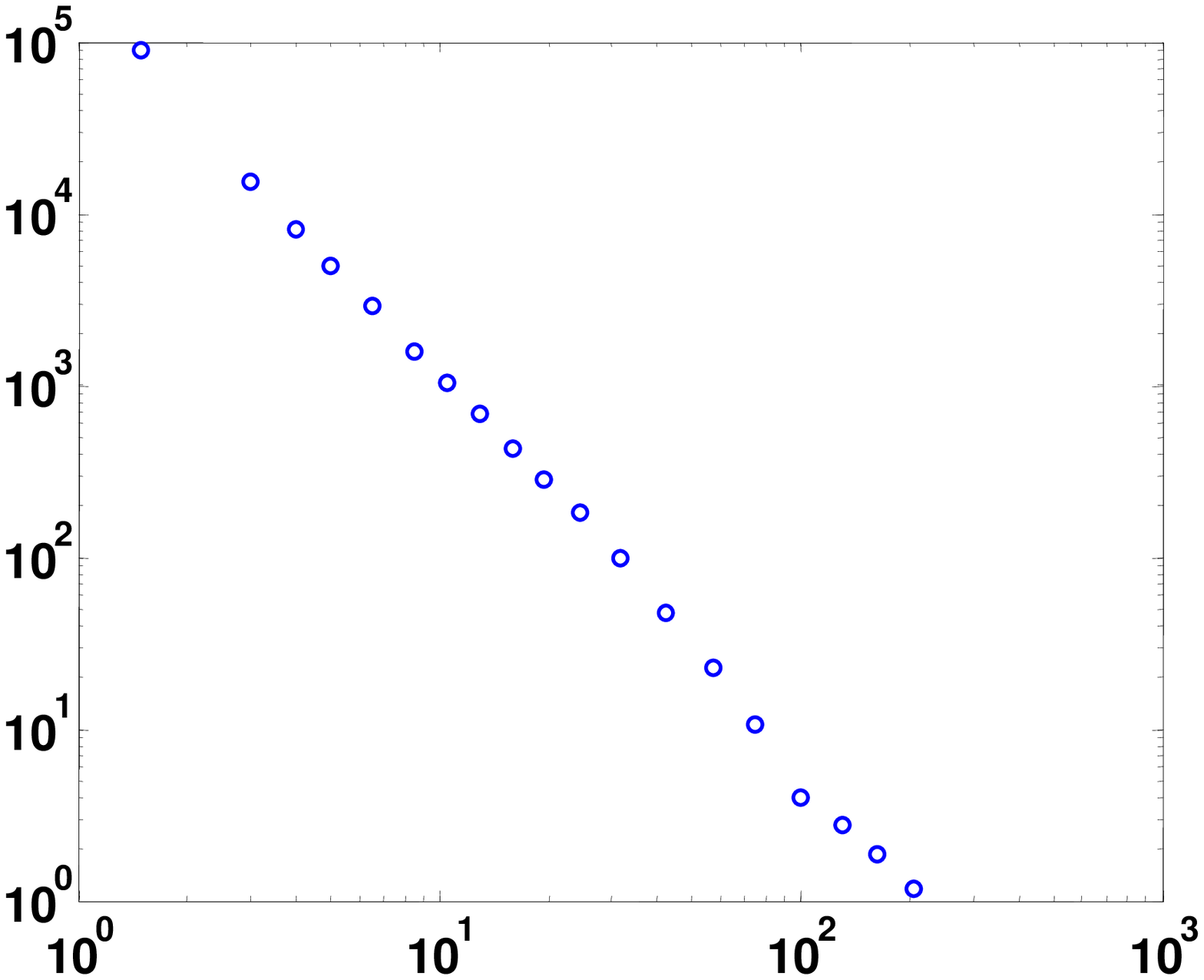}
  \hspace{-2mm}
  \includegraphics[width=.25\linewidth]{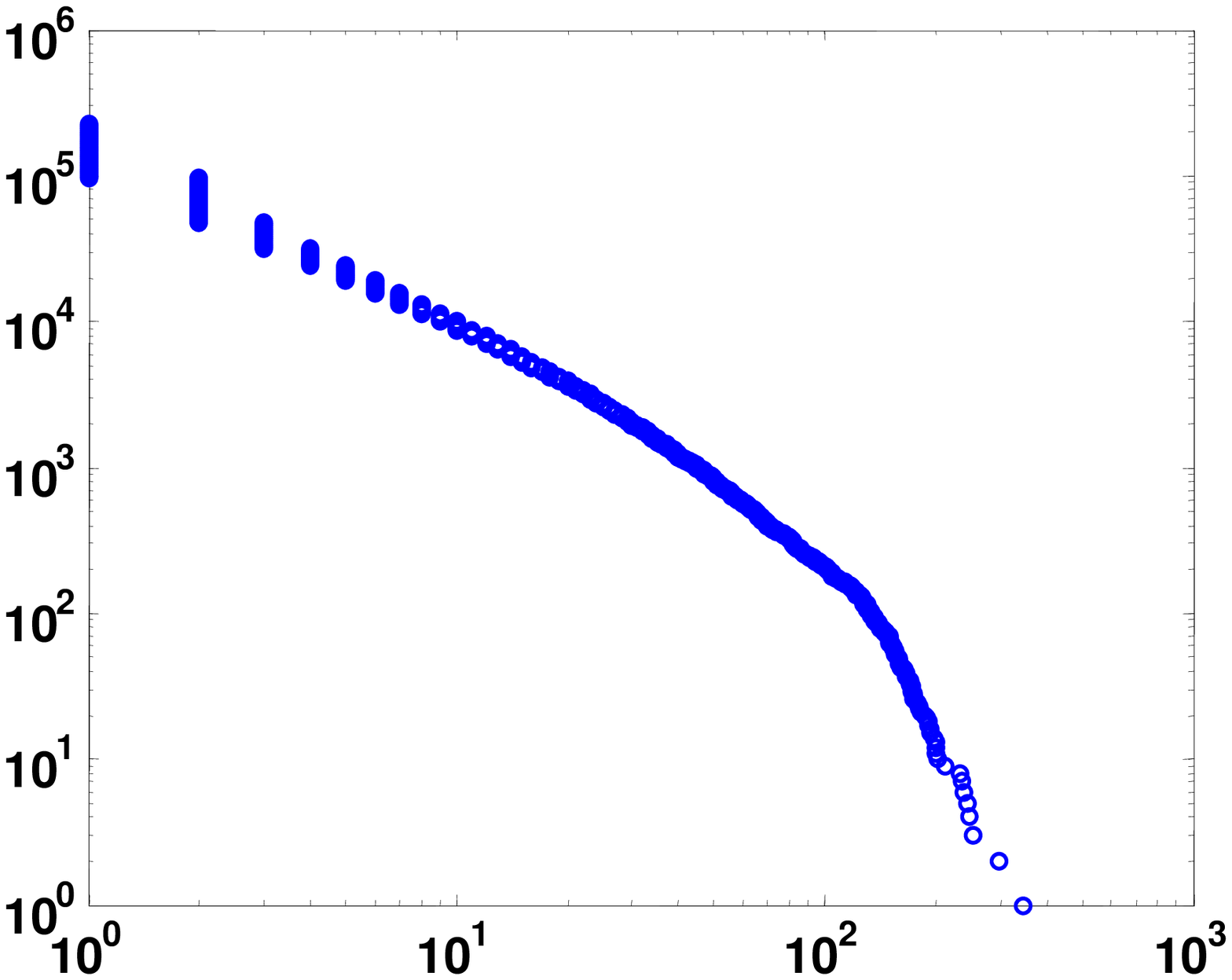}
  \hspace{-2mm}
  \includegraphics[width=.25\linewidth]{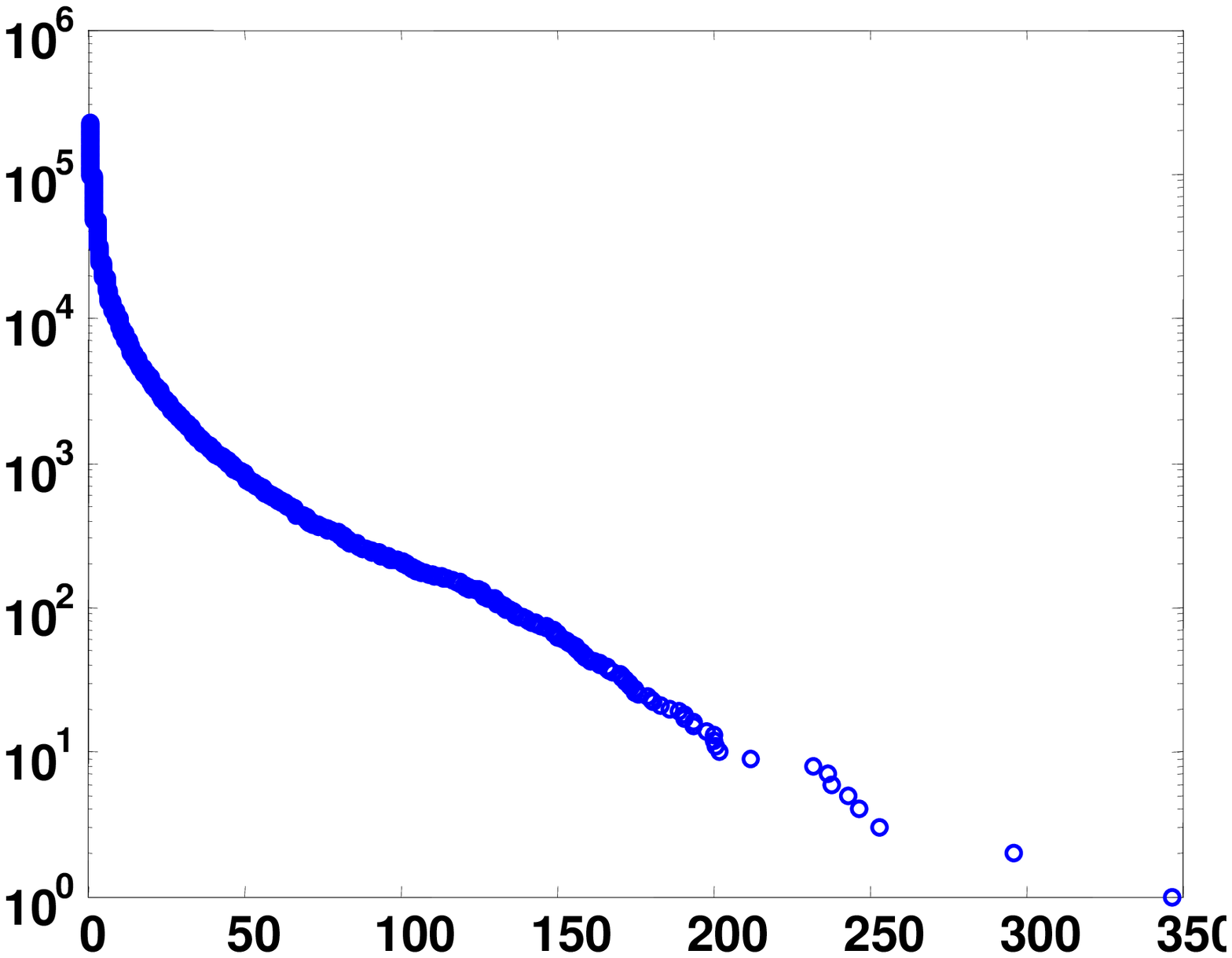}
  \vspace{-2mm}
  \caption{\footnotesize   \label{fig:mercator}
   {\sc Traceroute-derived router-level connectivity data from the
   Mercator project~\cite{GovTang00}.}
  \textbf{(a) Doubly logarithmic size-frequency plot:} Raw data.
  \textbf{(b) Doubly logarithmic size-frequency plot:} Binned data.
  \textbf{(c) Doubly-logarithmic size-rank plot:} Raw data with the
   2 extreme nodes (with connectivity $>$ 1,000) removed.
  \textbf{(d) Semi-logarithmic size-rank plot:} Raw data with the
   2 extreme nodes (with connectivity $>$ 1,000) removed.
  }
  \end{center}
\vspace{-4ex}
\end{figure*}


Ongoing research continues to reveal new idiosyncrasies of
traceroute-derived measurements and shows that their interpretation
or analysis requires great care and diligent mining of other
available data sources.  Although the challenges associated with
disambiguating the available measurements and identifying those
contributions that are relevant for the Internet's router-level
topology can be daunting, using these measurements at face value and
submitting them to commonly-used, black box-type statistical
analyses---as is common in the complex systems literature---is ill-advised
and bound to result in erroneous conclusions.  To illustrate,
Figure~\ref{fig:mercator}(a) shows the size-frequency plot for
the raw traceroute-derived router-level connectivity data obtained
by the Mercator project~\cite{GovTang00}, with
Figure~\ref{fig:mercator}(b) depicting a smoothed version of the
plot in (a), obtained by applying a straightforward binning operation
to the raw measurements, as is common practice in the physics literature.
In fact, Figures~\ref{fig:mercator}(a)--(b) are
commonly used in the SF literature (e.g., see~\cite{albert_barabasi2002})
as empirical evidence that the router-level topology of the Internet
exhibits power-law degree distributions.  However, in view of the
above-mentioned ambiguities of traceroute-derived measurements,
it is highly likely that the two extreme points with node degrees
above 1,000 are really instances where the high IP-level connectivity
is an illusion created by an underlying Layer 2 technology and
says nothing about the actual connectivity at the physical level.
When removing the two nodes in question and relying on the
statistically more robust size-rank plots in Figures~\ref{fig:mercator}
(c) and (d), we notice that neither the doubly-logarithmic nor
semi-logarithmic plots support the claim of a power law-type
node degree distribution for the Internet's router-level topology.
In fact, Figures~\ref{fig:mercator}(c) and (d) strongly suggest that the
actual router-level connectivity is more consistent with an
exponentially-fast decaying node degree distribution, in stark
contrast to what is typically claimed in the existing SF literature.

\subsection{(Re)Defining ``Scale-Free'' Graphs}
\label{sec:sf-def}

While it is unlikely that the Internet as a whole has
scaling degree sequences, it would not be in principle
technologically or economically infeasible to build a
network which did.  It would, however, be utterly
infeasible to build a large network with high-degree SF
hubs, or more generally one that had both high variability
in node degree and large $s(g)$. Thus in making precise the
definition of scale-free, there are essentially two
possibilities.  One is to define scale-free as simply
having a scaling degree sequence, from which no other
properties follow.  The other is to define scale-free more
narrowly in such a way that a rich set of properties are
implied. Given the strong set of self-similarity properties
of graphs $g$ having high $s(g)$, we propose the following
alternate definition of what it means for a graph to be
``scale-free''.

\begin{definition} \label{def:scalefree}
For graphs $g \in G(D)$ where $D$ is scaling, we measure the
{\it extent to which the graph $g$ is scale-free} by the metric
$s(g)$.
\end{definition}

\noindent This definition for ``scale-free graphs'' is
restricted here to simple, connected graphs having scaling
$D$, but $s(g)$ can obviously be computed for any graphs
having any degree sequence, and thus defining $s(g)$ as a
measure of ``scale-free'' might potentially be overly
narrow.  Nonetheless, in what follows, for degree sequences
$D$ that are scaling, we will informally call graphs $g\in
G(D)$ with low $s(g)$-values {\it ``scale-rich''}, and
those with high $s(g)$-values {\it ``scale-free.''} Being
structural in nature, this alternate definition has the
additional benefit of not depending on a stochastic model
underlying the set of graphs of interest.  It does not rely
on the statistical physics-inspired approach that focuses
on random ensembles and their most likely elements and is
inherent, for example, in the original Barab\'{a}si-Albert
construction procedure.

Our proposed definition for scale-free graphs requires that
for a graph $g$ to be called scale-free, the degree
sequence $D$ of $g$ must be scaling (or, more generally,
highly variable) {\it and} self-similar in the sense that
$s(g)$ must be large. Furthermore, $s(g)$ gives a
quantitative measure of the extent to which a scaling
degree graph is scale-free.  In addition, this definition
captures an explicit and obvious relationship between
graphs that are ``scale-free'' and have a ``hub-like core''
of highly connected centrally-located nodes. More
importantly, in view of Step 2 of the above-mentioned
logic, the claim that scale-free networks have ``SF hubs''
is true with scale-free defined as scaling degree sequence
{\it and} high $s(g)$, but false if scale-free were simply
to mean scaling degree sequence, as is commonly assumed in
the existing SF literature.

With a concise measure $s(g)$ and its connections with rich
self-similarity/self-dissimilar properties and likelihood,
we can look back and understand how both the appeal and
failure of the SF literature is merely a symptom of much
broader and deeper disconnects within complex networks
research. First, while there are many possible equivalent
definitions of scale-free, all nontrivial ones would seem
to involve combining scaling degree with self-similarity or
high likelihood and appear to be equivalent. Thus defined,
models that generate scale-free graphs are easily
constructed and are therefore not our main focus here.
Indeed, because of the strong invariance properties of
scaling distributions alone, it is easy to create limitless
varieties of randomizing generative models that can
``grow'' graphs with scaling degree $D$. Preferential
growth is perhaps the oldest of such models
\cite{Yule25,LurDel43,Simon55}, so it is no surprise that
it resurfaces prominently in the recent SF literature. No
matter how scaling is generated however, the high
likelihood and rewiring invariance of high-$s(g)$ graphs
make it further easy---literally highly likely--to insure
that these scaling graphs are also scale-free.

Thus secondly, the equivalence between ``high $s$'' and
``highly likely'' makes it possible to define scale-free as
the likely or generic outcome of a great variety of random
growth models. In fact, that ``low $s$'' or ``scale-rich''
graphs are vanishingly unlikely to occur at random explains
why the SF literature has not only ignored their existence
and missed their relevance but also conflated scale-free
with scaling.  Finally, since scaling and high $s$ are both
so easily and robustly generated, requiring only few simple
statistical properties, countless variations and
embellishments of scale-free models have been proposed,
with appealing but ultimately irrelevant details and
discussions of emergence, self-organization, hierarchy,
modularity, etc.  However, their additional self-similarity
properties, though still largely unexplored, have made the
resulting scale-free networks intuitively appealing,
particularly to those who continue to associate complexity
with self-similarity.

The practical implication is that while our proposed
definition of what it means for a graph to be ``scale-free''
recovers many claims in the existing SF literature, some
aspects cannot be salvaged.  As an alternate approach,
we could accept a definition of scale-free that is
equivalent to scaling, as is implicit in most of the SF
literature.  However, then the notion of ``scale-free'' is
essentially trivial, and almost all claims in the existing
literature about SF graphs are false, not just the ones
specific to the Internet. We argue that a much better
alternative is a definition of scale-free, as we propose,
that implies the existence of ``hubs'' and other emergent
properties, but is more restrictive than scaling. Our
proposed alternative, that scale-free is a special case of
scaling that further requires high $s(g)$, not only provides
a quantitative measure about the extent to which a graph is
scale-free, but also already offers abundant emergent
properties, with the potential for a rigorous and rich
theory.

In summary, notwithstanding the errors in the
interpretation and analysis of available network
measurement data, even if the Internet's router-level graph
were to exhibit a power law-type node degree distribution,
we have shown here and in other papers (e.g.,
see~\cite{sigcomm04,imc04}) that the final conclusion in
Step 4 is necessarily wrong for today's Internet. No matter
how scale-free is defined, the existing SF claims about the
Internet's router-level topology cannot be salvaged.  
Adopting our definitions, the router topology at least for
some parts of the Internet could in principle have high
variability and may even be roughly scaling , but it is
certainly nowhere scale-free. It is in fact necessarily
extremely ``scale-rich'' in a sense we have made rigorous
and quantifiable, although the diversity of scale-rich
graphs means that much more must be said to describe which
scale-rich graphs are relevant to the Internet. A main lesson
learned from this exercise has been that in the context of
such complex and highly engineered systems as the Internet,
it is largely impossible to understand any nontrivial
network properties while ignoring all domain-specific
details such as protocol stacks, technological or economic
constraints, and user demand and heterogeneity, as is
typical in SF treatments of complex networks.

\subsection{Towards a Rigorous Theory of SF Graphs}

Having proposed the quantity $s(g)$ as a
structural measures of the extent to which a given graph is
``scale-free'',  we can now review the characteristics of
scale-free graphs listed in Section \ref{sec:ConvSF} and
use our results to clarify what is true if scale-free is
taken to mean scaling degree sequence and large $s(g)$:
\begin{enumerate}
    \item SF networks have scaling (power law) degree sequence
    (follows by definition).
    \item SF networks are the likely outcome of various
    random growth processes (follows from the equivalence
    of $s(g)$ with a natural measure of graph likelihood).
    \item SF networks have a hub-like core structure (follows
    directly from the definition of $s(g)$ and the betweeness
    properties of high-degree hubs).
    \item SF networks are generic in the sense of being preserved
    by random degree-preserving rewiring (follows from the
    characterization of rewiring invariance of self-similarity).
    \item SF networks are universal in the sense of not depending
    on domain-specific details (follows from the structural nature
    of $s(g)$).
    \item SF networks are self-similar (is now partially clarified
    in that high $s(g)$ trees are preserved under both appropriately
    defined link trimming and coarse graining, as well as restriction
    to small motifs).
\end{enumerate}
Many of these results are proven only for special cases and
have only numerical evidence for general graphs, and thus
can undoubtedly be improved upon by proving them in greater
generality. However in most important ways the proposed
definition is entirely consistent with the spirit of
``scale-free'' as it appears in the literature, as noted by
its close relationship to previously defined notions of
betweeness, assortativity, degree correlation, and so on.
Since a high $s(g)$-value requires high-degree nodes to
connect to other high-degree nodes, there is an explicit
and obvious equivalence between graphs that are scale-free
(i.e., have high $s(g)$-value) and have a ``hub-like core''
of highly connected nodes. Thus the statement ``scale-free
networks have hub-like cores''---while incorrect under the
commonly-used original and vague definition (i.e., meaning
scaling degree sequence)---is now true almost by definition
and captures succinctly the confusion caused by some of the
sensational claims that appeared in the scale-free
literature.  In particular, the consequences for network
vulnerability in terms of the ``Achilles' heel'' and a zero
epidemic threshold follow immediately.

When normalized against a proper background set, our
proposed $s(g)$-metric provides insight into the diversity
of networks having the same degree sequence.  On the one
hand, graphs having $s(g) \approx s_{\max}$ are scale-free
and self-similar in the sense that they appear to exhibit
strong invariance properties across different scales, where
appropriately defined coarse-graining operations (including
link trimming) give rise to the different scales or levels
of resolution. On the other hand, graphs having $s(g) <<
s_{\max}$ are scale-rich and self-dissimilar; that is, they
display different structure at different levels of
resolution.  While for scale-free graphs, degree-preserving
random rewiring does not significantly alter their
structural properties, even a modest amount of rewiring
destroys the structure of scale-rich graphs.  Thus, we
suggest that a heuristic test as to whether or not a given
graph is scale-free is to explore the impact of
degree-preserving random rewiring. Recent work on the
Internet~\cite{sigcomm04} and metabolic networks
\cite{tanaka2005} as well as on more general complex
networks \cite{wolpert:macready:2000} demonstrates that
many important large-scale complex systems are scale-rich
and display significant self-dissimilarity, suggesting that
their structure is far from scale-free and the opposite of
self-similar.

\subsection{SF Models and the Internet?}
\label{sec:virtual-internet}

For the Internet, we have shown that no
matter how scale-free is defined, the existing SF claims
about the ``robust, yet fragile'' nature of these systems
(particularly any claims of an ``Achilles' heel'' type of
vulnerability) are wrong no matter how scale-free is
defined. By tracing through the reasoning behind these SF
claims, we have identified the source of this error
in the application of SF models to domains like engineering
(or biology)
where design, evolution, functionality, and constraints are
all key ingredients that simply cannot be ignored. In
particular, by assuming that scale-free is defined as
scaling (or, more generally, highly variable) plus high
$s(g)$, and further using $s(g)$ as a quantitative measure
of how scale-free a graph is, the failure of SF models to
correctly and usefully apply in an Internet-related context
has been limited to errors due to ignoring domain-specific
details, rather than to far more serious and general
mathematical errors about the properties of SF graphs
themselves. In fact, with our definition, there is the
potential for a rich and interesting theory of SF graphs,
looking for relevant and useful application domains.

One place where SF graphs may be appropriate and practically
useful in the study of the Internet is at the higher levels
of network abstraction, where interconnectivity is increasingly
unconstrained by physical limitations.
That is, while the lowest layers of the Internet protocol
stack involving the physical infrastructure such as routers
and fiber-optic cables have hard technological and economic
constraints, each higher layer defines its own unique
connectivity, and the corresponding network topologies
become by design increasingly more virtual and
unconstrained.
For example, in contrast to routers and
physical links, the connectivity structure defined by the
documents (nodes) and hyperlinks (connections) in the
World Wide Web (WWW) is designed to be
essentially completely unconstrained.  While we have seen
that it is utterly implausible that SF models can capture
the essential features of the router-level connectivity in
today's Internet, it seems conceivable that they could
represent {\it virtual} graphs associated with the Internet such as,
hypothetically, the WWW or other types of overlay networks.

However, even in the case of more virtual-type graphs
associated with the Internet, a cautionary note about the
applicability of SF models is needed.  For example,
consider the Internet at the level of autonomous systems,
where an {\it autonomous system (AS)} is a subnetwork or
domain that is under its own administrative control.  In an
AS graph representation of the Internet, each node
corresponds to an AS and a link between two nodes indicates
the presence of a ``peering relationship'' between the two
ASes---a mutual willingness to carry or exchange traffic.
Thus, a single ``node'' in an AS graph (e.g., AS 1239 is the
Sprintlink network) represents potentially hundreds or thousands of
routers as well as their interconnections. Although most
large ASes have several connections (peering points) to
other ASes, the use of this representation means that one
is collapsing possibly hundreds of different physical
(i.e., router-level) connections into a single logical link
between two ASes. In this sense, the AS graph is
expressively not a representation of any physical aspect of
the Internet, but defines a virtual graph representing
business (i.e., peering) relationships among network
providers (i.e., ASes). Significant attention has been
directed toward discovering the structural aspects of AS
connectivity as represented by AS graphs and inferred from
BGP-based measurements (where the {\it Border Gateway
Protocol} or {\it BGP} is the de facto standard inter-AS
routing protocol deployed in today's Internet
\cite{stewart2000,RouteViews}) and speculating on what
these features imply about the large-scale properties of
the Internet.  However, the networking significance of
these AS graphs is very limited since AS connectivity alone
says little about how the actual traffic traverses the
different ASes.  For this purpose, the relevant information
is encoded in the link type (i.e., peering agreement such
as peer-to-peer or provider-customer relationship) and in
the types of routing policies used by the individual ASes
to enforce agreed-upon business arrangements between two or
more parties.

In addition, due to the infeasibility of measuring AS connectivity directly,
the measurements that form the basis for inferring AS-level maps
consist of BGP routing table snapshots collected, for example, by
the University of Oregon Route Views Project \cite{RouteViews}.
To illustrate the degree of ambiguity in
the inferred AS connectivity data, note for example that due to the
way BGP routing works, snapshots of BGP routing tables taken at a few
vantage points on the Internet over time are unlikely to uncover and
capture all existing connections between ASs.  Indeed, \cite{ChangEtAl02}
suggests that AS graphs inferred from the Route Views data typically
miss between 20-50\% or even more of the existing AS connections.
This is an example of the general problem of {\it vantage point}
mentioned in~\cite{vern-imc04}, whereby the location(s) of exactly
where the measurements are performed can significantly skew the
interpretation of the measurements, often in quite non-intuitive ways.
Other problems that are of concern in this context have to do with
ambiguities that can arise when inferring the type of peering relationships
between two ASes or, more importantly, with the dynamic nature of
AS-level connectivity, whereby new ASes can join and existing ASes can
leave, merge, or split at any time.

This dynamic aspect is even more relevant in the context of
the Web graph, another virtual graph associated with the
Internet that is expressively not a representation of any
physical aspect of the Internet structure but where nodes
and links represent pages and hyperlinks of the WWW,
respectively.  Thus in addition to the deficiencies mentioned
in the context of router-level Internet measurements, the
topologies that are more virtual and ``overlay'' the Internet's
physical topology exhibit an aspect of dynamic changes that
is largely absent on the physical level.  This questions
the appropriateness and relevance of a careful analysis or
modeling of commonly considered static counterparts of
these virtual topologies that are typically obtained by
accumulating the connectivity information contained in a
number of different snapshots taken over some time period
into a single graph.

When combined, the virtual nature of AS or Web graphs and their
lack of critical networking-specific information make them
awkward objects for studying the ``robust yet fragile''
nature of the Internet in the spirit of the ``Achilles'
heel'' argument~\cite{AlbJeongBar00} or largely inappropriate
structures for investigating the spread of viruses on the Internet
as in~\cite{BergerBorgs05}.
For example, what does it mean to ``attack and disable'' a node
such as Sprintlink (AS 1239) in a representation of business
relationships between network providers?  Physical attacks
at this level are largely meaningless.  On the other hand,
the economic and regulatory environment for ISPs remains
treacherous, so questions about the robustness (or lack
thereof) of the Internet at the AS-level to this type of
disruption seem appropriate. And even if one could make
sense of physically ``attacking and disabling'' nodes or
links in the AS graph, any rigorous investigation of its
``robust yet fragile'' nature would have to at least
account for the key mechanisms by which BGP detects and
reacts to connectivity disruptions at the AS level.  In
fact, as in the case of the Internet's router-level
connectivity, claims of scale-free structure exhibited by
inferred AS graphs fail to capture the most essential
``robust yet fragile'' features of the Internet because
they ignore any significant networking-specific information
encoded in these graphs beyond connectivity. Again, the
actual fragilities are not to physical attacks on AS nodes
but to AS-related components ``failing on,'' particularly
via BGP-related software or hardware components working
improperly or being misconfigured, or via malicious
exploitation or hijacking of BGP itself.

\subsection{The Contrasting Role of Randomness}

To put our SF findings in a broader context, we briefly
review an alternate approach to the use of randomness for
understanding system complexity that implicitly underpins
our approach in a way similar to how statistical physics
underpins the SF literature. Specifically, the notions of
{\it Highly Optimized Tolerance (HOT)}~\cite{CDo99}
or {\it Heuristically Organized Tradeoffs}~\cite{fkp}
has been recently introduced as a conceptual framework for
capturing the highly organized, optimized, and ``robust yet
fragile'' structure of complex highly evolved systems
\cite{CD}.
Introduced in the spirit of canonical models from
statistical physics---such as percolation lattices,
cellular automata, and spin glasses---HOT is an attempt to
use simple models that capture some essence of the role of
design or evolution in creating highly structured
configurations, power laws, self-dissimilarity,
scale-richness, etc.  The emphasis in the HOT view is on
``organized complexity'', which contrasts sharply with the
view of ``emergent complexity'' that is preferred within
physics and the SF community.
The HOT perspective is motivated by biology and technology,
and HOT models typically involve optimizing functional
objectives of the system as a whole, subject to constraints
on their components, usually with an explicit source of
uncertainty against which solutions must be tolerant, or
robust. The explicit focus on function, constraints,
optimization, and organization sharply distinguish HOT from
SF approaches.  Both consider robustness and fragility but
reach opposite and incompatible conclusions.

A toy model of the HOT approach to modeling the
router-level Internet was already discussed earlier.  The
underlying idea is that consideration of the economic and
technological factors constraining design by Internet
Service Providers (ISPs) gives strong incentives to
minimize the number and length of deployed links by
aggregating and multiplexing traffic at all levels of the
network hierarchy, from the periphery to the core.  In
order to efficiently provide high throughput to users,
router technology and link costs thus {\em necessitate}
that by and large link capacities increase and router
degrees decrease from the network's periphery to its more
aggregated core.  Thus, the toy model {\it HOTnet} in
Figure~\ref{fig:toynet}(d), like the real router-level
Internet, has a mesh of uniformly high-speed low
connectivity routers in its core, with greater variability
in connectivity at its periphery. While a more detailed
discussion of these factors and additional examples is
available from \cite{sigcomm04,PNAS05}, the result is that this
work has explained where within the Internet's router-level
topology the high degree nodes might be and why they might
be there, as well as where they can't possibly be.

The HOT network that results is not just different than the
SF network but completely opposite, and this can be seen
not only in terms relevant to the Internet application
domain, such as the performance measure (\ref{eq:perf}),
robustness to router and link losses, and the link costs, but
in the criteria considered within the SF literature
itself.  Specifically, SF models are generated directly
from ensembles and random processes, and have generic
microscopic features that are preserved under random
rewiring.  HOT models have highly structured, rare
configurations which are destroyed by random rewiring,
unless that is made a specific design objective.  SF models
are universal in ignoring domain details, whereas HOT is
only universal in the sense that it formulates everything
in terms of robust, constrained optimization, but with
highly domain-specific performance objectives and
constraints.

One theme of the HOT framework has been that engineering
design or biological evolution easily generates scaling in
a variety of toy models once functional performance,
component constraints, and robustness tradeoffs are
considered. Both SF and HOT models of the Internet yield
power laws, but once again in opposite ways and with
opposite consequences. HOT emphasizes the importance of
high variability over power laws per se, and provides a
much deeper connection between variability or scaling
exponents and domain-specific constraints and features. For
example, the HOT Internet model considered here shows that
if high variability occurs in router degree it can be
explained by high variability in end user bandwidth
together with constraints on router technology and link
costs.  Thus HOT provides a predictive model regarding how
different external demands or future evolution of
technology could change network statistics. The SF models
are intrinsically incapable of providing such predictive
capability in any application domain.  The resulting
striking differences between these two modeling approaches
and their predictions are merely symptomatic of a much
broader gap between the popular physics perspective on
complex networks versus that of mathematics and
engineering, created by a profoundly different perspective
on the nature and causes of high variability in real world
data. For example, essentially the same kind of contrast
holds for HOT and SOC models~\cite{CD}, where SOC is yet
another theoretical framework with specious claims about
the Internet~\cite{Sole-Valverde,BakBook}.

In contrast to the SF approach, the HOT models described
above as well as their constraints and performance measures
do not require any assumptions, implicit or explicit, that
they were drawn directly from some random ensemble.
Tradeoffs in the real Internet and biology can be explained
without insisting on any underlying random models.  Sources
of randomness are incorporated naturally where uncertainty
needs to be managed or accounted for, say for the case of
the router-level Internet, in a stochastic model of user
bandwidth demands and geographic locations of users,
routers, and links, followed by a heuristic or optimal
design. This can produce either an ensemble of network
designs, or a single robust design, depending on the design
objective, but all results remain highly constrained and
are characterized by low $s(g)$ and high $\mbox{\em
Perf}(g)$. This is typical in engineering theories, where
random models are common but not required, and where
uncertainty can be modeled with random ensembles or
worst-case over sets.  In all cases, uncertainty models are
mixed with additional hard constraints, say on component
technology.

In the SF literature, on the other hand, random graph
models and statistical physics-inspired approaches to
networks are so deep-rooted that an underlying ensemble is
taken for granted. Indeed, in the SF literature the phrase
``not random'' typically does not refer to a deterministic
process but means random processes having some non-uniform
or high variability distribution, such as scaling.
Furthermore, random processes are used to directly generate
SF network graphs rather than model uncertainty in the
environment, leading in this case to high $s(g)$ and low
$\mbox{\em Perf}(g)$ graphs. This particular view of
randomness also blurs the important distinction between
what is unlikely and what is impossible.  That is, what is
unlikely to occur in a random ensemble (e.g. a low $s(g)$
graph) is treated as impossible, while what is truly
impossible (e.g. an Internet with SF hubs) from an
engineering perspective is viewed as likely from an
ensemble point of view. Similarly, the relation between
high variability, scaling, and scale-free is murky in the
SF literature. These distinctions may all be irrelevant for
some scientific questions, but they are crucial in the study
of engineering and biology and also essential for
mathematical rigor.

\section{A HOT vs. SF View \\of Biological Networks}\label{sec:bio}

This section describes how a roughly parallel SF vs HOT
story exists in metabolic networks, which is another
application area that has been very popular in the SF and
broader ``complex networks'' literature \cite{BarOltBio04}
and is also discussed in more detain in \cite{fox-keller}.
Recent progress has clarified many features of the global
architecture of biological metabolic networks \cite{Marie2}.  
We argue here that they have highly organized and optimized
tolerances and tradeoffs for functional requirements of
flexibility, efficiency, robustness, and evolvability, with
constraints on conservation of energy, redox, and many
small moieties.  These are all canonical examples of HOT
features, and are largely ignored in the SF literature. One
consequence of this HOT architecture is a highly structured
modularity that is self-dissimilar and scale-rich, as in
the Internet example.  All aspects of metabolism have
extremes in homogeneity and heterogeneity, and low and high
variability, including power laws, in both metabolite and
reaction degree distributions.  We will briefly review the 
results in \cite{tanaka2005} which illustrate these features
using the well-understood stoichiometry of metabolic
networks in bacteria.

One difficulty in comparing SF and HOT approaches is that
there is no sense in which ordinary (not bipartite) graphs
can be used to meaningfully describe metabolism, as we will
make clear below.  Thus one would need to generalize our
definition of scale-free to bipartite graphs just to
precisely define what it would mean for metabolism to be
scale-free. While this is an interesting direction not
pursued here, the SF literature has many fewer claims about
bipartite graphs, and they are typically studied by
projecting them down onto one set of vertices. What is
clear is that no definition can possibly salvage the claim
that metabolism is scale-free, and we will pursue this
aspect in a more general way.  The rewiring-preserved
features of scale-free networks would certainly be a
central feature of any claim that metabolism is scale-free.
This is also a feature of the two other most prominent
``emergent complexity'' models of biological networks,
edge-of-chaos (EOC) and self-organized criticality (SOC).
While EOC adds boolean logic and SOC adds cellular automata
to the graphs of network connectivity, both are by
definition unchanged by random degree-preserving rewiring.
In fact, they are preserved under much less restrictive
rewiring processes. Thus while there are currently no SF,
EOC, or SOC models that apply directly to metabolic
networks, we can clearly eliminate them a priori as
candidate theories by showing that all important
biochemical features of real metabolic networks are
completely disrupted by rewirings that are far more
restrictive than what is by definition allowable in SF,
EOC, or SOC models.

\subsection{Graph Representation}

Cellular metabolism is described by a series of chemical
reactions that convert nutrients to essential components
and energy within the cell, subject to conservation
constraints of atoms, energy and small moieties. The
simplest model of metabolic networks is a stoichiometry
matrix, or s-matrix for short, with rows of metabolites and
columns of reactions. For example, for the set of chemical
reactions
\begin{equation}
\left\{
\begin{array}{*{20}c}
{S_1 + {\rm NADH} \to S_2 + {\rm NAD}}, \\
{S_2 + {\rm ATP} \leftrightarrow S_3 + {\rm ADP}}, \\
{S_4 + {\rm ATP} \to S_5 + {\rm ADP}}, \\
\end{array}
\right. \label{eq_react}
\end{equation}
we can write the associated {\it stoichiometry matrix}, or
s-matrix, as
\begin{equation}
\begin{array}{*{20}c}
{} & {{\rm{Reactions}}} \\
{} & {\begin{array}{*{20}c}
{R_1 } & {R_2 } & {R_3 } \\
\end{array}} \\
{\begin{array}{*{20}c} {{\rm{Substrates}}} \hfill &
{\left\{ {\begin{array}{*{20}c}
{S_1 } \\
{S_2 } \\
{S_3 } \\
{S_4 } \\
{S_5 } \\
\end{array}} \right.} \hfill \\
{{\rm{Carriers}}} \hfill & {\left\{ {\begin{array}{*{20}c}
{\rm ATP} \\
{\rm ADP} \\
{\rm NADH} \\
{\rm NAD} \\
\end{array}} \right.} \hfill \\
\end{array}} & {\left[ {\begin{array}{*{20}c}
{ - 1} & 0 & 0 \\
1 & { - 1} & 0 \\
0 & 1 & 0 \\
0 & 0 & { - 1} \\
0 & 0 & 1 \\
0 & { - 1} & { - 1} \\
0 & 1 & 1 \\
{ - 1} & 0 & 0 \\
1 & 0 & 0 \\
\end{array}} \right]}\\
\end{array}
\label{eq_stoich}
\end{equation}
with the metabolites in rows and reactions in columns. This
is the simplest model of metabolism and is defined
unambiguously except for permutations of rows and columns,
and thus makes an attractive basis for contrasting
different approaches to complex networks
\cite{BarOltBio04,dorogovtsev-book2003,JeongTombor00,hier2}.

Reactions in the entire network are generally grouped into
standard functional modules, such as catabolism, amino acid
biosynthesis, nucleotide biosynthesis, lipid biosynthesis
and vitamin biosynthesis. Metabolites are categorized
largely into carrier and non-carrier substrates as in rows
of (\ref{eq_stoich}). Carrier metabolites correspond to
conserved quantities, are activated in catabolism, and act
as carriers to transfer energy by phosphate groups,
hydrogen/redox, amino groups, acetyl groups, and one carbon
units throughout all modules. As a result, they appear in
many reactions. Non-carrier substrates are categorized
further into precursor and other (than precursor and
carrier) metabolites. The ~12 precursor metabolites are
outputs of catabolism and are the starting points for
biosynthesis, and together with carriers make up the
``knot'' of the ``bow-tie'' structure \cite{Marie2} of the
metabolism. The other metabolites occur primarily in
separate reaction modules.


\begin{figure*}
\begin{center}
 \includegraphics[width=.75\linewidth]{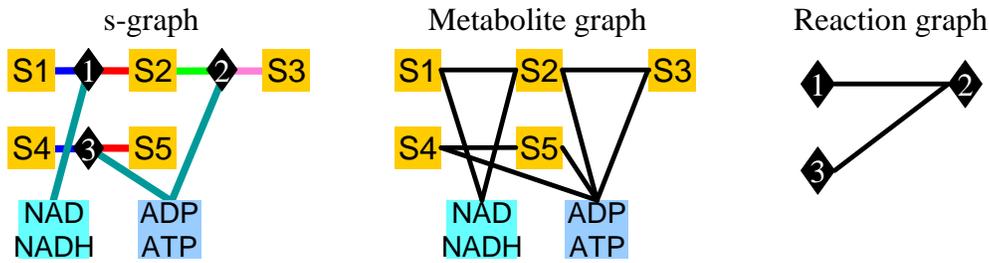}
 \caption{
   Graph representations of enzymatically catalyzed
   reactions from equation (\ref{eq_react}) having the s-matrix
   in equation (\ref{eq_stoich}).
   An s-graph consists of reaction nodes
   (black diamonds), non-carrier metabolite nodes (orange
   squares), and carrier metabolite nodes (light blue squares).
   Red and blue edges correspond to
   positive and negative elements in the stoichiometry matrix,
   respectively, for
   irreversible reactions, and pink and green ones correspond to
   positive and negative elements, respectively, for
   reversible reactions. All the
   information in the s-matrix appears schematically in the
   s-graph. Carriers which always occur in pairs (ATP/ADP,
   NAD/NADH etc.) are grouped for simplification.
   Corresponding metabolite graphs containing only metabolite
   nodes and reaction graph containing only reaction nodes lose
   important biochemical information. Note that all metabolites
   and reactions are ``close'' in the metabolite and reaction
   graphs, simply because they share common carriers,
   but could be arbitrarily far apart in any real biochemical
   sense.  For example, reactions 2 and 3 could be in amino acid
   and lipid biosynthesis, respectively, and thus
   would be far apart biochemically
   and in the s-graph. }
 \label{fig:s-graph}
\end{center}
\vspace{-4ex}
\end{figure*}

\begin{figure*}
\begin{center}
  \includegraphics[width=.95\linewidth]{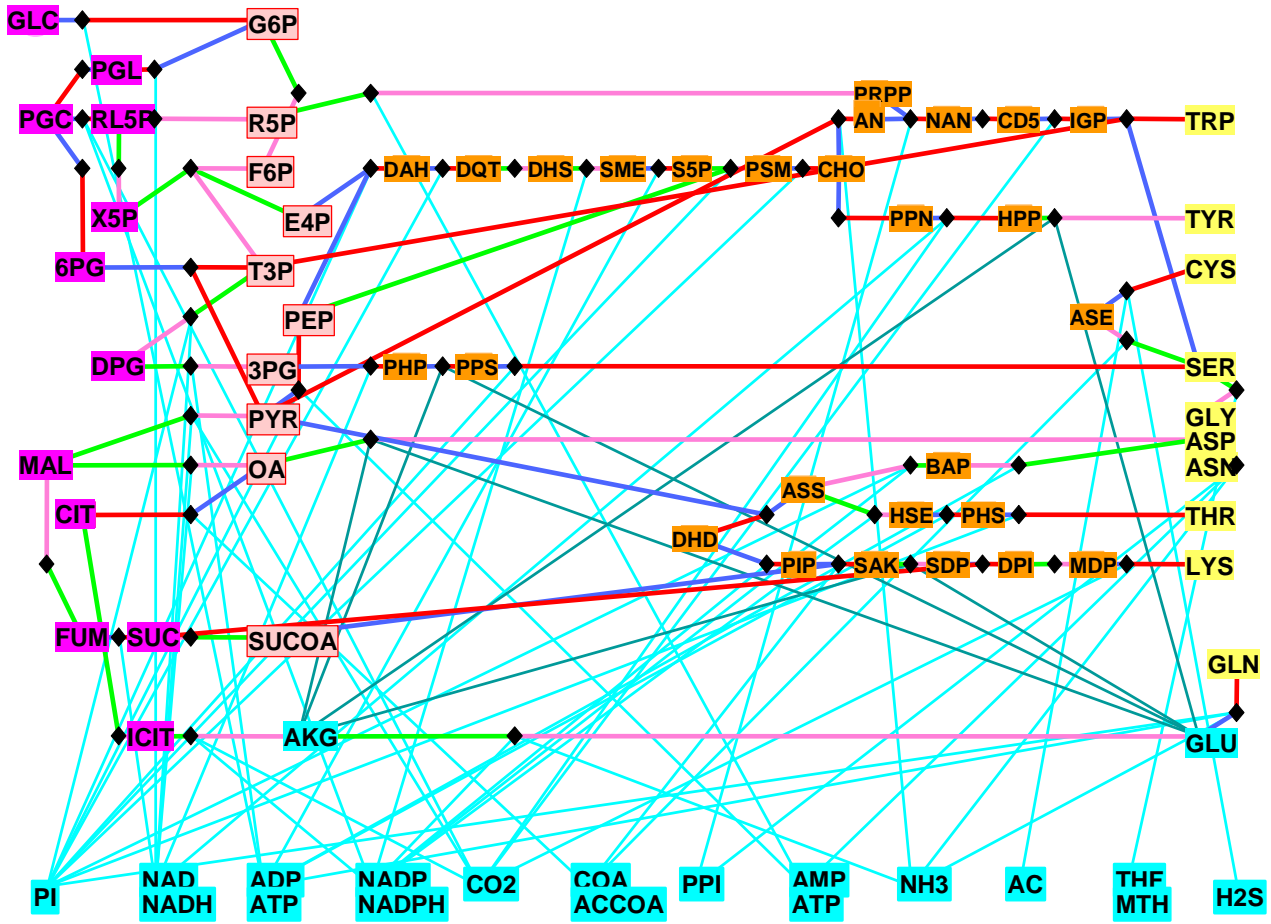}
  \caption{ An s-graph for part of the catabolism and
    amino acid biosynthesis module
    of {\it H.Pylori}. The conventions are the same as
    those in Figure~\ref{fig:s-graph}. This illustrates that
    long biosynthetic pathways build complex building blocks
    (in yellow on the right) from precursors (in orange on the
    left) in a series of simple reactions (in the middle),
    using shared common carriers (at the bottom). Each
    biosynthetic module has a qualitatively similar structure.}
  \label{fig:org_diag}
\end{center}
\end{figure*}

The information conveyed in the s-matrix can be represented
in a color-coded bipartite graph, called an {\it s-graph}
\cite{tanaka2005} (Figure~\ref{fig:s-graph}), where
both reactions and metabolites are represented as distinct
nodes and membership relationships of metabolites to
reactions are represented by links. With the color-coding
of links indicating the reversibility of reactions and the
sign of elements in the s-matrix, all the biochemical
information contained in the s-matrix is accurately
reflected in the s-graph. One of the most important
features of s-graphs of this type is the differentiation
between carrier (e.g.~ATP) and non-carrier metabolites that
help to clarify biochemically meaningful pathways. An
s-graph for a part of amino acid biosynthesis module of
{\it H.~Pylori} is shown in Figure~\ref{fig:org_diag}. The
objective of each functional module is to make output
metabolites from input metabolites through successive
reactions. The enzymes of core metabolism are highly
efficient and specialized, and thus necessarily have few
metabolites and involve simple reactions
\cite{tanaka2005}. As a result, long pathways are
required biochemically to build complex building blocks
from simpler building blocks within a function module. Long
pathways are evident in the s-graph in
Figure~\ref{fig:org_diag}.


\begin{figure*}
\begin{center}
  \includegraphics[width=.85\linewidth]{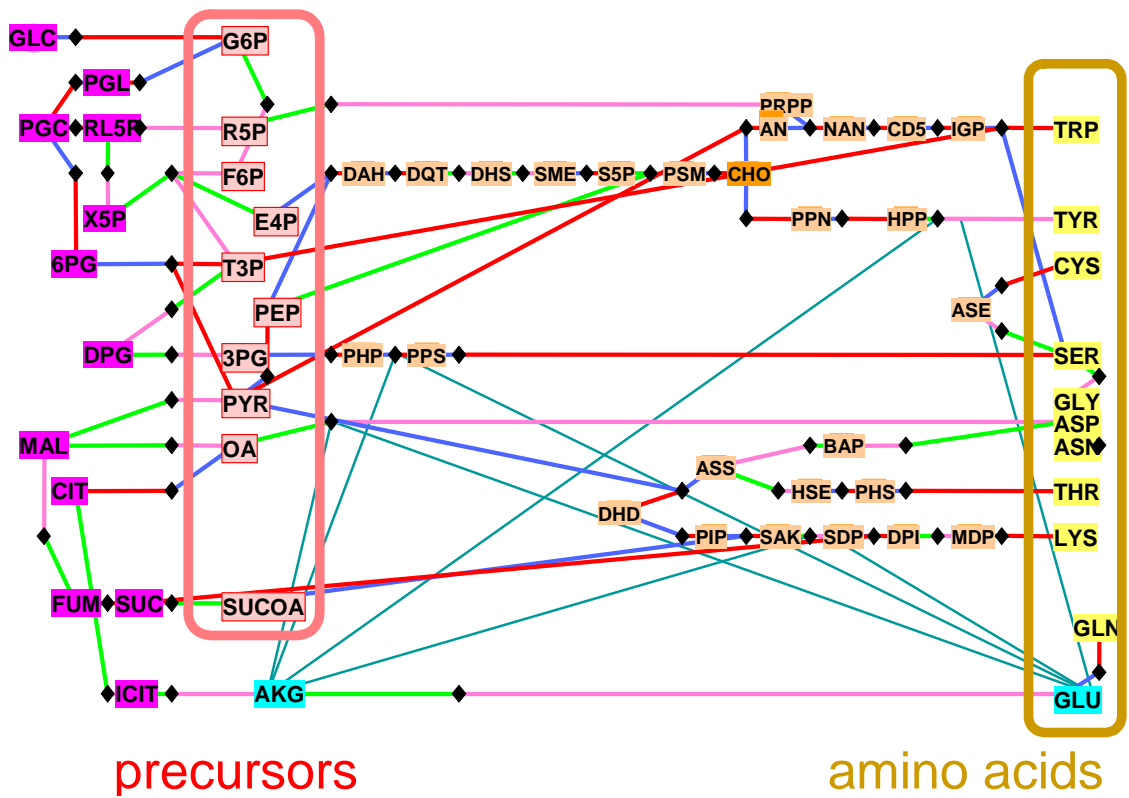}
  \vspace{-8pt}
  \caption{The s-graph in Figure~\ref{fig:org_diag}
    with carriers deleted to highlight the long assembly
    pathways. Note that there are no
    high degree ``hub'' nodes responsible for the global
    connectivity of this reduced s-graph. AKG and GLU are
    carriers for amino groups, and this role has been left
    in.}
  \label{fig:noCC}
\end{center}
\vspace{-8pt}
\end{figure*}



\begin{figure*}
  \begin{center}
   \includegraphics[width=0.90\linewidth]{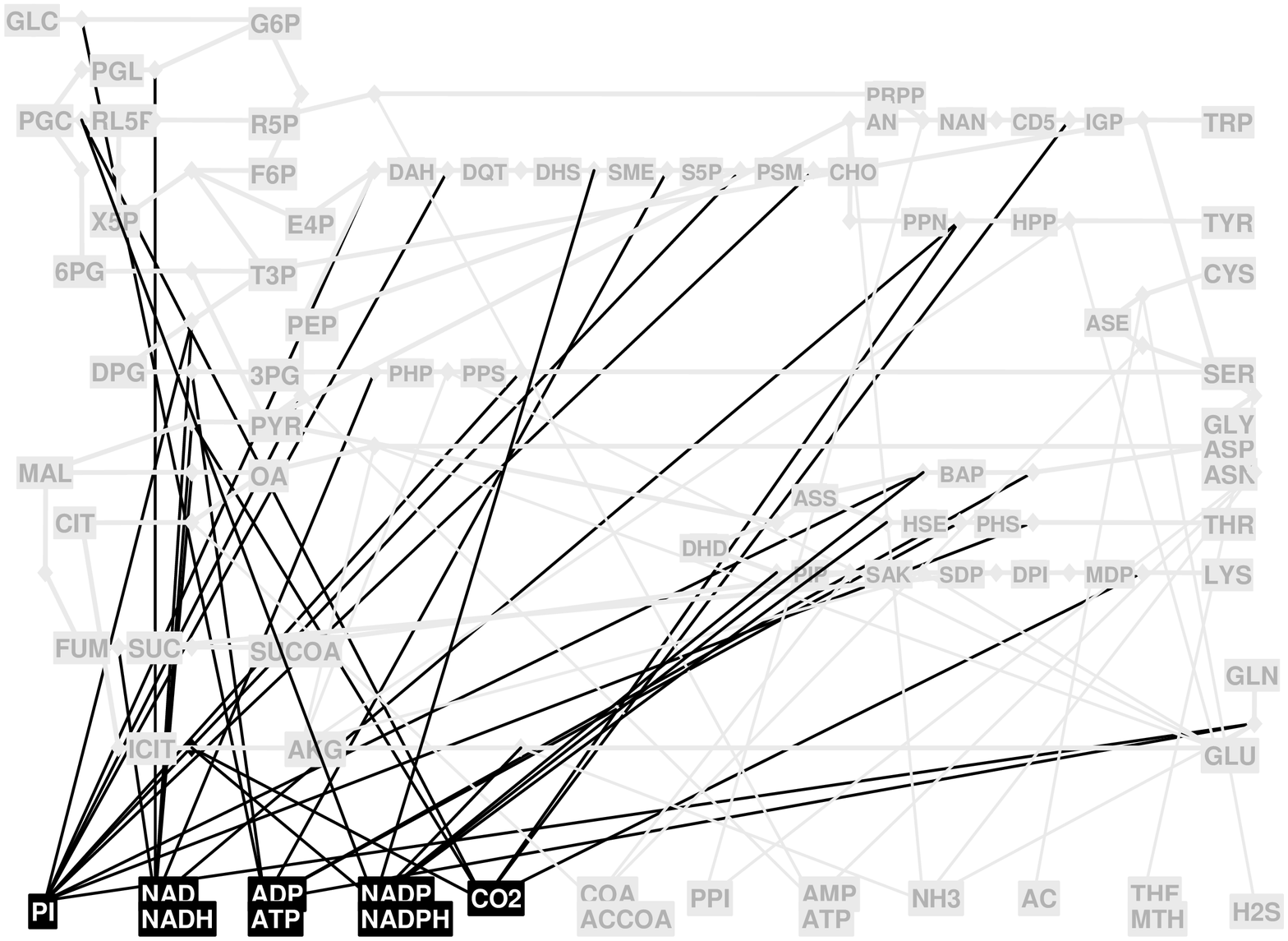}
    \caption{High connectivity metabolites in the s-graph in
      Figure~\ref{fig:org_diag} are these carriers
      which are not directly involved in the pathways. Because
      carriers are shared throughout metabolism, they are
      entirely responsible for the presence of high variability
      in metabolite degree, and thus the presence of scaling
      in metabolism.}
    \label{fig:CCpart}
  \end{center}
  \vspace{-4ex}
\end{figure*}


Simpler representations of the information in the s-graph
are possible, but only at a cost of losing significant
biochemical information. A metabolite graph in which nodes
represent only metabolites and are ``connected'' when they
are involved in the same reaction, or a reaction graph in
which nodes represent only reactions and are ``connected''
when they contain common metabolites (both shown in
Figure~\ref{fig:s-graph}) destroys much of the rich
structure and biochemical meaning when compared to the
s-graph.  (The metabolite graphs are sometimes further
reduced by deleting carriers.) Nonetheless, many recent
studies have emphasized the connectivity features of these
graphs, and reports of power laws in some of the degree
distribution have been cited as claims that (1) metabolic
networks are also scale-free \cite{BarOltBio04} and (2) the
presence of highly connected hubs and self-similar
modularity capture much of the essential details about
"robust yet fragile" feature of metabolism \cite{hier2}.
Here, highly connected nodes are carriers, which are shared
throughout metabolism.

We will first clarify why working with any of these simple
graphs of metabolism, rather than the full s-graph,
destroys their biochemical meaning and leads to a variety
of errors. Consider again the simple example of Equation
(\ref{eq_react}) and its corresponding s-matrix
(\ref{eq_stoich}). Here, assume that reactions $R_1$ and
$R_2$ are part of the pathways of a functional module, say
amino acid biosynthesis, and reaction $R_3$ is in another
module, say lipid biosynthesis. Then the metabolite and
reaction graphs both show that substrates $S_3$ and $S_4$,
as well as reactions $R_2$ and $R_3$ are ``close'' simply
because they share ATP/ADP. However, since they are in
different functional modules, they are not close in any
biologically meaningful sense. (Similarly, two functionally
different and geographically distant appliances are not
``close'' in any meaningful sense simply because both
happen to be connected to the US power grid.) Attempts to
characterize network diameter are meaningless in such
simplified metabolite graphs because they fail to extract
biochemically meaningful pathways. Additional work using
structural information of metabolites with carbon atomic
traces \cite{Arita04} has clarified that the average path
lengths between all pairs of metabolites in {\it E.~coli}
is much longer than has been suggested by approaches that
consider only simplistic connectivity in metabolite graphs.
``Achilles' heel'' statements \cite{AlbJeongBar00} for
metabolic networks are particularly misleading.
Eliminating, say, ATP from a cell is indeed lethal but the
explanation for this must involve its biochemical role, not
its graph connectivity.  Indeed, the ``Achilles' heel''
arguments suggest that removal of the highly connected
carrier ``hub'' nodes would fragment the graph, but
Figure~\ref{fig:noCC} shows that removing the carriers from
biologically meaningful s-graph in
Figure~\ref{fig:org_diag} still yields a connected network
with long pathways between the remaining metabolites. If
anything, this reduced representation highlights many of
the more important structural feature of metabolism, and
most visualizations of large metabolic networks use a
similar reduction. Attempts to ``fix'' this problem by a
priori eliminating the carriers from metabolite graphs
results in graphs with low variability in node degree and
thus are not even scaling, let alone scale-free. Thus the
failure of the SF graph methods to explain in any way the
features of metabolism is even more serious than for the
Internet.

\subsection{Scale-rich metabolic networks}

Recent work \cite{tanaka2005} has clearly shown the
origin of high variability in metabolic networks by
consideration of both their constraints and functional
requirements, together with biochemically meaningful
modular decomposition of metabolites and reactions shown
above. Since maintaining a large genome and making a
variety of enzymes is costly, the total number of reactions
in metabolism must be kept relatively small while providing
robustness of the cell against sudden changes, often due to
environmental fluctuations, in either required amount of
products or in available nutrients. In real metabolic
networks, scaling only arises in the degree distribution of
total metabolites.  The reaction node degree distribution
shows low variability because of the specialized enzymes
which allow only a few metabolites in each reaction. High
variability in metabolite node degrees is a result of the
mixture of a few high degree shared carriers with many
other low degree metabolites unique to each function
module, with the precursors providing intermediate degrees.
Thus, the entire network is extremely scale-rich, in the
sense that it consists of widely different scales and is
thus fundamentally self-dissimilar.

\begin{figure*}
  \begin{center}
   \includegraphics[width=0.5\linewidth]{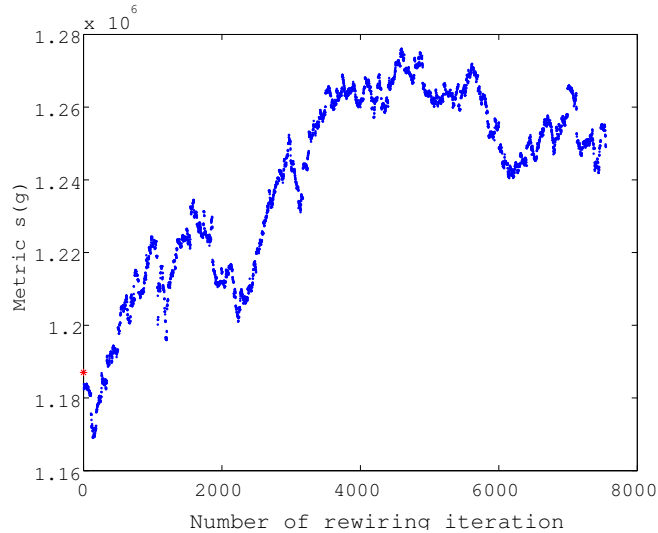}
    \caption{Metric $s(g)$ for real metabolite graph for
      {\it H.~Pylori} ($\ast$) and those for graphs obtained by
      degree-preserving random rewiring.}
    \label{fig:Pllh}
  \end{center}
  \vspace{-4ex}
\end{figure*}

Scale-richness of metabolic networks has been evaluated
quantitatively in \cite{tanaka-doyle:2004_2} by
degree-preserving rewiring of real stoichiometry matrices,
which severely alters their structural properties.
Preserving only the metabolite degrees gives much higher
variability in reaction node degree distribution than is
possible using simple enzymes, and rewiring also destroys
conservation of redox and moieties. The same kind of
degree-preserving rewiring on a simple HOT model \cite{FOSBE}, 
proposed with the essential feature of metabolism, such as simple
reactions, shared carriers, and long pathways, has
reinforced these conclusions but in a more analytical
framework \cite{tanaka-doyle:2004_2}. Even (biologically
meaningless) metabolite graphs have a low $S(g)$ value, and
are thus scale-rich, not scale-free. The simple reactions
of metabolism require that the high-degree carriers are
more highly connected to low-degree metabolites than to
each other, as is shown in the metabolite graph in
Figure~\ref{fig:s-graph}, yielding a relatively low $s(g)$
value. Figure~\ref{fig:Pllh} shows the $s(g)$ values for
the {\it H.~Pylori} metabolite graph compared with those
for graphs obtained by random degree-preserving rewiring of
that graph.


Even the most restrictive possible rewiring destroys the
structure of metabolism, showing that no SF, SOC, or EOC
models are possible, even in principle.  Suppose we freeze
the role of carriers in each reaction, and then allow only
rewiring of the remaining metabolites.  This would be
equivalent to Figure~\ref{fig:noCC} with the carrier for
amino group roles of AKG and GLU also frozen.  What then
remains is nearly a tree, and thus the rewiring counts from
Figure~\ref{fig:rewire} are approximately correct.  Note
that half of all rewirings disconnect a tree, and monte
carlo numerical experiments of successive rewirings
produces a jumble of futile cycles and short, dead end
pathways\cite{tanaka-doyle:2004_2}.  The long assembly
lines of real metabolism are extremely rare configurations
and highly scale-rich, and are vanishingly unlikely to
arise by any random ensemble model such as in SF, SOC, or
EOC theories.

Real metabolic networks are scale-rich in every conceivable
interpretation, and cannot be scale-free in any sense
consistent with the either the definitions in this paper or
with the spirit of the SF literature. In contrast to any
approach that treats metabolic networks as generic, a
biological perspective requires that the organization of
metabolic networks be discussed with emphasis on the
functional requirements of conversion of nutrients to
products with flexibility, efficiency, robustness, and
evolvability under the constraints on enzyme costs and
conservation of energy, redox, and many small moieties
\cite{Marie1,Stelling}. This structure is a natural
consequence of a highly optimized and structured tradeoff
(HOT) ``bow-tie'' structure, which facilitates great
robustness and efficiency but is also a source of
vulnerability, but primarily to hijacking and fail-on of
components \cite{Marie2,FOSBE,Stelling}. A power law in
metabolite degree is simply the natural null hypothesis
with any structure that exhibits high variability by shared
carriers, and thus is by itself not suggestive of any
further particular mechanism.

Another prominent example of biologic networks claimed to be SF
\cite{Jeong-Mason,Yook-Oltvai} is protein-protein interaction (PPI) networks.
This claim has lead to conclude that identifying high-degree "hub" proteins reveals 
important features of PPI networks.
However, recent analysis \cite{PPI} evaluating the claims that
PPI node degree sequences follow a power law, a necessary condition for
networks to be SF, shows that the node degree sequences of some 
published refined PPI networks do not have power laws when analyzed 
correctly using the cumulative plots as discussed in Section 2.1.
Thus these PPI networks are not SF networks.
It is in principle possible that the data studied in \cite{PPI} is
misleading because of the small size of the network and
potential experimental errors, and that real PPI networks
might have some features attributed to SF networks.
At this time we only can draw conclusions about (noisy)
subgraphs of the true PPI network since the data sets are
incomplete and presumably contain errors. If it is true
that appropriately sampled subraphs of a SF graph is SF as
was claimed in \cite{Yook-Oltvai}, they possess a power law
node degree sequence. That these subgraphs exhibit
exponential node degree sequences suggests that the entire
network is not SF. Since essentially all the claims that
biological networks are SF are based on ambiguous
frequency-degree analysis, this analysis must be redone to
determine the correct form of the degree sequences. 
Analysis in \cite{PPI} has provided clear examples that ambiguous plots of
frequency-degree could lead to erroneous conclusion on the
existence and parametrization of power law relationships.

As we have shown above, cell metabolism 
plausibly can have power laws for
some data sets, but have none of the other features
attributed to SF networks. Metabolic networks
have been shown to be scale-rich (SR), in the sense
that they are far from self-similar \cite{tanaka2005} despite some
power laws in certain node degree sequence. Their power law
node degree sequence is a result of the mixture of
exponential distributions in each functional module, with
carriers playing a crucial role. In principle, PPI networks
could have this SR structure as well, since their
subnetworks have exponential degree sequence, and perhaps
power laws could emerge at higher levels of organization.
This will be revealed only when a more complete network is
elucidated. Still, the most important point is not whether
the node degree sequence follows a power law, but whether
the variability of the node degree sequences is high or low, 
and the biological protocols that necessitate
this high or low variability. These issues will be explored
in future publications.

\section{Conclusions}\label{sec:concl}

The set $G(D)$ of graphs $g$ with fixed scaling degree $D$
is extremely diverse. However, most graphs in $G(D)$ are,
using our definition, scale-free and have high $s$-values.
This implies that these scale-free graphs are not diverse
and actually share a wide range of ``emergent'' features,
many of which are often viewed as both intriguing and
surprising, such as hub-like cores, high likelihood under a
variety of random generation mechanisms, preservation under
random rewiring, robustness to random failure but fragility
to attack, and various kinds of self-similarity. These
features have made scale-free networks overwhelmingly
compelling to many complex systems researchers and have
understandably given scale-free findings tremendous popular
appeal \cite{BarabasiAlbert99,
barabasi-internet,AlbJeongBar00,OttinoNature,BarabasiBook,BallBook}.
This paper has confirmed that these emergent features are
plausibly consistent with our definition, and we have
proven several connections, but much remains heuristic and
experimental.  Hopefully, more research will complete what
is potentially a rich graph-theoretic treatment of
scale-free networks.

Essentially all of the extreme diversity in $G(D)$ is in
its fringes that are occupied by the rare scale-rich small
$s$ graphs.  These graphs have little or nothing in common
with each other or with scale-free graphs beyond their
degree sequence so, unfortunately, $s$ is a nearly
meaningless measure for scale-rich graphs. We have shown
that those technological and biological networks which have functional
requirements and component constraints tend to be
scale-rich, and HOT is a theoretical framework aimed at
explaining in simplified terms the features of these
networks. In this context, scale-free networks serve at
best as plausible null hypotheses that typically collapse
quickly under scrutiny with real data and are easily
refuted by applying varying amounts of domain knowledge. 

At the same time, scale-free networks may still be relevant
when applied to social or virtual networks where
technological, economic, or other constraints play perhaps
a lesser or no role whatsoever. Indeed, a richer and more
complete and rigorous theory could potentially help
researchers working in such areas. For example, as
discussed in Section \ref{sec:rewire}, exploring the impact
of degree-preserving random rewiring of components can be
used as a simple preliminary litmus test for whether or not
a SF model might be appropriate. It takes little domain
expertise to see that randomly rewiring the internal
connections of, say, the microchips or transistors in a
laptop computer or the organs in a human body will utterly
destroy their function, and thus that SF models are
unlikely to be informative. On the other hand, one can
think of some technological (e.g.~wireless ad-hoc networks)
and many social networks where robustness to some kinds of
random rewiring is an explicitly desirable objective, and
thus SF graphs are not so obviously inapplicable. For
example, it might be instructive to apply this litmus test
to an AS graph that reflects AS connectivity only as
compared to the same graph that also provides information
about the type of peering relationships and the nature of
routing policies in place.

This paper shows that scale-free networks have the
potential for an interesting and rich theory, with most
questions, particularly regarding graphs that are not
trees, still largely open. Perhaps a final message of this
paper is that to develop a coherent theory for scale-free
networks will require adhering to more rigorous
mathematical and statistical standards than has been
typical to date.

\section*{Acknowledgments}

The authors are indebted to several colleagues for ongoing
conversations and valuable feedback, particularly David Aldous, Jean
Carlson, Steven Low, Chris Magee, Matt Roughan, Stanislav Shalunov.
This work was supported
by Boeing, AFOSR URI 49620-01-1-0365 ``Architectures for Secure and
Robust Distributed Infrastructures'', the Army Institute for
Collaborative Biotechnologies, NSF Award: CCF-0326635 ``ITR COLLAB:
Theory and Software Infrastructure for a Scalable Systems Biology,''
AFOSR Award: FA9550-05-1-0032 ``Bio Inspired Networks,'' and Caltech's
Lee Center for Advanced Networking.
Parts of this work were done at the Institute of Pure and Applied
Mathematics (IPAM) at UCLA as part of the 2002 annual program on
``Large-scale communication networks.''

\appendix

\section{Constructing an $s_{\max}$-graph}
\label{sec:smax-appendix}

As defined previously, the $s_{\max}$ graph is the element $g$ in
some background set $G$ whose connectivity maximizes the quantity
$s(g)= \sum_{(i,j)\in \mathcal{E}} d_i d_j$, where $d_i$ is the degree
of vertex $i \in \mathcal{V}$, $\mathcal{E}$ is the set of links
that define $g$, and $D = \{ d_1, d_2, \dots d_n \}$ is the
corresponding degree sequence.  Recall that since $D$ is ordered
according to $d_1 \geq d_2 \geq \dots \geq d_n$, there will usually
be many different graphs with vertices satisfying $D$.  The purpose of
this Appendix is to describe how to construct such an element for
different background sets, as well as to discuss the importance of
choosing the ``right'' background set.

\subsection{Among ``Unconstrained'' Graphs}

As a first case, consider the set of graphs having degree sequence $D$,
with only the requirement that $\sum_{i=1}^n d_i$ be {\it even}.  That
is, we do not require that these graphs be simple (i.e., they can have
self-loops or multiple links between vertices) or that they even be
connected, and we accordingly call this set of graphs ``unconstrained''.
Constructing the $s_{\max}$ element among these graphs can be achieved
trivially, by applying the following two-phase process.
First, for each vertex $i$:
if $d_i$ is even, then attach $d_i / 2$ self-loops;
if $d_i$ is odd, then attach $(d_i -1)/2$ self-loops, leaving one
available ``stub''.
Second, for all remaining vertices with ``stubs'', connect them in
pairs according to decreasing values of $d_i$.
Obviously, the resulting graph is not unique as the $s_{\max}$ element
(indeed, two vertices with the same degree could replace their
self-loops with connections among one another).  Nonetheless, this
construction does maximize $s(g)$, and in the case when $d_i$ is even
for all $i \in \mathcal{V}$, one achieves an $s_{\max}$ graph with
$s(g) = \sum_{i=1}^n (d_i/2)\cdot d_i^2$.
As discussed in Section~\ref{sec:assort}, against this background of
unconstrained graphs, the $s_{\max}$ graph is the perfectly
assortative (e.g., $r(g) =1$) graph.
In the case when some $d_i$ are odd, then the $s_{\max}$ graph will
have a value of $s(g)$ that is somewhat less and will depend on the
specific degree sequence.  Thus, the value
$\sum_{i=1}^n (d_i/2)\cdot d_i^2$ represents an idealized upper bound
for the value of $s_{\max}$ among unconstrained graphs, but it can
only be realized in the case when all vertex degrees are even.

\subsection{Among Graphs in $G(D)$}

A significantly more complicated situation arises when
constructing elements of the space $G(D)$, that
is, simple connected graphs having $n$ vertices and a particular degree
sequence $D$.
Even so, not all sequences $D$ will allow for the connection of $n$
vertices, i.e.~the set $G(D)$ may be empty.
In the language of discrete mathematics, one says that a
sequence of integers $\{d_1, d_2, \dots, d_n\}$ is {\it
graphical} if it satisfies the degree sequence of some
simple, connected graph, that is if $G(D)$ is nonempty.
One characterization of whether or not a sequence $D$ corresponds to a
simple, connected graph is due to Erd\"{o}s and Gallai~\cite{EG60}.

\begin{theorem}[Erd\"{o}s and Gallai~\cite{EG60}] \label{eq:erdos}
A sequence of positive integers $d_1, d_2, \dots, d_n$ with
$d_1 \geq d_2 \geq \dots \geq d_n$ is graphical if and only if
$\sum_{i=1}^n d_i$ is even and for each integer $k$, $1 \leq k \leq
n-1$,
\[ \sum_{j=1}^k d_j \leq k(k-1) + \sum_{j=k+1}^n \min(k,d_j). \]
\end{theorem}

As already noted, one possible problem is
that the sequence may have ``too many'' or ``too few'' degree-one
vertices.  For example, since the total number of links $l$ in any graph
will be equal to $l = \sum_{i=1}^n d_i /2$, a connected graph cannot
have an odd $\sum_{i=1}^n d_i$, but if this happens then adding or
subtracting a degree-one vertex to $D$ would ``fix'' this problem.
Theorem \ref{eq:erdos} further states that additional conditions are
required to ensure a simple connected graph, specifically that the
degree of any vertex cannot be ``too large''.  For example, the sequence
$\{10, 1, 1, 1\}$ cannot correspond to a simple graph.  We will not
attempt to explain all such conditions, except to note that
improvements have been made to Theorem \ref{eq:erdos} that reduce the
number of sufficient conditions to be checked~\cite{Tripathi2003} and also
that several algorithms have been developed to test for the existence
of a graph satisfying a particular degree sequence $D$ (e.g., see the
section on ``Generating Graphs'' in~\cite{skiena1997}).

Our approach to constructing the $s_{\max}$ element of $G(D)$ is via
a heuristic procedure that incrementally builds the network in a
greedy fashion, by iterating through the set of all potential links
$\mathcal{O} = \{(i,j): i<j; i,j=1,2,\dots,n\}$, which we order
according to decreasing values of $d_i d_j$.  In what follows we refer
to the value $d_i d_j$ as the {\it weight} of link $(i,j)$.
We add links from the ordered list of elements in $\mathcal{O}$
until all vertices have been added and the corresponding links satisfy
the degree sequence $D$. To facilitate the exposition of this
construction, we introduce the following notation.
Let $\mathcal{A}$ be the set of vertices that have been added to the
partial graph $\tilde{g}_\mathcal{A}$, such that
$\mathcal{B = V \backslash A}$ is the set of remaining vertices to be added.
At each stage of the construction, we keep track of the
{\it current degree} for vertex $i$, denoted $\tilde{d}_i$,
so that it may be compared with its {\it intended degree} $d_i$
(note that $\tilde{d}_i = 0$ for all $i \in \mathcal{B}$).
Define $\tilde{w}_i = d_i - \tilde{d}_i$ as the number of
remaining {\it stubs}, that is, the number of connections still to be
made to vertex $i$.  Note that values of $\tilde{d}_i$ and
$\tilde{w}_i$ will change during the construction process, while the
intended degree $d_i$ remains fixed.
For any point during the construction,
define $\tilde{w}_{\mathcal{A}} = \sum_{i \in \mathcal{A}} \tilde{w}_i$
to be the total number of remaining stubs in $\mathcal{A}$
and $d_{\mathcal{B}} = \sum_{i \in \mathcal{B}} d_i$
to be the total degree of the unattached vertices in $\mathcal{B}$.
The values $\tilde{w}_{\mathcal{A}}$ and  $d_{\mathcal{B}}$
are critical to ensuring that the final graph is connected and
has the intended degree sequence.  In particular, our algorithm
will make use of several conditions.

\vspace{1ex}

\noindent
{\bf Condition A-1: (Disconnected Cluster).}
If at any point during the incremental construction the partial graph
$\tilde{g}_\mathcal{A}$ has $\tilde{w}_\mathcal{A} = 0$ while
$|\mathcal{B}|>0$, then the final graph will be disconnected.

\vspace{1ex}

\noindent
{\bf Proof:} By definition $\tilde{w}_\mathcal{A}$ is the number of
stubs available in the partial graph $\tilde{g}_\mathcal{A}$.
If there are additional nodes to be added to the graph but
no more stubs in the partial graph, then any incremental growth
can occur only by forming an additional, separate cluster. \qed

\vspace{1ex}

\noindent
{\bf Condition A-1a: (Disconnected Cluster).}
If at any point during the construction algorithm the partial graph
$\tilde{g}_\mathcal{A}$ has $\tilde{w}_\mathcal{A} = 2$ with
$|\mathcal{B}|>0$, then adding a link between the two stubs in
$\tilde{g}_\mathcal{A}$ will result in a disconnected graph.

\vspace{1ex}

\noindent
{\bf Proof:} Adding a link between the two stubs will yield
$\tilde{w}_\mathcal{A} = 0$ with $|\mathcal{B}|>0$, thus resulting
in Condition A-1. \qed

\vspace{1ex}

\noindent
{\bf Condition A-2: (Tree Condition).}
If at any point during the construction
\begin{equation} \label{eq:treecond}
d_{\mathcal{B}} = 2|\mathcal{B}| - \tilde{w}_{\mathcal{A}},
\end{equation}
then the addition of all remaining vertices and links to the graph
must be {\it acyclic} (i.e., tree-like, without loops) in order to
achieve a single connected graph while satisfying the degree sequence.

\vspace{1ex}

\noindent {\bf Proof:}
To see this more clearly, suppose that for some intermediate point in
the construction process that $\tilde{w}_{\mathcal{A}} = m$.
That is, there are exactly $m$ remaining stubs in the connected
component to which the remaining vertices in $\mathcal{B}$ must attach.
We can prove that, in order to satisfy the degree sequence while
maintaining a single connected graph, each of these $m$ stubs must
become the root of a tree.
First, recall from basic graph theory that an acyclic graph connecting
$n$ vertices will have exactly $l = n-1$ links.
Define $\mathcal{B}_j \subset \mathcal{B}$ for $j=1, \dots, m$ to be
the subset of remaining vertices to be added to stub $j$, where
$\bigcup_{j=1}^m \mathcal{B}_j = \mathcal{B}$.  Further assume for the
moment that $\bigcap_{j=1}^m \mathcal{B}_j = \emptyset$, that is, each
vertex in $\mathcal{B}$ connects to a subgraph rooted at one and only
one stub.  Connecting the vertices in $\mathcal{B}_j$ to a subgraph
rooted at stub $j$ will require a
minimum of $|\mathcal{B}_j|$ links (i.e.~$|\mathcal{B}_j|-1$ links to
form a tree among the $|\mathcal{B}_j|$ vertices plus one additional
link to connect the tree to the stub).  Thus, in order
to connect the vertices in the set $\mathcal{B}_j$ as a tree rooted at
stub $j$, we require $\sum_{k \in \mathcal{B}_j} d_k = 2|\mathcal{B}_j|-1$,
and to attach all vertices in $\mathcal{B}$ to the $m$ stubs we have
\begin{eqnarray*}
d_{\mathcal{B}}
& = & \sum_{i \in \mathcal{B}} d_i
= \sum_{j=1}^m \sum_{k \in \mathcal{B}_j} d_k
\\
& = &  \sum_{j=1}^m \left( 2|\mathcal{B}_j|-1 \right) \nonumber \\
& = & 2|\mathcal{B}| - m \nonumber \\
& = & 2|\mathcal{B}| -  \tilde{w}_{\mathcal{A}}. \nonumber
\end{eqnarray*}
Thus, at the point when (\ref{eq:treecond}) occurs, only trees can be
constructed from the remaining vertices in $\mathcal{B}$.  \qed

\subsubsection*{The Algorithm}

Here, we introduce the algorithm for our heuristic construction and
then discuss the conditions when this construction is guaranteed to
result in the $s_{\max}$ graph.

\vspace{1ex}
\noindent
\begin{itemize}
\item
{\sc \underline{Step 0} (Initialization):} \\
Initialize the construction by adding vertex 1 to the partial graph;
that is, begin with $\mathcal{A} = \{ 1 \}$,
$\mathcal{B} = \{2,3, \dots, n\}$, and $\mathcal{O} = \{(1,2), \dots \}$.
Thus, $\tilde{w}_\mathcal{A} = d_1$ and $d_\mathcal{B} = \sum_{i=2}^n d_i$.

\item {\sc \underline{Step 1} (Link Selection)}: Check to see if there
  are any {\it admissible} elements in the ordered list $\mathcal{O}$.
  \renewcommand{\labelenumi}{(\alph{enumi})}
  \begin{enumerate}
  \item If $|\mathcal{O}|=0$, then {\sc Terminate}.  Return the
    graph $\tilde{g}_\mathcal{A}$. 

  \item If $|\mathcal{O}|>0$, select the element(s), denoted here
    as $(i,j)$, having the largest weight $d_i d_j$, noting that there
    may be more than one of them.  For each such
    link $(i,j)$, check $\tilde{w}_i$ and $\tilde{w}_j$:
    If either $\tilde{w}_i=0$ {\it or} $\tilde{w}_j=0$ then remove
    $(i,j)$ from $\mathcal{O}$.

  \item If no admissible links remain, return to {\sc Step 1}(a).

  \item Among all remaining links having {\it both} $\tilde{w}_i>0$ and
    $\tilde{w}_j>0$, select the element $(i,j)$ with the largest value
    $\tilde{w}_i$ (where for each $(i,j)$ $\tilde{w}_i$ is the {\it
    smaller} of $\tilde{w}_i$ and $\tilde{w}_j$), and proceed to
    {\sc Step 2}.
  \end{enumerate}

\item {\sc \underline{Step 2} (Link Addition)}: For the link $(i,j)$
  to be added, consider two types of connections.

  \begin{itemize}
  \item Type I: $i \in \mathcal{A}, j \in \mathcal{B}$.  Here, vertex $i$ is
    the highest-degree vertex in $\mathcal{A}$ with non-zero hubs
    (i.e., $d_i = \max_{k \in \mathcal{A}} d_k$ and $\tilde{w}_i > 0$)
    and $j$ is the highest-degree vertex in $\mathcal{B}$.
    Add link $(i,j)$ to the partial graph $\tilde{g}_\mathcal{A}$:
    remove vertex $j$ from $\mathcal{B}$ and add it to
    $\mathcal{A}$, decrement $\tilde{w}_i$ and $\tilde{w}_j$, and
    update both $\tilde{w}_\mathcal{A}$ and $d_\mathcal{B}$
    accordingly. Remove $(i,j)$ from the ordered list
    $\mathcal{O}$.

  \item Type II: $i \in \mathcal{A}, j \in \mathcal{A}, i \neq j$.  Here,
    $i$ and $j$ are the largest vertices in $\mathcal{A}$ for which
    $\tilde{w}_i > 0$ and $\tilde{w}_j > 0$.

    \begin{itemize}
      \item Check the {\it Tree Condition}: \\
    If $d_{\mathcal{B}} = 2|\mathcal{B}| - \tilde{w}_{\mathcal{A}},$
    then Type II links are not permitted.  Remove the link $(i,j)$
    from $\mathcal{O}$ {\it without adding it to the partial
    graph}.

      \item Check the {\it Disconnected Cluster Condition}: \\
    If $\tilde{w}_\mathcal{A} = 2$, then adding this link would
    result in a disconnected graph.  Remove the link
    $(i,j)$ from $\mathcal{O}$ {\it without adding it to the partial
    graph}.

      \item Else, add the link
    $(i,j)$ to the partial graph: decrement $\tilde{w}_i$ and
    $\tilde{w}_j$, and update $\tilde{w}_\mathcal{A}$ accordingly.
        Remove $(i,j)$ from the ordered list $\mathcal{O}$.

    \end{itemize}

  \end{itemize}

  Note: There is potentially a third case in which
  $i \in \mathcal{B}, j \in \mathcal{B}, i \neq j$; however this can
  only occur if there are no remaining stubs in the partial graph
  $\tilde{g}_\mathcal{A}$.  This is precluded by the test for the
  Disconnection Condition among Type II link additions; however if the
  algorithm were modified to allow this, then this third case would
  represent the situation where graph construction continues with a
  new (disconnected) cluster.  Adding link $(i,j)$ to the graph would
  require moving both vertices $i$ and $j$ from $\mathcal{B}$ to
  $\mathcal{A}$, decrementing $\tilde{w}_i$ and $\tilde{w}_j$,
  updating both $\tilde{w}_\mathcal{A}$ and $d_\mathcal{B}$
  accordingly, and removing $(i,j)$ from the ordered list
  $\mathcal{O}$.

\item {\sc \underline{Step 3} (Repeat)}: Return to {\sc Step 1}.

\end{itemize}

\noindent
Each iteration of the algorithm either adds a link from the list
in $\mathcal{O}$ or removes it from consideration.  Since there are a
finite number of elements in $\mathcal{O}$, the algorithm is
guaranteed to terminate in a finite number of steps.  Furthermore,
the ordered nature of $\mathcal{O}$ ensures the following property.

\vspace{1ex}

\noindent
{\bf Proposition A-3:}
At each point during the above construction, for any vertices $i \in
\mathcal{A}$ and $j \in \mathcal{B}$, $d_i \geq d_j$.

\vspace{1ex}

\noindent {\bf Proof:} By construction, if $i \in \mathcal{A}$
and $j \in \mathcal{B}$, then for some previously added vertex
$k \in \mathcal{A}$, it must have been the case that
$d_k d_i \geq d_k d_j$.  Since $d_k > 0$, it follows that $d_i \geq
d_j$. \qed

\vspace{1ex}

A less obvious feature of this construction is whether or not the
algorithm returns a simple connected graph satisfying degree sequence
$D$ (if one exists).  While this remains an open question, we show
that if the Tree Condition is ever reached, then the algorithm is
guaranteed to return a graph satisfying the intended degree sequence.

\vspace{1ex}

\noindent
{\bf Proposition A-4: (Tree Construction)}.
Given a graphic sequence $D$, if at {\it any} point during the above
algorithm the Tree Condition is satisfied, then
\renewcommand{\labelenumi}{(\alph{enumi})}
\setlength{\parskip}{0.5ex}
\begin{enumerate}
\addtolength{\itemsep}{-1ex}
\item the Tree Condition will remain satisfied through all
  intermediate construction, and
\item the final graph will exactly satisfy the intended degree
  sequence.
\end{enumerate}

\vspace{1ex}

\noindent {\bf Proof:} To show part (a), assume that
$d_{\mathcal{B}} = 2|\mathcal{B}| - \tilde{w}_{\mathcal{A}}$ and
observe that as a result only a link satisfying Type I can be
added next by our algorithm.  Thus, the next link $(i,j)$ to be added
will have $i \in \mathcal{A}$ and $j \in \mathcal{B}$, and in doing
so we will move vertex $j$ from the working set $\mathcal{B}$ to
$\mathcal{A}$.  As a result of this update, we will have
$\Delta d_\mathcal{B} = - d_j$, $\Delta |\mathcal{B}| = -1$,
and $\Delta \tilde{w}_\mathcal{A} = d_j -2$.  Thus, we have
updated the following values.
%
%
\begin{eqnarray*}
d_\mathcal{B}' & \equiv & d_\mathcal{B} + \Delta d_\mathcal{B} \\
& = & d_\mathcal{B} - d_j \\
\\
2|\mathcal{B}'| - \tilde{w}_{\mathcal{A}}' & \equiv &
2(|\mathcal{B}| + \Delta|\mathcal{B}|)
- (\tilde{w}_\mathcal{A} + \Delta \tilde{w}_\mathcal{A}) \\
& = & 2(|\mathcal{B}| -1) - (\tilde{w}_\mathcal{A} + d_j -2) \\
& = & 2|\mathcal{B}|  - \tilde{w}_\mathcal{A} - d_j \\
& = & d_\mathcal{B} - d_j
\end{eqnarray*}
Thus, $d_{\mathcal{B}}' = 2|\mathcal{B}'| - \tilde{w}_{\mathcal{A}}'$,
and the Tree Condition will continue to hold after the addition of
each subsequent Type I link $(i,j)$.

To show part (b), observe that after $|\mathcal{B}|$ Type I
link additions (each of which results in $\Delta |\mathcal{B}| = -1$)
the set $\mathcal{B}$ will be empty, thereby implying
also that $d_\mathcal{B} = 0$.  Since the relationship
$d_{\mathcal{B}} = 2|\mathcal{B}| - \tilde{w}_{\mathcal{A}}$
continues to hold after each Type I link addition,
then it must be that $|\mathcal{B}|=0$ and $d_\mathcal{B} = 0$
collectively imply $\tilde{w}_{\mathcal{A}} = 0$.  Furthermore,
since $\tilde{w}_{\mathcal{A}} = \sum_{i \in \mathcal{A}} \tilde{w}_i$
and $\tilde{w}_i = d_i - \tilde{d}_i \geq 0$ for all $i$, then
$\tilde{w}_i = 0$ for all $i$, and the degree sequence is satisfied.
\qed

\vspace{1ex}

An important question is under what conditions the Tree Condition is
met during the construction process.  Rewriting this condition as
$ d_{\mathcal{B}} - \left[ 2|\mathcal{B}| - \tilde{w}_{\mathcal{A}}
  \right] = 0$,
observe that when the algorithm is initialized in {\sc Step 0},
we have $d_\mathcal{B} = \sum_{i=2}^n d_i$,
$\tilde{w}_\mathcal{A} = d_1$ and that $|\mathcal{B}| = n-1$.
This implies that after initialization, we have
\[ d_{\mathcal{B}} - \left[ 2|\mathcal{B}| - \tilde{w}_{\mathcal{A}}
\right] = \sum_{i=2}^n d_i -2|\mathcal{B}| + d_1
=  \sum_{i=1}^n d_i - 2(n-1) \]
Note that minimal connectivity among $n$ nodes is achieved by a tree
having total degree $\sum_{i=1}^n d_i = 2(n-1)$, and this corresponds
to the case when the Tree Condition is met at initialization.
However, if the sequence $D$ is graphical and the Tree Condition is
not met at initialization, then
$d_{\mathcal{B}} - \left[ 2|\mathcal{B}| - \tilde{w}_{\mathcal{A}}
  \right] = 2z > 0$, where $z = \left(\sum_{i=1}^n d_i /2\right) - (n-1)$
is the number of ``extra'' links above what a tree would require.
Assuming $z > 0$, consider the outcome of subsequent
{\sc Link Addition} operations, as defined in {\sc Step 2}:
\begin{itemize}
\item As already noted, when a Type I connection is made (thus adding
  a new vertex $j$ to the graph), we have
  $\Delta d_{\mathcal{B}} = - d_j$,
  $\Delta \tilde{w}_{\mathcal{A}} = d_j - 2$, and
  $\Delta |\mathcal{B}| = -1$, which in turn means that
  Type I connections result in $\Delta \left( d_{\mathcal{B}} - \left[
  2|\mathcal{B}| - \tilde{w}_{\mathcal{A}} \right] \right) = 0$.

\item Accordingly, when a Type II connection is made between two stubs
  in $\mathcal{A}$, we have $\Delta \tilde{w}_{\mathcal{A}} = - 2$, and
  both $|\mathcal{B}|$ and $d_\mathcal{B}$ remain unchanged.  Thus,
  $\Delta \left( d_{\mathcal{B}} - \left[
  2|\mathcal{B}| - \tilde{w}_{\mathcal{A}} \right] \right) = -2$.
\end{itemize}
So if $d_{\mathcal{B}} - \left[ 2|\mathcal{B}| - \tilde{w}_{\mathcal{A}}
\right] = 2z > 0$, then subsequent link additions will cause this
value to either decrease by 2 or remain unchanged, or in other words,
adding additional links can only bring the algorithm closer to the
Tree Condition.
Nonetheless, our algorithm is {\it not} guaranteed to reach the Tree
Condition for all graphic sequences $D$
(i.e., we have not proved this), although we have not found any
counter-examples in which the algorithm fails to achieve the desired
degree sequence.
If that were to happen, however, the algorithm would
terminate with $\tilde{w}_i > 0$ for some vertex $i \in \mathcal{A}$,
even though $|\mathcal{B}|=0$.
Nonetheless, in the case where the graph resulting from our construction
does satisfy the intended degree sequence $D$, we can prove that it
is indeed the $s_{\max}$ graph.

\vspace{1ex}

\noindent
{\bf Proposition A-5: (General Construction)}.
If the graph $g$ resulting from our algorithm is a connected,
simple graph satisfying the intended degree sequence $D$, then this
graph is the $s_{\max}$ graph of $G(D)$.

\vspace{1ex}

\noindent
{\bf Proof:} Observe that, in order to satisfy the degree sequence
$D$, the graph $g$ contains a total of $l = \sum_{i=1}^n d_i/2$ links
from the ordered list $\mathcal{O}$.
Since elements of $\mathcal{O}$ are ordered by decreasing weight $d_i
d_j$, it is obvious that, in the absence of constraints that require
the final graph to be connected or satisfy the sequence $D$, a graph
containing the first $l$ elements of $\mathcal{O}$ will maximize
$\sum_{(i,j) \in \mathcal{E}} d_i d_j$.
However, in order to ensure that $g$ is an element of the space
$G(D)$, when selecting the $l$ links it is usually necessary to
``skip'' some elements of $\mathcal{O}$, and Conditions A-1 and A-2
identify two simple situations where skipping a potential link is required.
While skipping links under other conditions may be necessary to
guarantee that the resulting graph satisfies $D$ (indeed, the current
algorithm is not guaranteed to do this), our argument is that
{\it if these are the only conditions} under which elements of
$\mathcal{O}$ have been skipped during construction {\it and} the
resulting graph does satisfy $D$, then the resulting graph maximizes
$s(g)$.

To see this more clearly, consider a second graph $\tilde{g} \neq g$
also constructed from the ordered list $\mathcal{O}$.  Let
$\mathcal{E} \subset \mathcal{O}$ be the (ordered) list of links in
the graph $g$, and let $\tilde{\mathcal{E}} \subset \mathcal{O}$ be
the (ordered) list of links in the graph $\tilde{g}$.  Assume
that these two lists differ by only a single element, namely
$e \in \mathcal{E}, e \not\in \tilde{\mathcal{E}}$ and
$\tilde{e} \not\in \mathcal{E}, \tilde{e} \in \tilde{\mathcal{E}}$,
where
$\mathcal{E} \backslash e = \tilde{\mathcal{E}} \backslash \tilde{e}$.
By definition, both $e$ and $\tilde{e}$ are elements of $\mathcal{O}$,
and there are two possible cases for their relative position within
this ordered list (here, we use the notation ``$\prec$'' to mean
``proceeds in order'').
\begin{itemize}
\item
If $e \prec \tilde{e}$, then $\tilde{g}$ uses in place of $e$ a link
that occurs ``later'' in the sequence $\mathcal{O}$.  However, since
$\mathcal{O}$ is ordered by weight, using $\tilde{e}$ cannot result
in a higher value for $s(\tilde{g})$.
\item
If $\tilde{e} \prec e$, then $\tilde{g}$ uses in place of $e$ a link
that occurs ``earlier'' in the sequence $\mathcal{O}$---one that
had been ``skipped'' in the construction of $g$.  However,
the ``skipped'' elements of $\mathcal{O}$ will correspond to
instances of Conditions A-1 and A-2, and using them must necessarily result
in a graph $\tilde{g} \not\in G(D)$ because it is either disconnected
or because its degree sequence does not satisfy $D$.
\end{itemize}
Thus, for any other graph $\tilde{g}$, it must be the case that either
$s(\tilde{g}) \leq s(g)$ or $\tilde{g} \not\in G(D)$, and therefore
we have shown that $g$ is the $s_{\max}$ graph. \qed

\subsection{Among Connected, Acyclic Graphs}

In the special case when $\sum_{i=1}^n d_i = 2(n-1)$, there exists
only one type of graph structure that will connect all $n$ nodes,
namely an acyclic graph (i.e., a tree).  All connected acyclic graphs
are necessarily simple.
Because acyclic graphs are a special case of elements in $G(D)$,
generating $s_{\max}$ trees is achieved by making the appropriate
Type I connections in the aforementioned algorithm.  In effect, this
construction is essentially a type of deterministic preferential
attachment, one in which we iterate through all vertices in the
ordered list $D$ and attach each to the highest-degree vertex with
a remaining stub.

In the case of trees, the arguments underlying the $s_{\max}$ proof
can be made more precise.
Observe that the incremental construction of a tree
is equivalent to choosing for each vertex in $\mathcal{B}$ the single
vertex in $\mathcal{A}$ to which it becomes attached.  Consider the
choices available for connecting two vertices $k,m \in \mathcal{B}$
to vertices $i,j \in \mathcal{A}$
where $d_i \geq d_j$, $d_k \geq d_m$, and observe that
\( d_i d_k + d_i d_m \geq d_i d_k + d_j d_m \geq d_j d_k + d_i d_m
\geq d_j d_k + d_j d_m,\)
where second inequality follows from Proposition 3 while
the first and last inequalities are by assumption.
There are two cases of interest.
First, if $\tilde{w}_i > 1$ and $\tilde{w}_j \geq 1$, then it is clear
that it is optimal to connect {\it both} vertices $k,m \in
\mathcal{B}$ to vertex $i \in \mathcal{A}$.
Second, if $\tilde{w}_i = 1$ and $\tilde{w}_j \geq 1$, then it is clear
that it is optimal to connect
$k \in \mathcal{B}$ to $i \in \mathcal{A}$ and
$m \in \mathcal{B}$ to $j \in \mathcal{A}$.
All other scenarios can be decomposed into these two cases, thus
proving that the algorithm's incremental construction for a tree is
guaranteed to result in the $s_{\max}$ graph.

There are many important properties of $s_{\max}$ trees that are
discussed in Section~\ref{sec:struct}, which we now prove.

\subsubsection{Properties of $s_{\max}$ Acyclic Graphs}

Recall that our working definition of so-called {\it betweenness}
(also known as {\it betweenness centrality}) for a vertex
$v \in \mathcal{V}$ in an acyclic graph is given by
\[ C_b(v) = \frac{\sum_{s < t \in \mathcal{V}} \sigma_{st}(v)}
                 {\sum_{s < t \in \mathcal{V}}\sigma_{st}}
= \frac{ \bar{\sigma} (v) } { n(n-1)/2 },
\]
where we use the notation $\bar{\sigma}(v)$ to denote the
number of unique paths in the graph passing through node
$v$, and where the total number of unique paths between vertex
pairs $s$ and $t$ is $n(n-1)/2$.

For a given node $v \in \mathcal{V}$, let $\mathcal{N}(v)$
denote the set of neighboring nodes, where by definition
$|\mathcal{N}(v)| = d_v$.  For all nodes that are not the
root of the tree, exactly one of these neighbors will be
``upstream'' while the rest will be ``downstream''
(in contrast, the root node has only downstream neighbors).
Define $b_j$ to be the total number of nodes ``connected''
through the $j^{th}$ neighbor.  Our convention will be to
denote the ``upstream'' neighbor with index 0 (if it exists);
thus for all nodes $v$ other than the root, one has
$\sum_{j=0}^{d_v-1} b_j = n-1$ (for the root node $r$, the
appropriate summation is $\sum_{j=1}^{d_r} b_j = n-1$).
Using this notation, it becomes clear that, for each node $v$
other than the root of the tree, we can express
\[
\bar{\sigma}(v) = \sum_{j,k=0 \atop j<k}^{d_v-1} b_j b_k
=  b_0 \sum_{k=1}^{d_v-1} b_k +
             \sum_{{j,k=1} \atop {j < k}}^{d_v-1} b_j b_k. \]
Thus, $\bar{\sigma}(v)$ decomposes into two components:
the first measures the number of paths between upstream
and downstream nodes that pass through node $v$, and
the second measures the number of paths passing through
node $v$ that are between downstream nodes only.
Since $\sum_{s < t \in \mathcal{V}} \sigma_{st}$ is a
constant for trees containing $n$
nodes, when comparing the centrality for two nodes $u$ and $v$,
we work directly with $\bar{\sigma}(u)$ and $\bar{\sigma}(v)$.
In so doing, for nodes $u$ and $v$ we will denote $b_j^u$, $b_j^v$
as the number of nodes connected to each via their respective
$j^{th}$ neighbor.

One property of the $s_{\max}$ graph that will be useful
for showing that there exists monotonicity between node
centrality and node degree is given by the following Lemma.
\begin{lemma} \label{lemma:smax}
Let $g$ be the $s_{\max}$ acyclic graph for degree sequence $D$,
and consider two nodes $u,v \in \mathcal{V}$ satisfying $d_u > d_v$.
Then, it necessarily follows that
\begin{equation}
\sum_{j,k=1 \atop j<k}^{d_u-1} b_j^u b_k^u >
\sum_{j,k=1 \atop j<k}^{d_v-1} b_j^v b_k^v.
\end{equation}
\end{lemma}
\noindent
Note that the summation is over {\it downstream} nodes only,
thus Lemma~\ref{lemma:smax} states that, for $s_{\max}$
trees, the contribution to centrality from paths between
downstream nodes is greater for nodes with higher degree.

\vspace{1ex}

\noindent
{\bf Proof of Lemma~\ref{lemma:smax}:} Recalling from
Proposition~\ref{prop:smax-tree-ordering} that $b_j^u \geq b_j^v$
for all $j=1, 2, \dots, d_v-1$, and noting that $d_u > d_v$,
\begin{eqnarray*}
\sum_{j,k=1 \atop j<k}^{d_u-1} b_j^u b_k^u & = &
\sum_{j,k=1 \atop j<k}^{d_v-1} b_j^u b_k^u
+ \sum_{j=1}^{d_v-1} \sum_{k=d_v}^{d_u-1}
b_j^u b_k^u
+ \sum_{j,k=d_v \atop j<k}^{d_u-1} b_j^u b_k^u \\
& > &
\sum_{j,k=1 \atop j<k}^{d_v-1} b_j^v b_k^v
+ \sum_{j=1}^{d_v-1} \sum_{k=d_v}^{d_u-1}
b_j^u b_k^u +
\sum_{j,k=d_v \atop j<k}^{d_u-1} b_j^u b_k^u \\
& > &
\sum_{j,k=1 \atop j<k}^{d_v-1} b_j^v b_k^v.
\end{eqnarray*}
Thus, the proof is complete. \qed

\vspace{2ex}

Lemma~\ref{lemma:smax} in turn facilitates a proof of the
more general statement regarding the centrality
of nodes in the $s_{\max}$ acyclic graph, as stated in
Proposition~\ref{prop:centrality-tree}.

\begin{figure*}[t]
  \begin{center}
  \includegraphics[width=\linewidth]{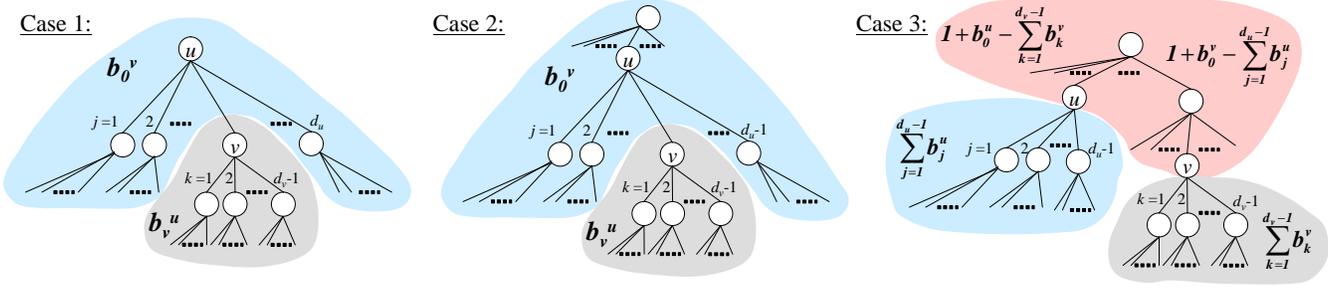}
  \caption{\footnotesize   \label{fig:centrality}
   {Centrality of high-degree nodes in the $s_{\max}$ tree.}
  }
  \end{center}
\vspace{-4ex}
\end{figure*}

\vspace{1ex}
\noindent
{\bf Proof of Proposition~\ref{prop:centrality-tree}:}
We proceed in two parts.  First, we show that if
node $v$ is downstream from node $u$, then
$\bar{\sigma}(u) > \bar{\sigma}(v)$.  Second, we show that
if $v$ is in a different branch of the tree from $u$
(i.e., neither upstream nor downstream from $u$) but
$d_u > d_v$, then $\bar{\sigma}(u) > \bar{\sigma}(v)$.

Starting first with the scenario where $v$ is downstream from $u$,
there are two cases that need to be addressed.

\vspace{1ex}
\noindent \underline{Case 1:} node $v$ is directly downstream
from node $u$, and node $u$ {\it is the root of the tree}.
Observe that we can represent $\bar{\sigma}(v)$ as
\begin{eqnarray}
\bar{\sigma}(v) & = & b_0^v \sum_{k=1}^{d_v-1} b_k^v +
      \sum_{{j,k=1} \atop {j<k}}^{d_v-1} b_j^v b_k^v \nonumber \\
& = & \bigg( \sum_{j=1 \atop j \neq v}^{d_u} b_j^u \bigg)
      \bigg( b_v^u - 1 \bigg)
    + \sum_{j,k=1 \atop j<k}^{d_v-1} b_j^v b_k^v, \ \ \ \ \label{eq:2}
\end{eqnarray}
since $b_0^v = \sum_{j=1;j \neq v}^{d_u} b_j^u$ and
also that $b_v^u = 1 + \sum_{k=1}^{d_v-1} b_k^v$.
For node $u$, we have
\begin{eqnarray}
\bar{\sigma}(u) & = & \sum_{{j,k=1} \atop {j<k}}^{d_u} b_j^u b_k^u
\nonumber \\
& = & b_v^u \sum_{k=1 \atop k \neq v}^{d_u} b_k^u +
             \sum_{j,k=1 \atop j<k; j,k\neq v}^{d_u} b_j^u b_k^u.
         \label{eq:3}
\end{eqnarray}
Comparing $\bar{\sigma}(u)$ and $\bar{\sigma}(v)$, we observe
that the first term of (\ref{eq:3}) is clearly greater than
the first term of (\ref{eq:2}).  Furthermore, by
Lemma~\ref{lemma:smax}, we also observe that the second term
of (\ref{eq:3}) is also greater than the second term of (\ref{eq:2}).
Thus, we conclude for this case that $\bar{\sigma}(u) > \bar{\sigma}(v)$.

\vspace{1ex}
\noindent \underline{Case 2:} node $v$ is directly downstream
from node $u$, but node $u$ {\it is not} the root of the tree.
Recognizing for any node $i$ that
$\sum_{j=1}^{d_i-1} b_j = (n-1) - b_0$,
we write
\begin{eqnarray*}
\bar{\sigma}(u) & = & b_0^u \big( n - 1 - b_0^u \big) +
             \sum_{j,k=1 \atop j<k}^{d_u-1} b_j^u b_k^u \\
\bar{\sigma}(v) & = & b_0^v \big( n - 1 - b_0^v \big) +
             \sum_{j,k=1 \atop j<k}^{d_v-1} b_j^v b_k^v
\end{eqnarray*}
As before, we observe from Lemma~\ref{lemma:smax} that
\( \sum_{j,k=1;j<k}^{d_u-1} b_j^u b_k^u >
\sum_{j,k=1;j<k}^{d_v-1} b_j^v b_k^v,  \)
so proving that $\bar{\sigma}(u) > \bar{\sigma}(v)$ in this case
requires simply that we show
\begin{equation} \label{eq:4}
b_0^u \Big( (n - 1) - b_0^u \Big) > b_0^v \Big( (n - 1) - b_0^v \Big).
\end{equation}
Observe that $b_0^v = b_0^u + 1 + \sum_{j=1;j \neq v}^{d_u-1} b_j^u$.
As a result, we have
\begin{eqnarray*}
&& b_0^v \Big((n - 1) - b_0^v \Big) \\
&= & \Big( b_0^u + 1 + \sum_{j=1 \atop j \neq v}^{d_u-1} b_j^u \Big) \\
&& \Big((n - 1) - b_0^u - 1 - \sum_{j=1 \atop j \neq v}^{d_u-1} b_j^u \Big)\\
& = & b_0^u \Big((n-1) - b_0^u \Big)
+ \Big(1 + \sum_{j=1 \atop j \neq v}^{d_u-1} b_j^u \Big) \\
&& \bigg((n-1) - 2 b_0^u - \Big(1 + \sum_{j=1 \atop j \neq v}^{d_u-1} b_j^u
\Big) \bigg)
\end{eqnarray*}
Since $1 + \sum_{j=1;j \neq v}^{d_u-1} b_j^u > 0$,
(\ref{eq:4}) is true if and only if
\begin{eqnarray*}
(n-1) - 2 b_0^u - \Big(1 + \sum_{j=1 \atop j \neq v}^{d_u-1} b_j^u \Big)
& < & 0
\end{eqnarray*}
which is equivalent to
\begin{eqnarray*}
(n-1) - b_0^u & < & b_0^u + 1 + \sum_{j=1 \atop j \neq v}^{d_u-1}
  b_j^u \\
\sum_{k=1}^{d_u-1} b_k^u & < & b_0^u + 1 + \sum_{j=1;j \neq v}^{d_u-1}
  b_j^u \\
b_v^u & < & b_0^u + 1.
\end{eqnarray*}
This final statement will always be true for the $s_{\max}$ tree,
since the ``upstream'' branch from node $u$ will always contain
at least as many nodes as the downstream branch corresponding to node
$v$.

These two cases prove that any ``upstream'' node in the $s_{\max}$
tree is always more central than any ``downstream'' node, since
by extension if $u$ is directly upstream from $v$ then
$\bar{\sigma}(u) > \bar{\sigma}(v)$, and if
$v$ is directly upstream from $w$ then
$\bar{\sigma}(v) > \bar{\sigma}(w)$.  It therefore follows
that $\bar{\sigma}(u) > \bar{\sigma}(w)$, and,
by induction, that the ``root'' node of the $s_{\max}$ tree
(having highest degree) is the most central within the entire
tree.

\vspace{1ex}
\noindent \underline{Case 3:}
Now we turn to the case where node $v$ is not directly downstream
(or upstream) from node $u$.  As before, we write
\begin{eqnarray*}
\bar{\sigma}(u) & = & b_0^u \sum_{k=1}^{d_u-1} b_k^u +
             \sum_{j,k=1;j<k}^{d_u-1} b_j^u b_k^u, \\
\bar{\sigma}(v) & = & b_0^v \sum_{k=1}^{d_v-1} b_k^v +
             \sum_{j,k=1;j<k}^{d_v-1} b_j^v b_k^v.
\end{eqnarray*}
As with the previous cases, by Lemma~\ref{lemma:smax} we know that
\( \sum_{j,k=1;j<k}^{d_u-1} b_j^u b_k^u >
\sum_{j,k=1;j<k}^{d_v-1} b_j^v b_k^v,  \)
so proving that $\bar{\sigma}(u) > \bar{\sigma}(v)$ in this case
requires simply that we show that
\begin{equation} \label{eq:5}
b_0^u \sum_{k=1}^{d_u-1} b_k^u > b_0^v \sum_{k=1}^{d_v-1} b_k^v.
\end{equation}
We rewrite each of these as
\begin{eqnarray*}
b_0^u & = & \sum_{j=1}^{d_v -1} b_j^v + \Big( b_0^u - \sum_{j=1}^{d_v -1}
b_j^v \Big)\\
b_0^v & = & \sum_{j=1}^{d_u -1} b_j^u + \Big( b_0^v - \sum_{j=1}^{d_u -1}
b_j^u \Big)
\end{eqnarray*}
so that we have
\begin{eqnarray*}
b_0^u \sum_{k=1}^{d_u-1} b_k^u
& = & \bigg(\sum_{j=1}^{d_v -1} b_j^v + \Big( b_0^u - \sum_{j=1}^{d_v -1}
b_j^v \Big) \bigg) \sum_{k=1}^{d_u-1} b_k^u \\
b_0^v \sum_{k=1}^{d_v-1} b_k^v
& = & \bigg(\sum_{j=1}^{d_u -1} b_j^u + \Big( b_0^v - \sum_{j=1}^{d_u -1}
b_j^u \Big)\bigg) \sum_{k=1}^{d_v-1} b_k^v
\end{eqnarray*}
and observe that
\[ b_0^u - \sum_{j=1}^{d_v -1} b_j^v = b_0^v - \sum_{j=1}^{d_u -1} b_j^u,
\]
which is a non-negative constant, that we denote $\kappa$.  Thus,
\begin{eqnarray*}
b_0^u \sum_{j=1}^{d_u-1} b_j^u - b_0^v \sum_{j=1}^{d_v-1} b_j^v
& = & \kappa
\bigg( \sum_{j=1}^{d_u-1} b_j^u  - \sum_{j=1}^{d_v-1} b_j^v \bigg),
\end{eqnarray*}
which is also non-negative since
\( \sum_{j=1}^{d_u-1} b_j^u  > \sum_{j=1}^{d_v-1} b_j^v, \)
and so (\ref{eq:5}) also holds.  Thus, we have shown that
$\bar{\sigma}(u) > \bar{\sigma}(v)$ in the $s_{\max}$ tree whenever
$d_u > d_v$, thus completing the proof. \qed

\section{The $s(g)$-Metric and Assortativity}
\label{sec:assort-appendix}

Following the development of Newman~\cite{Newman02},
let $P(\{D_i = k\})=P(k)$ be the node degree
distribution over the ensemble of graphs and define $Q(k) =
(k+1)P(k+1)/\sum_{j\in D} j P(j)$ to be the normalized
distribution of {\it remaining degree} (i.e., the number of
``additional'' connections for each node at either end of
the chosen link). Let $\bar{D}=\{d_1-1, d_2-2, \cdots,
d_n-1\}$ denote the remaining degree sequence for $g$. This
remaining degree distribution is $Q(k) = \sum_{k'\in
\bar{D}} Q(k,k')$, where $Q(k,k')$ is the {\it joint
probability distribution} among remaining nodes, i.e.,
$Q(k,k') = P(\{D_i= k+1, D_j = k'+1| (i,j) \in
\mathcal{E}\})$.
In a network where the remaining degree of any two vertices
is independent, i.e. $Q(k,k')=Q(k)Q(k')$, there is no
degree-degree correlation, and this defines a network that
is neither assortative nor disassortative (i.e., the
``center'' of this view into the ensemble).
In contrast, a network with $Q(k,k') = Q(k) \delta[k - k']$
defines a perfectly assortative network.
Thus, graph assortivity $r$ is quantified by the {\it
average} of $Q(k,k')$ over all the links
\begin{equation} r = \frac{\sum_{k,k'\in \bar{D}} k k' (Q(k,k')-Q(k)Q(k'))}
{\sum_{k,k' \in \bar{D}} k k' (Q(k)\delta[k -
k']-Q(k)Q(k'))}, \label{assort_sto}
\end{equation}
with proper centering and normalization according to the
value of perfectly assortative network, which ensures that
$-1 \leq r \leq 1$. Many stochastic graph generation
processes can be understood directly in terms of the
correlation distributions among these so-called remaining
nodes, and this functional form facilitates the direct
calculation of their assortativity. In particular,
Newman~\cite{Newman02} shows that both Erd\"{o}s-Reny\'{i}
random graphs and Barab\'{a}si-Albert preferential
attachment growth processes yield ensembles with zero
assortativity.

Newman~\cite{Newman05} also develops the following
sample-based definition of assortativity
\[ r(g) =
\frac{\left[  \sum_{(i,j)\in \mathcal{E}}d_i d_j / l
\right] - \left[ \sum_{(i,j)\in \mathcal{E}}
\frac{1}{2}(d_i+ d_j) / l \right]^2}
 {\left[  \sum_{(i,j)\in \mathcal{E}} \frac{1}{2}(d_i^2+ d_j^2) / l
\right] - \left[  \sum_{(i,j)\in \mathcal{E}}
\frac{1}{2}(d_i+ d_j) / l \right]^2 } , \label{assort_tmp}
\]
which is equivalent to (\ref{assort_str}).

While the ensemble-based notion of assortativity in
(\ref{assort_sto}) has important differences from the
sample-based notion of assortativity in (\ref{assort_str}),
their relationship can be understood by viewing a given
graph as a singleton on an ensemble of graphs (i.e., where
the graph of interest is chosen with probability 1 from the
ensemble).  For this graph, if we define the number of
nodes with degree $k$ as $N(k)$, we can derive the degree
distribution $P(k)$ and the remaining degree distribution
$Q(k)$ on the ensemble as
\[ P(k)=\frac{N(k)}{n} \]
and
\[ Q(k)=\frac{(k+1)P(k+1)}{\sum_{j\in D} jP(j)}
=\frac{(k+1)N(k+1)}{\sum_{j\in D} jN(j)}.
\]
Also it is easy to see that
\begin{eqnarray*}
\sum_{i\in \mathcal{V}} d_i& = & \sum_{k \in D} k N(k) =2l, \\
\sum_{i\in \mathcal{V}} d_i^2 & = & \sum_{k \in D} k^2  N(k), \\
& \vdots & \\
\sum_{i\in \mathcal{V}} d_i^m & = & \sum_{k \in D} k^m
N(k),
\end{eqnarray*}
where $m$ is a positive integer.

Equations (\ref{assort_sto}) and (\ref{assort_str}) can be
related term-by-term in the following manner.  The first term of
the numerator, $Q(k,k')$, represents the joint probability
distribution of the (remaining) degrees of the two nodes at
either end of a randomly chosen link. For a given graph,
let $l(k,k')$ represent the number of links connecting
nodes with degree $k$ to nodes with degree $k'$.  Then, we
can write \( Q(k,k')= l(k,k') / l \), and hence
\[
\sum_{k,k' \in \bar{D}} k k' Q(k,k') = \frac{1}{l}
\sum_{(i,j)\in \mathcal{E}} d_i d_j.
\]
\noindent The first term of the denominator of $r$ in
equation (\ref{assort_sto}) can be written as
\begin{eqnarray}
\sum_{k,k' \in \bar{D}} k k' Q(k)\delta[k - k']
\label{eq:ens1}
& = & \sum_{k \in \bar{D}} k^2 Q(k) \\
&= & \frac{\sum_{k \in D} (k+1)^3 N(k+1)}{\sum_{i\in D}
    jN(j)} \nonumber \\
& =& \frac{\sum_{i \in \mathcal{V}} d_i^3 }{2l},
\label{eq:str1}
\end{eqnarray}
and the ``centering''  term (in both the numerator and the
denominator) is
\begin{eqnarray}
\sum_{k,k' \in \bar{D}} k k' Q(k) Q(k')
& = & \left( \sum_{k\in \bar{D}} k Q(k)\right)^2 \label{eq:ens2} \\
& = & \left(
    \frac{\sum_{k \in D} (k+1)^2 N(k+1)}{\sum_{i\in D} jN(j)}
    \right)^2 \nonumber \\
& =& \left( \frac{\sum_{i \in \mathcal{V}} d_i^2
}{2l}\right)^2. \label{eq:str2}
\end{eqnarray}
In both of these cases, the offset of a constant in
representing the degree sequence as  $D$ versus $\bar{D}$
does not effect the overall calculation. The relationships
between the ensemble-based quantities (LHS of
\ref{eq:ens1}) and (LHS of \ref{eq:ens2}) and their
sample-based (i.e., structural) counterparts
(\ref{eq:str1}) and (\ref{eq:str2}) holds (approximately)
when the expected degree equals the actual degree.

To see why (\ref{eq:str2}) can be viewed as the ``center'',
we consider the following thought experiment: {\it what is
the structure of a deterministic graph with degree sequence
$D$ and having zero assortativity?}  In principle, a node
in such a graph will connect to any other node in
proportion to each node's degree.
While such a graph may not exist for general $D$, one can
construct a deterministic {\it pseudograph} $\tilde{g}$
having zero assortativity in the following manner.  Let $A
= [a_{ij}]$ represent a (directed) node adjacency matrix of
non-negative real values, representing the ``link weights''
in the pseudograph. That is, links are not constrained to
integer values but can exist in fractional form.  The zero
assortative pseudograph will have symmetric weights given
by
\[ a_{ij} = \left( \frac{d_j}{\sum_{k \in \mathcal{V}} d_k} \right)
\left( \frac{d_i}{2} \right) = \left( \frac{d_i}{\sum_{k
\in \mathcal{V}} d_k} \right) \left( \frac{d_j}{2} \right)
= a_{ji}. \] Thus, the weight $a_{ij}$ for each link
emanating out of node $i$ is in proportion to the degree of
node $j$, in a manner that is relative to the sum of all
node degrees. In general, the graphs of interest to us are
undirected, however here it is notationally convenient to
consider the construction of directed graphs.  Using these
weights, the total weight among all links entering and
exiting a particular node $i$ equals
\[
\sum_{j \in \mathcal{V}} a_{ij} + \sum_{k \in \mathcal{V}}
a_{ki} = d_i /2 + d_i /2 = d_i. \] Accordingly, the total
``link weights'' in the pseudograph are equal to
\[ \sum_{i,j \in \mathcal{V}} a_{ij} =
\sum_{j \in \mathcal{V}} d_j/2 =  l, \] where $l$
corresponds to the total number of links in a traditional
graph. The $s$-metric for the pseudograph $\tilde{g}_A$
represented by matrix $A$ can be calculated as
\begin{eqnarray*}
s(\tilde{g}_A) & = & \sum_{j \in \mathcal{V}} \sum_{i \in
\mathcal{V}}
d_i a_{ij} d_j \\
& = & \sum_{j \in \mathcal{V}} \left[ \sum_{i \in
\mathcal{V}} d_i \left( \frac{d_j}{\sum_{k \in \mathcal{V}}
d_k} \right)
\left( \frac{d_i}{2} \right) \right] d_j\\
& = & \frac{ \Big( \sum_{j \in \mathcal{V}} d_j^2 \Big)
\Big( \sum_{i \in \mathcal{V}} d_i^2 \Big)}
{2 \Big( \sum_{k \in \mathcal{V}} d_k \Big) } \\
& = & \frac{ \Big( \sum_{j \in \mathcal{V}} d_j^2 \Big)^2}
{4 l },
\end{eqnarray*}
and we have
\begin{eqnarray*}
\frac{s(\tilde{g}_A)}{l} & = & \left( \frac{\sum_{i \in
\mathcal{V}} d_i^2 }{2l}\right)^2,
\end{eqnarray*}
which is equal to (\ref{eq:str2}).

In principle, one could imagine a deterministic procedure
that uses the structural pseudograph $\tilde{g}_A$ to
generate the zero assortativity graph among an
``unconstrained'' background set $G$.  That is, graphs
resulting from this procedure could have multiple links
between any pair of nodes as well as multiple self-loops
and would not necessarily be connected.  The challenge in
developing such a procedure is to ensure that the resulting
graph has degree sequence equal to $D$, although one can
imagine that in the limit of large graphs this becomes less
of an issue. By extension, it is not hard to conceive a
stochastic process that uses the structural pseudograph
$\tilde{g}_A$ to generate a statistical ensemble of graphs
having expected assortativity equal to zero. In fact, it is
not hard to see why the GRG method is very close to such a
procedure.

Note that the total weight in the pseudograph between nodes
$i$ and $j$ equals $a_{ij} + a_{ji} = d_i d_j / 2l$.
Recall from Section~\ref{sec:likelihood} that
the GRG method described is based on the choice of a
probability $p_{ij} = \rho d_i d_j$ of connecting two
nodes $i$ and $j$, and also that in order to ensure that
$E(d_i) = d_i$ one needs $\rho = 1 /2l$, provided that
$\max_{i \neq j \in \mathcal{V}} d_i d_j \leq 2l$.
Thus, the GRG method can be
viewed as a stochastic procedure that generates real graphs
from the pseudograph $\tilde{g}_A$, with the one important
difference that the GRG method always results in simple
(but not necessarily connected) graphs.
Thus, the zero assortativity pseudograph $\tilde{g}_A$ can
be interpreted as the ``deterministic outcome'' of a
GRG-like construction method.  Accordingly, one expects
that the statistical ensemble of graphs resulting from the
stochastic GRG method could have zero assortativity, but
this has not been proven.

In summary, graph assortativity captures a fundamental
feature of graph structure, one that is closely related to
our $s$-metric. However, the existing notion of
assortativity for an individual graph $g$ is implicitly
measured against a background set of graphs $G$ that is
{\it not} constrained to be either simple or connected.
The connection between the sample-based and ensemble-based
definitions makes it possible to calculate the
assortativity among graphs of different sizes and having
different degree sequences, as well as for different graph
evolution procedures. Unfortunately, because this metric is
computed relative to an unconstrained background set, in
some cases this normalization (against the $s_{\max}$
graph) and centering (against the $\tilde{g}_A$
pseudograph) does a relatively poor job of distinguishing
among graphs having the {\it same} degree sequence, such as
those in Figure~\ref{fig:toynet}.


\end{multicols}

\end{document}